# Ternary superhydrides under pressure of Anderson's theorem: Near-record superconductivity in (La, Sc)$H_{12}$

Dmitrii V. Semenok[1, †, *], Ivan A. Troyan[2, †], Di Zhou[1, †, *], Andrei V. Sadakov[3], Kirill S. Pervakov[3], Oleg A. Sobolevskiy[3], Anna G. Ivanova[2], Michele Galasso[4], Frederico Gil Alabarse[5], Wuhao Chen[6,7], Chuanying Xi[8], Toni Helm[9], Sven Luther[9], Vladimir M. Pudalov[3,10] and Viktor V. Struzhkin[11,*]

[1] *Center for High Pressure Science & Technology Advanced Research, Bldg. #8E, ZPark, 10 Xibeiwang East Rd, Haidian District, Beijing, 100193, China*

[2] *A.V. Shubnikov Institute of Crystallography of the Kurchatov Complex of Crystallography and Photonics, 59 Leninsky Prospekt, Moscow 119333, Russia*

[3] *V. L. Ginzburg Center for High-Temperature Superconductivity and Quantum Materials, Moscow, 119991 Russia*

[4] *Institute of Solld State Physics, University of Latvla, 8 Kengaraga str., LV-1063 Riga, Latvla*

[5] *Elettra Sincrotrone Trieste, Strada Statale 14 km163,5 in AREA Science Park, Basovizza, Trieste 34149, Italy*

[6] *Department of Physics, Southern University of Science and Technology, Shenzhen 518055, China*

[7] *Quantum Science Center of Guangdong–Hong Kong–Macao Greater Bay Area (Guangdong), Shenzhen, China*

[8] *Anhui Province Key Laboratory of Condensed Matter Physics at Extreme Conditions, High Magnetic Field Laboratory of the Chinese Academy of Science, Hefei 230031, Anhui, China.*

[9] *Hochfeld-Magnetlabor Dresden (HLD-EMFL) and Würzburg-Dresden Cluster of Excellence, Helmholtz-Zentrum Dresden-Rossendorf (HZDR), Dresden 01328, Germany*

[10] *National Research University Higher School of Economics, Moscow 101000, Russia*

[11] *Shanghai Key Laboratory of Material Frontiers Research in Extreme Environments (MFree), Shanghai Advanced Research in Physical Sciences (SHARPS), Pudong, Shanghai 201203, China*

*Corresponding authors: Dmitrii Semenok (dmitrii.semenok@hpstar.ac.cn), Di Zhou (di.zhou@hpstar.ac.cn) and Viktor Struzhkin (viktor.struzhkin@hpstar.ac.cn).

†These authors contributed equally.

## Abstract

Lanthanum-hydrogen system and its derivatives remain among the most promising for achieving room-temperature superconductivity. In this study, we examined the formation of ternary lanthanum-scandium superhydrides at pressures up to 220 GPa. The primary product of the LaSc alloy's reaction with hydrogen is newly discovered cubic (La,Sc)$H_{12}$, demonstrating clear superconducting transition in all six channels of the van der Pauw scheme at 244-248 K. In this compound with an unusually large unit cell volume, virtually no magnetoresistance was observed in fields up to 68 Tesla. Synthesized samples of (La,Sc)$H_{12}$ demonstrate pronounced superconducting diode and SQUID-like effects at a record high temperature of 233 K. Furthermore, our analysis revealed the possible formation of lower hexagonal polyhydrides (La,Sc)$H_{6-7}$, which could potentially account for the drop in electrical resistance observed near 274 K. This anomaly between 265-290 K also appears in the radio-frequency transmission measurements and may be of a superconducting nature.

## 1. Introduction

Pressure-stabilized polyhydrides are a new rapidly growing class of high-temperature superconductors [1], [2]. Remarkable properties of $H_3S$ (critical temperature, $T_c$ = 200 K) [3], $YH_6$ ($T_c$ = 224 K) [4] and $LaH_{10}$ ($T_c$ = 250 K) [5,6] at 130-200 GPa catalyzed the search for superconductivity (SC) in compressed ternary (X,Y)-H polyhydrides that can be obtained by pulsed laser heating of various alloys and intermetallics with hydrogen or ammonia borane (AB, $NH_3BH_3$) in diamond anvil cells (DACs). The uniquely high critical temperatures, upper critical magnetic fields up to 300 T [7] and critical current densities of superhydrides are very attractive for the creation of new electronic devices based on high-pressure DACs. The general and as yet unsolved problem of ternary superhydrides boils down to the question: "Can $T_c$ in ternary polyhydrides be higher than in binary ones and reach room-temperature values?". As the analysis of the thermodynamic stability of the SC state shows, from a fundamental point of view there is no reason why room-temperature superconductivity cannot be achieved in compressed polyhydrides [8-10].

Anderson's theorem [11,12] exerts considerable pressure on the experimental search for new high-temperature superconductors in (pseudo)ternary and high-entropy[7] hydride systems, which are perhaps its best illustration[13]. The problem is that the synthesis of hydrides via the laser heating inevitably leads to random mixing of heavy atoms in their sublattice with the formation of solid solutions [14]. A rather long list of already experimentally investigated ternary systems can be given, where the introduction of a third element just slightly changes the critical temperature: $(La,Y)H_{10}$ [14], $(La,Y)H_4$ [15], $(La,Ca)H_{10}$ [16], $(Y,S)H_9$ and $(Y,S)H_6$ [17], $(La,Al)H_{10}$ [18], $(La,Be)H_8$ [19]... Moreover, in the case of introduction of atoms with *f*-electrons a significant decrease of $T_c$ is observed: $(La,Nd)H_{10}$ [13], $(Y,Ce)H_x$ [20] and $(La,Ce)H_{9\text{-}10}$ [21,22] serve as examples. The recent synthesis of the ordered (truly) ternary hydride $LaB_2H_8$ [23] also does not solve the problem, since its $T_c$ is only slightly higher than that of $LaH_4$ [21] or $La_4H_{23}$ [24,25]. A solution may be a "cold" synthesis from pure $H_2$ and intermetallics, in which the heavy atoms remain in their positions and the hydrogen slowly diffuses and dissociates in the metallic sublattice.

In this work, we turn our attention to the La-Sc-H system at pressures up to 230 GPa. In this system a room-temperature superconductivity has recently been predicted for both the binary phase $Pm\bar{3}$-$ScH_{12}$ [26], and for the ternary one $P6/mmm$-$LaSc_2H_{24}$ [27] ($XH_8$ type). Partial substitution of La atoms by Sc in $LaH_{10}$ was discussed by Kostrzewa et al. [28], and also indicates a significant increase in $T_c$ up to 294 K. The experiment indeed shows the formation of a $(La,Sc)H_{12}$ with $T_c$ = 244-248 K , however, with a different symmetry of the cubic lattice (*fcc*). Samples of $(La,Sc)H_{12}$ exhibit a pronounced diode effect and SQUID-like resistance oscillations [29] accompanied by the disappearance of residual resistance below 215 K at all six van der Pauw channels. Moreover, we found a partial drop in electrical resistance at 274 K in one of the $(La,Sc)H_x$ samples, and anomalies, typical for superconductivity, in the radio-frequency transmission of lanthanum-scandium hydrides in the range of 265-290 K.

## 2. Results and Discussion

We studied six samples of La-Sc hydrides under pressure: DACs LS-1, LS-2 were prepared for the room- and low-temperature X-ray diffraction (XRD) measurements, DACs LS-3, LS-5 were designed for pulsed magnetic fields, DAC LS-4 was intended to study the dependence of critical temperature on pressure. Finally, DAC LS-6 contained two Lenz lenses and was prepared for radio-frequency transmission measurements. In all cases except LS-5, a La-Sc 1:1 alloy, obtained by sintering of La and Sc metal powders in an argon arc, was used. To load the DAC LS-5, we used La:Sc 1:2 alloy, prepared by the same method. The composition and homogeneous distribution of elements in the LaSc alloy were confirmed by X-ray fluorescent (XRF) and energy dispersive X-ray (EDX) analysis (Figure 2b, Supporting Figures S3-S5). Ammonia borane (AB, $NH_3BH_3$) was used as a source of hydrogen. Laser heating of samples was carried out through a series of ~0.3 s infrared laser pulses ($\lambda$ = 1.06 μm), in which the sample temperature increased to at least 1000-1500 K leading to the decomposition of AB. In general, the DACs preparation and transport measurements are similar to our previous works [30]. A detailed description of high-pressure DACs is given in Supporting Table S1.

*2.1 Synthesis of La-Sc polyhydrides at 189-196 GPa*

Powder X-ray diffraction (XRD) combined with first-principles calculations of the unit cell volume, thermodynamically stable structures, and electron-phonon mediated superconductivity is the standard method for establishing the structure of superhydrides [1,31]. This led to the choice of XRD as the first step of our research.

X-ray diffraction study of a DAC LS-1 sample synthesized at 196 GPa shows the presence of a cubic ($Fm\bar{3}m$) phase as the main reaction product (Figure 1a). But what is surprising is the unit cell volume of the resulting compound: almost 32 Å$^3$ per metal atom exceeds the expected volume of the $LaH_{10}$ at 196 GPa. But given that 50 ± 5 % of the alloy is much smaller Sc atoms, we have to attribute a larger amount of hydrogen to the chemical formula of this compound. It should be (La,Sc)$H_{12\pm x}$, where x is about 1. For simplicity, we will further use the chemical formula (La,Sc)$H_{12}$ for this compound. Indeed, *fcc*-$ScH_3$ at 140 GPa has a unit cell volume 15.05 Å$^3$/Sc [32], whereas *fcc*-$LaH_3$ at the same pressure has the volume of 20.5 Å$^3$/La [5]. Therefore, difference in the volumes V(La) –V(Sc) is 5.45 Å$^3$ per atom at 140 GPa. If we take this difference into account and consider a slight decrease in cell volumes during compression to 2 Mbar, the amount of hydrogen in the resulting compound will be equivalent to the same in $LaH_{12}$ [5,33]. This also corresponds to the unusually large unit cell volume of "$LaH_{10+\delta}$" obtained from single-crystal X-ray analysis of lanthanum polyhydrides [34]. Moreover, we believe that La superhydride with the canonical formula "$LaH_{10}$" is rare in experiments. Instead, since the earliest works, it demonstrates variability in composition within the range of $XH_{9-12}$ [35], at the same maintaining a maximum critical temperature of at least 240-250 K. This amazing constancy of the maximum $T_c$, which does not depend on randomly distributed impurities (B, N, Al, Ca, Y, Sc …), nor on the method of obtaining La superhydride, nor on variations in the composition of hydrogen (±1–2 H atoms), is, in our opinion, one of the most striking manifestations of Anderson's theorem for the Bardeen–Cooper–Schrieffer (BCS) superconductors [11].

Due to the absence of additional X-ray reflections from the scandium sublattice, we will further consider the (La,Sc) solid solution model for the heavy atom sublattice. This situation is common for many ternary polyhydrides [13,14,16,21,22]. There is no doubt that several other phases are present in the sample, as the number of peaks exceeds that for both the $Fm\bar{3}m$-(La,Sc)$H_{12}$, as well as the number of reflections for a distorted phase *F(P)mmm*-(La,Sc)$H_{12}$ (see Supporting Figures S7-S8). The latter one is likely formed due to the large difference in volumes and properties of La and Sc atoms. Existence of side phases in the sample also follows from the presence of several steps in the observed SC transitions (Figures 3,4). Additional XRD reflections may appear from Ta/Au electrodes and from the lower hydride $Fm\bar{3}m$-(La,Sc)$H_2$ (Figure 1c).

The X-ray diffraction study of the sample in the DAC LS-3 at 189 GPa was limited by the small diameter (d = 15.3 mm) of this highly asymmetric cell intended for studies in pulsed magnetic fields. The result of the XRD measurements was unexpected: instead of the usual $Fm\bar{3}m$-like pattern we discovered a hexagonal pattern that can be refined as *P6/mmm* (Figure 1e). Recently, a similar structure, *P6/mmm*-$LaSc_2H_{24}$ ($XH_8$ type), was predicted to be a room-temperature superconductor with $T_c$ = 316 K below 200 GPa [27]. This compound has a calculated unit cell volume of 24.15 Å$^3$/(La, Sc) at 200 GPa [27] which is significantly less than the expected volume of $LaH_8$ due to the smaller volume of the Sc atom. The experimental cell volume of the phase (La,Sc)$H_x$ that we obtained in DAC LS-3 is 22.4 Å$^3$/(La, Sc) at 189 GPa despite the higher content of La. This indicates that the hydrogen content is less than in the predicted $LaSc_2H_{24}$ and its chemical formula is close to (La,Sc)$H_{6-7}$. As we will see below, presence of this hexagonal phase in DAC LS-3 may be reflected in the temperature dependence of the electrical resistance, demonstrating a pronounced drop at ≈ 274 K, noticeably higher than $T_c$ of cubic (La,Sc)$H_{12}$ phase (Figures 3, 4a). However, it should be noted that an additional experiment with $LaSc_2$ alloy, conducted in DAC LS-5 at 188 GPa, showed only traces of a series of superconducting transitions (Supporting Figure S28).

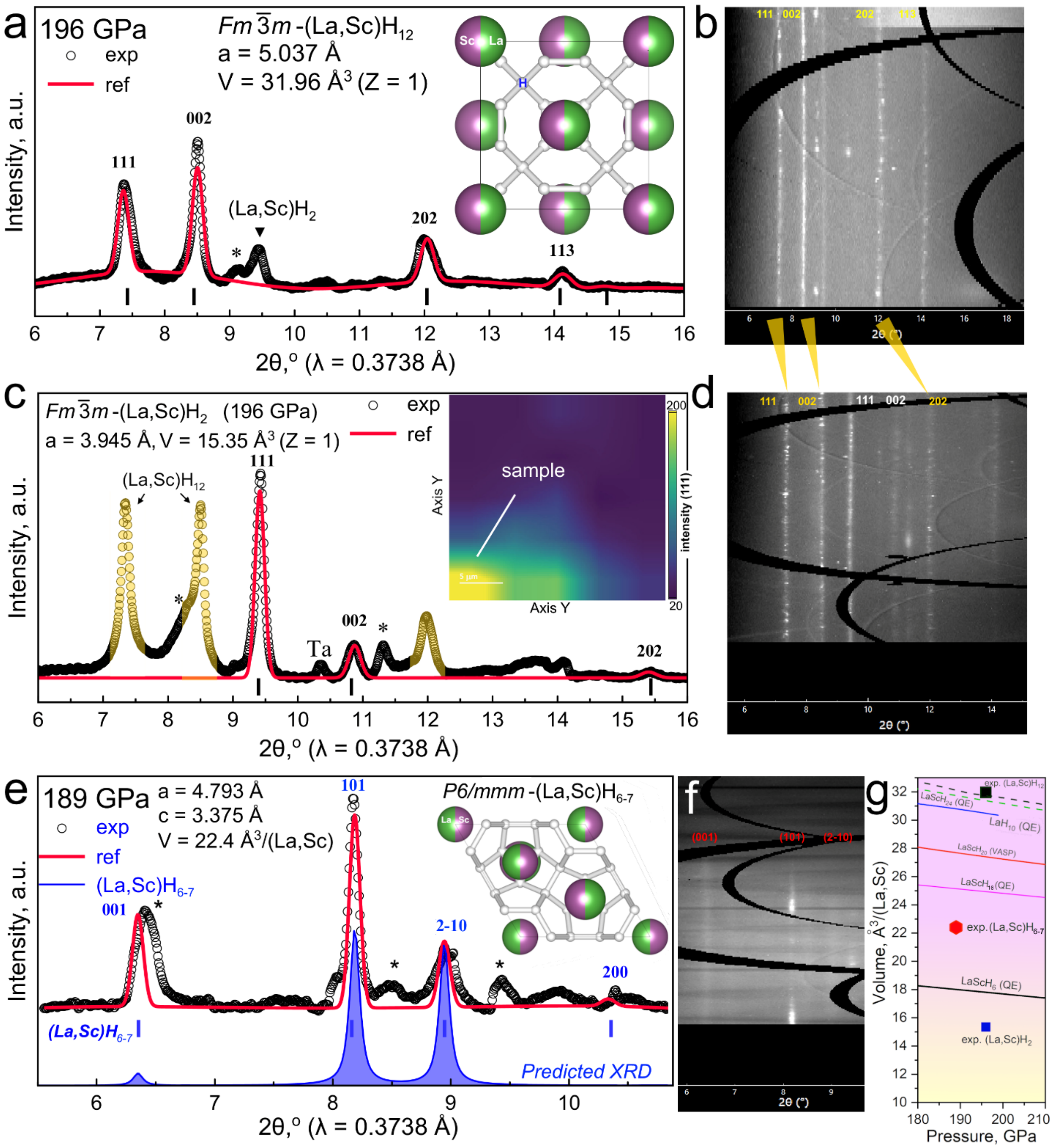

**Figure 1.** Powder X-ray diffraction analysis and the Le Bail refinements of the unit cell of (La,Sc)$H_{12}$, (La,Sc)$H_7$ and (La,Sc)$H_2$ phases in DACs LS-1 and LS-3 at 189-196 GPa. (a) Le Bail refinement of the unit cell parameters of $Fm\bar{3}m$-(La,Sc)$H_{12}$ (JANA2006) in DAC LS-1: black points mark experimental data, the red line – is the refinement. Inset: structural model of this compound based on a cubic La sublattice of $LaH_{10}$. Asterisks mark unexplained reflections. (b) X-ray diffraction image ("cake") of the sample in DAC LS-1. (c) Experimental XRD and Le Bail refinement of the unit cell parameters of $Fm\bar{3}m$-(La,Sc)$H_2$ synthesized in the DACs LS-1 (JANA2006). Yellow color marks reflections from $Fm\bar{3}m$-(La,Sc)$H_{12}$. "Ta" indicates to a possible peak from Ta/Au electrodes. The inset shows spatial distribution of (La,Sc)$H_{12}$ 111 reflection. (d) Typical X-ray diffraction image ("cake") which contains reflections from both (La,Sc)$H_2$ and (La,Sc)$H_{12}$, DAC LS-1. (e) Experimental XRD pattern and Le Bail refinement of the unit cell parameters of (La,Sc)$H_{6\text{-}7}$ found in DAC LS-3. Filled blue plot – is calculated XRD pattern for predicted $P6/mmm$-$LaSc_2H_{24}$ with solid solution of (La, Sc) in metal sublattice. (f) X-ray diffraction image ("cake") of the sample in DAC LS-3. (g) Calculated

equations of state (QE, VASP codes) and experimental unit cell volumes used to estimate the H-content in obtained compounds.

*2.2 Synthesis of La-Sc polyhydrides at 150-153 GPa. Low-temperature XRD.*

In 2020-2021, it was suggested that compressed hydrides may not be superconductors, and resistive transitions in them are a consequence of structural phase transitions [36]. This was facilitated by the absence of X-ray diffraction data near observed transition temperatures. The lack of structural information motivated our low-temperature X-ray diffraction experiment with (La,Sc) hydrides at 153-169 GPa performed at the Xpress beamline of the Elettra synchrotron radiation facility. Gold piece was used as the pressure gauge in these experiments.

In this part of the work, we synthesized La-Sc ternary hydrides at 150-153 GPa (Figure 2a) in DAC LS-2 from the corresponding 1:1 La-Sc alloy, the XRF and EDX analysis of which is shown in Figures 2b, Supporting Figures S3-S5. The main product of the reaction at this pressure can be refined using hexagonal structures, for example, via $P6_3/mmc$-$ScH_6$, predicted by Hou et al. [37] However, the ratio $c/a$ = 1.26 is unusual for this space group (Figure 2c). The unit cell volume of obtained compound is 21.85 Å$^3$/(La, Sc) which indicates composition of (La, Sc)$H_{6-x}$ similar to the predicted hexagonal $ScH_6$ [37]. We believe that the proposed $P6_3/mmc$ structure describes well the resulting diffraction pattern except for an unexpectedly low intensity of the (101) reflection (Figure 2b), and can serve as a first approximation for further structural analysis. Comparison of the experimental unit cell volume, 21.85 Å$^3$/(La, Sc), with the theoretical calculation ($V_{DFT}$ = 23.64 Å$^3$/(La, Sc)) shows that the hydrogen content in this hexagonal phase is between 5 and 6 per metal atom. It should be noted, that the intensity of the XRD pattern in low-temperature cryostat appears insufficient to accurately refine the unit cell parameters at all studied pressure points (Supporting Table S3).

The DAC LS-2 was cooled to a temperature of 150 K while low-temperature powder XRD was recorded. No obvious phase transitions were observed between 150-300 K and 153-169 GPa (Figure 2e). However, there are some signs of small distortion of the crystal lattice at 200-250 K: splitting of the 100 reflection and anomaly of the $c/a$ ratio (Figure 2e, Supporting Table S3). When cooled to 100 K, the pressure in DAC LS-2 increased above 170 GPa that caused collapse of diamond anvils followed by the pressure drop to 30 GPa. The XRD pattern obtained at 30 GPa and 300 K indicates a significant change in the structure and H-content of the (La,Sc) hydrides, and can be described as a two-phase mixture consisting of tetragonal $I4/mmm$-(La,Sc)$H_{3-x}$ (phase I) and thermodynamically stable, pseudocubic $P4_1$-(La,Sc)$H_{\sim 3}$ (phase II, Supporting Tables S3 and S7). The deviation from the calculated unit cell volumes of (La,Sc)$H_{3-x}$ and (La,Sc)$H_3$ indicates the non-stoichiometric composition of the obtained compounds. We should again note that the X-ray beam intensity is not sufficient for a qualitative interpretation of the diffraction data. The main conclusion from this low-temperature XRD experiment is that high-$T_c$ superconductivity in the La-Sc-H system is not accompanied by pronounced phase transitions at temperature above 150 K.

Therefore, the X-ray diffraction study of the La-Sc-H system indicates possible presence of polyhydrides which are similar to the predicted ones: these are hexagonal $LaSc_2H_{24}$ [27], $ScH_6$ [37] , and cubic $ScH_{12}$ [26]. Indeed, in the experiment we see the cubic phase (La,Sc)$H_{12}$, the hexagonal phases (La,Sc)$H_{6-7}$ and (La,Sc)$H_{6-x}$ at 150 GPa. It was already noticed (for instance, in Refs. [30,38]) that structural search programs such as USPEX [39-41], CALYPSO [42,43], AIRSS [44,45] are good at calculating the hydrogen content of stable polyhydrides, whereas predicting of correct crystal symmetry is much more difficult. And in our experiments we are faced with exactly this situation: we see La-Sc hydrides with the predicted stoichiometry, but with an unexpected lattice symmetry. Despite this drawback, first-principles methods provide important clues to the interpretation of products of high-pressure synthesis.

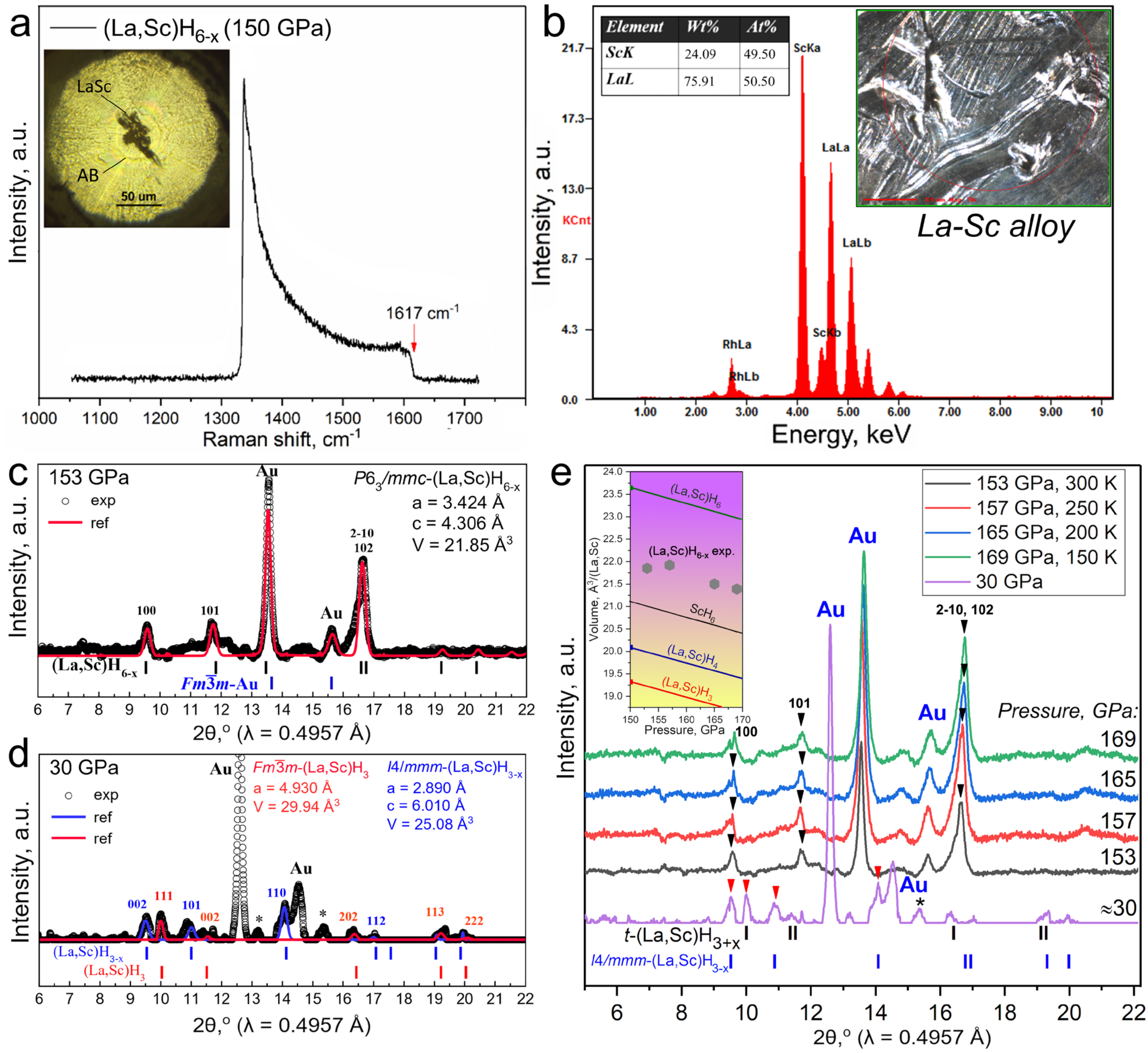

**Figure 2.** Synthesis and X-ray diffraction of La-Sc hydrides in DAC LS-2 at 150-153 GPa. (a) Raman spectrum of the sample in DAC LS-2 in the region of C-C vibrations after laser heating. Inset: photo of the sample. (b) X-ray fluorescence analysis of initial La-Sc alloy. Composition is close to 1:1 (see Table in the inset). (c) Experimental X-ray diffraction patterns and Le Bail refinements of unit cell parameters of hexagonal $(La,Sc)H_{6-x}$ at 153 GPa (300 K). "Au" - denotes reflections from gold used as the internal pressure sensor. Asterisks indicate uninterpreted peaks. (d) The same for tetragonal $(La, Sc)H_{3-x}$ and pseudocubic $(La, Sc)H_3$ at 30 GPa after collapse of the DAC LS-2. (e) A series of XRD patterns of the DAC LS-2 sample taken during decreasing temperature from 300 to 150 K with a step of 50 K. As it cooled, the pressure in the DAC spontaneously increases with $dP/dT \approx -0.11$ GPa/K. The pattern at 30 GPa corresponds to a broken DAC.

### *2.3 Superconductivity of La-Sc polyhydrides*

Transport properties of La-Sc hydrides at 196 GPa were studied in high-pressure DACs LS-1 and LS-3. We used diamond anvils with a culet diameter of 50 μm, equipped with four sputtered (Ta/Au) electrodes and tungsten gasket covered by $CaF_2$/epoxy insulating layer. For a detailed description of the experiment, see Supporting Information. The studies were performed in both direct current (DC, Figure 3a), and alternating current (AC) modes. Regardless of the test method chosen, $(La,Sc)H_{12}$ sample in DAC LS-1 shows a disappearance of electrical resistance below 235 K (Figure 3a). Transport measurements (Figures 3a, d) indicate the possible presence of two phases in the sample, with close critical temperatures: $T_c$(onset) ≈ 247-248 and 242 K (see also Figure 5f). The broadening of superconducting transitions in magnetic fields up to 16 T is insignificant, as for many other polyhydrides [46] which is associated with a high concentration of defects, pinning centers and very high depinning potential barriers reaching in hydrides 1-2×$10^5$ K [47,48]. This correlates with virtually zero magnetoresistance (MR) over the studied range of temperatures (170-223 K) and magnetic fields (Figure 3e).

In some cases, non-uniform distribution of sample conductivity in the van der Pauw circuit leads to "false" voltage (resistance) drops, which actually correspond to a local increase in the resistivity [36]. Such voltage drops may be incorrectly interpreted as superconducting transitions. In this situation, it is necessary to examine all combinations of electrodes. For a four-electrode circuit there are six different channels, two of them are "diagonal" (or the Hall type). Results of measuring the voltage drop across the (La,Sc)$H_{12}$ sample at 196 GPa on four off-diagonal electrode combinations vdp1-4 are presented in Figures 3a-b. In the region of the SC transition, the resistance decreases from ≈ 0.1 Ω to 1-2 μΩ, that is, by a factor of $10^5$. The thermal noise of the measuring circuit is symmetrical on all four channels. Moreover, the diagonal channels (*diag1,3* and *diag2,4*, see Supporting Figure S11) also show the disappearance of the sample's electrical resistance and symmetrical residual thermal noise. The obtained result excludes the possibility of any metal-insulator transitions in the sample.

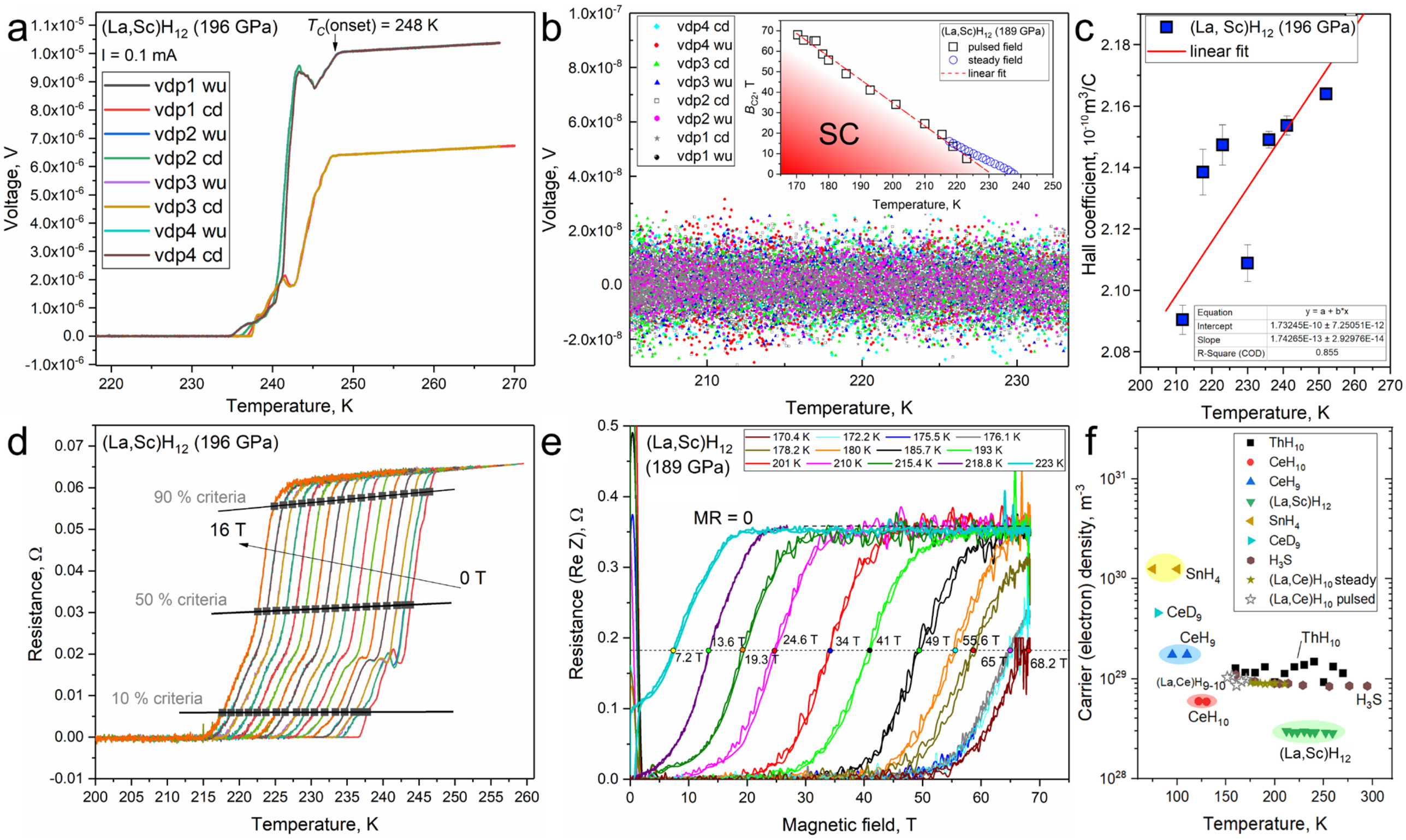

**Figure 3.** Transport properties of (La, Sc) hydrides at 189 and 196 GPa in steady and pulsed magnetic fields. (a) Dependence of the voltage drop across the DAC LS-1 sample (La, Sc)$H_{12}$ on temperature in the warming ("wu") and cooling ("cd") cycles for four direct channels of the van der Pauw circuit (vdp1-4). Here we used the delta mode with a DC excitation current of 0.1 mA. (b) Residual voltage on the sample after transition to the superconducting state on four direct channels (vdp1-4) in the cooling and warming cycles. Average residual resistance of the sample is ~1-2 μΩ. Inset: the boundary of the SC and non-SC regions on the phase diagram, as well as the $B_{C2}(T)$ temperature dependence in magnetic fields up to 68 T, the $R_{50\%}$ criterion was applied. (c) Temperature dependence of the absolute value of Hall coefficient in (La, Sc)$H_{12}$ at 196 GPa. Steady fields ± 33 T were used for measurements. Red line – is the linear fit. (d) Temperature dependence of the electrical resistance of the (La, Sc)$H_{12}$ in various constant magnetic fields from 0 to 16 Tesla. The criteria $R_{10\%}$, $R_{50\%}$, $R_{90\%}$ were applied to determine $B_{C2}(T)$. (e) Dependence of the real component of the electrical resistance of the sample in DAC LS-3 on a pulsed magnetic field up to 68 T at various temperatures from 223 K to 170.4 K. "MR" denotes magnetoresistance. (f) Carrier (electron) concentration in various superhydrides according to the Hall coefficient measurements: $n_e = 1/eR_H$. Thickness of samples is assumed 1 ± 0.5 μm based on the visible light interference pattern between diamond anvils.

We studied the behavior of La-Sc hydrides in steady (0 – 33 T, DAC LS-1) and pulsed magnetic fields with a pulse duration of 150 ms and maximum field induction of 68 T (DAC LS-3). Since the primary superconducting transition in DAC LS-3 sample is observed at the $T_c \approx 244$ K (Figure 4a), we will henceforth consider (La,Sc)$H_{12}$ to be the main content of the LS-3 sample despite the fact that we were unable to detect this phase using XRD analysis due to the limited opening angle in this asymmetric DAC. Such a strong pulsed magnetic field was used to establish the shape of the superconducting phase diagram (inset in Figure 3b). In this case, measurements of the dependence of the electrical resistance (real part, Re Z) on the magnetic field were carried out in AC mode at frequencies of 16.66 kHz and 33.33 kHz in the temperature range from 170.4 K to 223 K (inset in Figure 3e). In this range of temperatures and fields, the magnetic phase diagram of (La,Sc)$H_{12}$ is linear ($dB_{C2}/dT \approx -1.15$ T/K), as for many other hydrides [13,49,50,51]. Very high upper critical field $B_{C2}(0)$ of La-Sc superhydrides is remarkable. The extrapolated value of $B_{C2}(0)$ can exceed 200-230 T (Figure 3b). Such high upper critical field is characteristic of dirty ("hard") superconductors and high-entropy polyhydrides [7].

Metal superhydrides synthesized at high pressure are, in most cases, fine powders [52] with a large number of defects in microcrystals [1,13,30,53]. This determines many physical properties of hydrides. For example, wide diffusive XRD patterns [38,54,55], very low RRR (residual-resistance ratio) about 1.1–1.5 [4], and often near zero magnetoresistance. Typically, the magnetoresistance of metals is positive and $\propto \mu_e^2 B^2$, where $\mu_e$ is the electron mobility (in the absence of holes). However, hydrides are different. The existence of a region of negative magnetoresistance was established in $La_4H_{23}$ [24,25] and $CeH_{10}$ [56], whereas magnetoresistance of (La,Sc)$H_{12}$ in DAC LS-3 is practically equal to zero like in graphene [57].

One of the most likely explanations of this fact is a very defective and finely dispersed structure of the sample containing B, and N impurities from ammonia borane, fluctuations of Sc and La concentration like in NbTi [58], presence of $H_2$ molecules and H vacancies. In this case, the electron mean free path ($l_e$) is significantly less than the Larmor radius ($r_g$) even at 70 T

$$l_e \ll r_g = \frac{m_e^* v_F}{eB}, \qquad (1)$$

where the Fermi velocity $v_F$ can be estimated as $\sim 3\times10^5$ m/s [59], and $m_e^*$ is the effective electron mass. If we take $m_e^* = m_e$ and magnetic field induction $B = 50$ T, then $r_g \sim 38$ nm. This value should be comparable to the crystallite size and the average distance between defects in (La, Sc)$H_{12}$. In this case, scattering on defects and grain boundaries will be the dominant process, and the influence of the magnetic field will be insignificant. Of course, it cannot be completely ruled out that the positive magnetoresistance ($\propto \mu_e^2 B^2$), observed due to the distortion of electron trajectories in a magnetic field, can compensate for negative magnetoresistance related to the preformation of Cooper pairs, giving an ultimately zero result in a certain range of magnetic fields.

Finally, we have studied the Hall effect and its sign in the non-superconducting state of (La, Sc)$H_{12}$ in DAC LS-1. In the Hall measurement scheme, the superconducting transition begins at the same $T_c$ around 245 K, but is accompanied by a more complex behavior of the electrical resistance $R(T, B)$. Supporting Figures S12-S14 show that the Hall voltage depends on magnetic field linearly. The Hall coefficient $R_H$ is found negative. This indicates that electrons are the primary charge carriers. $|R_H| = (2.13 \pm 0.03)\times10^{-10}$ m$^3$/C in (La, Sc)$H_{12}$ slightly decreases along with temperature (Figure 3c). This corresponds to increase of carrier concentration by 3 % during cooling from 259 to 212 K: $n_e$((La,Sc)$H_{12}$) $= 1/eR_H = (2.95 \pm 0.05)\times10^{28}$ m$^{-3}$ (Figure 3f). In general, almost temperature independent carrier concentration is typical for metals. The found values of electron concentration $n_e$((La,Sc)$H_{12}$), within the accuracy of the error in determining the sample thickness, correspond to the concentration of carriers in other superhydrides (Figure 3f).

*2.4 Metastable resistive transitions, SQUID-like, and diode effects in (La, Sc)$H_{12}$*

When studying the Hall effect in the DAC LS-3, we unexpectedly found that the onset of the resistance drop at a certain stage of the experiment shifted to higher temperatures, reaching unusually high value of 274 K (Figure 4a, Supporting Figure S24). At the same time, the main superconducting transition, probably corresponding to the (La, Sc)$H_{12}$, was still observed at $T_c$ = 244 K. High-temperature superconductivity with $T_c$ > 270 K was predicted in both cubic $ScH_{12}$ [26] and hexagonal $LaSc_2H_{24}$ [27]. Despite the fact that detected resistance anomaly at 274 K correlates with the presence of a similar hexagonal phase in DAC LS-3 (Figure 1e), consideration of the recent NMR and radio-frequency transmission data on lanthanum polyhydrides[60] makes the hexagonal crystalline modifications of (La,Sc)$H_{12}$ a more preferred explanation. Unfortunately, after 8 cooling/heating cycles (Supporting Figure S21), this "metastable" anomaly almost disappeared. However, some residual traces of it in the Hall measurements were found even a month after the initial detection (Supporting Figure S25).

Synthesized samples of La-Sc hydrides are heterogeneous. Trapping of magnetic flux by inhomogeneities in a superconducting sample [50] may lead to appearance of SQUID (superconducting quantum interference device) [29] and asymmetric conductivity effects [61] in hydrides. The later one, also known as diode effect, can be detected in (La,Sc)$H_{12}$ superhydride (DAC LS-1) with the help of a current containing a DC offset of ± 1 mA and small sinusoidal signal of 0.1 mA (RMS) with a frequency of 100 Hz ($I(t) = \pm 1 + 0.1\times\sqrt{2}\,\sin(100\times t/2\pi)$ mA, where $t$ – is the time). The signal amplitude at the second pair of electrodes of the 4-contact van der Pauw circuit was detected using a lock-in amplifier SR830 and used to calculate the sample resistance (Figures 4d-e). In this approach, all constant voltage contributions associated with thermoelectric phenomena (thermo EMF) are filtered out by the lock-in amplifier. A similar result may be obtained by direct subtraction the resistance data obtained for different directions of DC current (Supporting Figures S21-S22).

As a result, we found that the dependence of the electrical resistance of the sample on temperature $R(T)$ has a significant hysteresis (Figures 4b-c) in the region of 220-230 K, which completely disappears above 235 K. Below this temperature there is a difference not only between curves with different directions of DC current (± 1 mA), but also between the cooling and warming curves ($\Delta T$ > 2 K), which indicates the presence of a "memory" effect of the sample[62]. When a direct current flows through the sample, it creates vortices in SC grains. These frozen vortices [47,48] lead to the symmetry-breaking effects and significant asymmetry ($R_+/R_-$ = 1.23, at 215 K this ratio is much higher) in the sample conductivity as it observed in superconducting diodes [61]. At the same time, the effect depends on the prehistory of the sample, which stores a kind of "vortex memory" of the type of cycle (cooling or warming). The discovered diode effect is not large but is noteworthy because it opens up the possibility of creating SC diodes based on superhydrides made from $LaH_{10}$ at record-high operating temperatures above 220-230 K.

Finally, we observed interference oscillations of electrical resistance in DAC LS-3 using a previously developed setup (Figure 4f). Applying of a weak modulating magnetic field of $B_{max}$ ~10 G with a frequency of $\nu_0$ = 0.1 Hz led to a generation of higher harmonics up to $14\nu_0$ already at a temperature of 226 K (Figure 4g), which is 47 K higher than the previous result achieved with (La,Ce)$H_{10+x}$ [29]. The temperature range where the generation of higher harmonics is observed coincides with the region of the diode effect. Thus, these effects are interrelated. Moreover, measurements of the voltage drop on the DAC LS-1 sample in a magnetic field of ±300 Gauss at a temperature of 233.4 K demonstrate a pronounced periodic dependence of $R(B)$, characteristic of the DC SQUID effect in a ring with a diameter of ≈ 600 nm (Supporting Figure S27).

The SQUID-like effect in (La,Sc)$H_{12}$ is observed at a small bias current $I_{bias}$ = 400-600 μA, which improves the situation with heating of the sample. This is significantly less than the 1-2 mA we had to use in the case of (La,Ce)$H_{10+x}$ [29] and brings us closer to the currents applied in commercial SQUIDs (~1–10 μA). Unfortunately, the size of the randomly formed SQUID-like contours was about 0.6–1 μm or less as indicated by the low sensitivity of the sample to applied magnetic field (Supporting Figures S27).

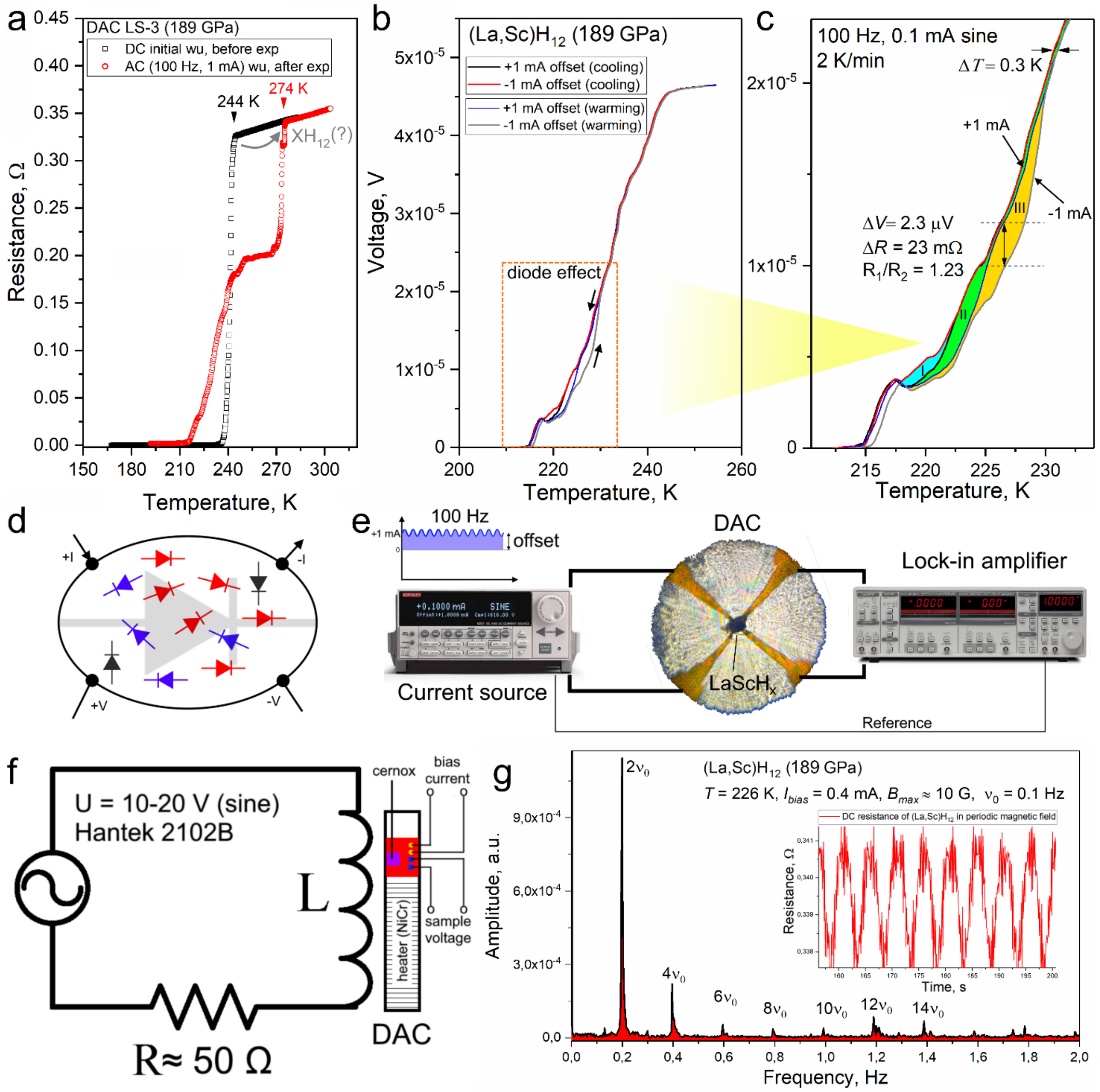

**Figure 4.** Diode and DC SQUID effects in (La,Sc)$H_{12}$ (DAC LS-3). (a) Stable (black circles) and "metastable" (red circles) resistive transitions in the La-Sc superhydride at 189 GPa. The arrows indicate a possible electrochemical process of hydrogen migration between electrodes, leading to broadening of the SC transition and the appearance of new metastable phases. (b) Diode effect observed in the sample when applying offset DC current of +/– 1 mA. Area I (azure) corresponds to the hysteresis of the sample resistance when the offset current changes its sign in a cooling cycle (2 K/min). Area II (green) corresponds to the resistance hysteresis at a constant offset +1 mA, occurring between heating and cooling cycles. Area III (yellow) corresponds to the hysteresis of the sample resistance when the current offset sign changes in the warming cycle (2 K/min). (d) Schematic illustration of the random arrangement of asymmetric conductivity regions in the sample. (e) Experimental scheme of measurements of the diode effect, which excludes the contribution of thermo EMF. (f) Scheme of studying the SQUID effect in the sample when a weak modulated magnetic field (~ 10 G) is applied. (g) Generation of higher harmonics when the (La,Sc)$H_{12}$ SQUID resistance changes in response to a periodic weak external magnetic field. Inset: complex periodic pattern of the electrical resistance versus time.

*2.5 Radio-frequency transmission measurements*

Considering the signs of a possible high-$T_c$ superconducting transition around 274 K in the La-Sc-H system (DAC LS-3), we performed an additional study using the radio-frequency (RF) transmission method. This highly sensitive method [63], which we used earlier to study the La-H system in combination with high-pressure $^1$H NMR and transport measurements [60], allows us to detect even small-volume superconducting phases that are not in contact with the electrodes.

For this experiment, three-stage Ta/Au Lenz lenses were sputtered and cut by Ga FIB on both diamond anvils of the DAC LS-6 equipped with a 50 μm anvil culet size. These lenses allow to inductively couple the high-frequency signal to the sputtered LaSc ring (Figure 5a). During laser heating at 174 GPa, we observed a crack formation and a pressure drop in the DAC to 155 GPa (Figure 5b, c, and Supporting Figure S29e). To prevent the further damage of anvils, the laser heating was stopped at ~15-20% of the sample area. However, the obtained volume of superconducting phases was sufficient for further study. The experimental setup is shown in Supporting Figure S1.

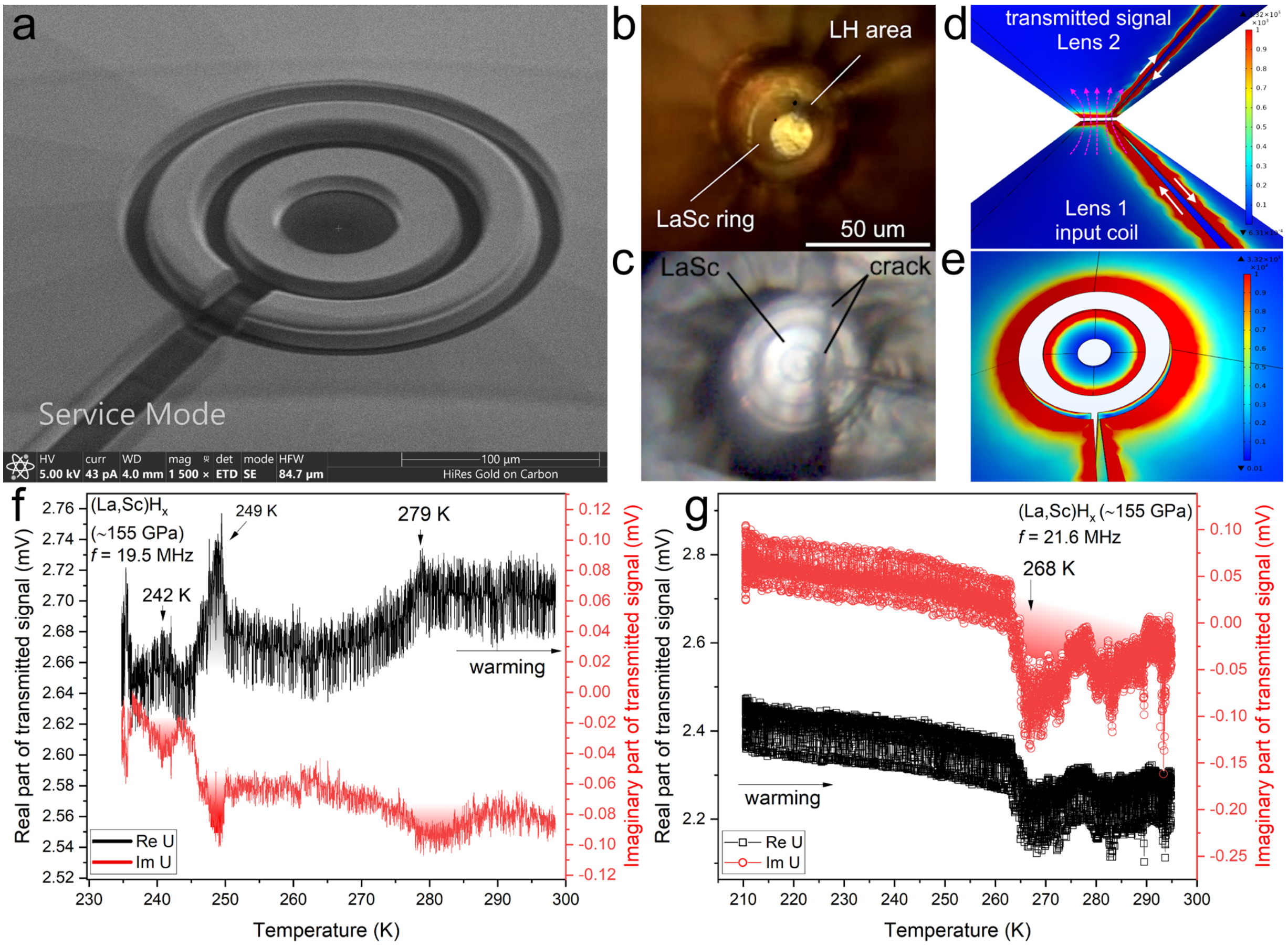

**Figure 5.** Radio-frequency transmittion measurements of the $(La,Sc)H_x$ sample in the DAC LS-6. (a) Scanning electron microscopy of the ring LaSc sample in the center of the Lenz lens system. (b) Optical image of the sample after partial laser heating (LH), the LaSc ring is slightly displaced from the culet center. (c) Photograph of the sample after partial cracking of an anvil and pressure drop to 155 GPa. (d) Sketch of the RF signal transmission from Lens 1 to Lens 2 through the hydride sample. Colormap corresponds to the distribution of the surface loss density (SLD) on metal surfaces. (e) SLD distribution on a 60 μm ring sample at an excitation current frequency of 200 MHz. (f) Real (Re U, black) and imaginary (Im U, red) parts of the transmitted signal in the warming cycle at a carrier frequency of 19.5 MHz and (g) 21.6 MHz. There are pronounced features in the transmission near the known superconducting transition (240-250 K) in $(La,Sc)H_{12}$ and in the vicinity of the previously found resistive anomaly (268-279 K).

The essence of the RF method in combination with the use of Lenz lenses is that the induced current flows along the surface and edges of the deposited lenses, concentrating in the area of the anvil culet with the ring sample placed on it (Figure 5d, e). At the point of the superconducting transition of the sample, a sharp change in its surface impedance is observed, which is expressed in a change in the voltage on the receiving coil and the appearance of a feature in the form of a step, peak or kink in the temperature dependence of a sample transmission. The specific shape of the feature depends on the carrier frequency (i.e., the depth of field penetration), the morphology of the sample and the distribution of superconducting phases in it, as well as on the width of the vortex liquid region in the hydride under study [47,48], in which intense absorption of electromagnetic radiation occurs.

We investigated the temperature interval from 200 to 310 K at frequencies of 1, 10, 19.5, 21.6 and 200 MHz in the warming and cooling modes. In all the studied cases, there is a pronounced feature in the sample transmission in the range from 265 to 290 K (Figure 5f,g and Supporting Figure S29), which we previously noted in the transport measurements of DAC LS-3. At some frequencies, such as 19.5 MHz, pronounced RF transmission signals are observed at 242 K and 249 K, which are characteristic of superconductivity in $Fm\bar{3}m$-(La,Sc)$H_{12}$ (Figure 3a). It also manifests itself as a weak signal at the second harmonic ($2F$ = 66 Hz) of the low-frequency ($F$ = 33 Hz) modulating field in which the sample is placed. Thus, the study of the RF transmission of the (La,Sc)$H_x$ sample in DAC LS-6 supports the conclusions of transport measurements about the presence of at least two fractions in the sample: 1) superconducting phase-I with $T_c$ = 240-250 K, and 2) phase-II with $T_c$ = 265-290 K. It is important to note that the obtained results are consistent with the previous reports on the possible presence of superconducting phases with the $T_c$ of 265-280 K in the La-H system under pressure [6,64,65] (Supporting Figure S30)

## Conclusions

A low-temperature (150-300 K, 153-169 GPa) X-ray diffraction study of superconducting lanthanum-scandium superhydrides ruled out phase transitions with significant symmetry changes as a possible reason for the sharp drop of electrical resistance in the La-Sc hydrides. At 189-196 GPa we discovered new cubic polyhydride (La,Sc)$H_{12}$ which demonstrates the disappearance of electrical resistance on all six channels of the van der Pauw four-contact circuit below 230 K. The onset critical temperature of this phase, which is dominant in our experiment, reaches $T_c$ = 248 K, which is in agreement with Anderson's theorem for doped BCS-ME superconductors. In addition, La-Sc hydrides demonstrates a pronounced DC SQUID-like interference oscillations of electrical resistance and diode effect at an unprecedented high temperature of about 233 K.

Investigation in strong steady and pulsed magnetic fields has shown that magnetoresistance of (La,Sc)$H_{12}$ is practically absent. Extrapolated upper critical magnetic field in some heterogeneous samples may exceed 200-230 T at 0 K. Measured Hall coefficient is negative, has no features near the $T_c$ and in absolute value ($\sim 2\times10^{-10}$ $m^3$/C) is close to the values known for other superhydrides.

One of the (La,Sc)$H_x$ samples at 189 GPa demonstrated unusual “metastable” drop in electrical resistance around 274 K. This feature, together with a peak at 240-250 K corresponding to the cubic (La,Sc)$H_{12}$, was also found in the radio-frequency transmission measurements at 155 GPa in the temperature interval of 265-290 K. The probable presence of superconductivity above 0 °C makes the La-Sc-H system very attractive for further research.

## Acknowledgments

D. S. and D. Z. thank Beijing Natural Science Foundation (grant No. IS23017) and National Natural Science Foundation of China (NSFC, grant No. 12350410354) for support of this research. D. Z. thanks China Postdoctoral Science Foundation (Certificate No. 2023M740204) and finical support from HPSTAR. D.S. and D.Z. are grateful for financial support from the EMFL-ISABEL project. The high-pressure experiments were supported by the Russian Science Foundation (Project No. 22-12-00163). XRF analysis was performed using the equipment of the Shared Research Center of the Kurchatov Complex of Crystallography and Photonics. We acknowledge the European Synchrotron

Radiation Facility (ESRF) for provision of synchrotron radiation facilities and Momentum Transfer for facilitating the measurements. Jakub Drnec is thanked for assistance and support in using beamline ID31. The measurement setup was developed with funding from the European Union's Horizon 2020 research and innovation program under the STREAMLINE project (grant agreement ID 870313). Measurements performed as part of the MatScatNet project were supported by OSCARS through the European Commission's Horizon Europe Research and Innovation programme under grant agreement No. 101129751. A portion of this work was performed on the Steady High Magnetic Field Facilities, High Magnetic Field Laboratory (CAS). All authors thank the staff scientists of the SPring-8, Elettra, Xpress station (proposals No. 20215034 and No. 20220288) and the ESRF, station ID27 (proposal MA-5924), synchrotron radiation facilities for their help with X-ray diffraction measurements, especially Dr. Anna S. Pakhomova (ESRF), Dr. Saori Kawaguchi (SPring-8), Dr. Boby Joseph (Elettra), Prof. Katsuya Shimizu and Dr. Yuki Nakamoto (Osaka University).

## Contributions

I.A.T., D.V.S., D.Z., M.G., A.V.S., O.A.S. A.G.I., A.S.P., F.G.-A.,C.X., T.H., S.L. and K.P. performed the experiments. D.Z., M.G. and D.S. prepared theoretical part of the paper. K.S.P. prepared the La–Sc alloys. A.G.I. analyzed composition of La-Sc alloys. I.A.T. prepared diamond anvil cells for experiments. A.V.S., O.A.S., and V.M.P. performed the magnetotransport experiments in magnetic fields below 16 T and participated in the data processing and discussions. W.C. wrote scripts for critical current measurements. C.X. helped with Hall measurements in steady magnetic fields up to 33 T. T.H. and S.L. assisted in research in pulsed magnetic fields up to 68 T. A.S.P. and F.G.-A. performed X-ray diffraction studies. D.V.S., V.M.P. and V.V.S. wrote the manuscript. All the authors discussed the results and offered useful inputs.

## Data availability

Authors declare that the main data supporting our findings of this study are contained within the paper and Supporting Information. All relevant data are available from the corresponding authors upon request.

## Code availability

Quantum ESPRESSO code is free for academic use and available after registration on https://www.quantum-espresso.org/. Vienna ab-initio Simulation Package (VASP) code is available after registration on https://www.vasp.at/. USPEX code is free for academic use and available after registration on https://uspex-team.org/en/.

SUPPORTING INFORMATION

# Ternary superhydrides under pressure of Anderson's theorem: Near-record superconductivity in (La, Sc)$H_{12}$

Dmitrii V. Semenok[1,†,*], Ivan A. Troyan[2,†], Di Zhou[1,†,*], Andrei V. Sadakov[3], Kirill S. Pervakov[3], Oleg A. Sobolevskiy[3], Anna G. Ivanova[2], Michele Galasso[4], Frederico Gil Alabarse[5], Wuhao Chen[6,7], Chuanying Xi[8], Toni Helm[9], Sven Luther[9], Vladimir M. Pudalov[3,10] and Viktor V. Struzhkin[11,*]

*[1] Center for High Pressure Science & Technology Advanced Research, Bldg. #8E, ZPark, 10 Xibeiwang East Rd, Haidian District, Beijing, 100193, China*

*[2] A.V. Shubnikov Institute of Crystallography of the Kurchatov Complex of Crystallography and Photonics (KKKiF), 59 Leninsky Prospekt, Moscow 119333, Russia*

*[3] V. L. Ginzburg Center for High-Temperature Superconductivity and Quantum Materials, Moscow, 119991 Russia*

*[4] Institute of Solld State Physics, University of Latvla, 8 Kengaraga str., LV-1063 Riga, Latvla*

*[5] Elettra Sincrotrone Trieste, Strada Statale 14 km163,5 in AREA Science Park, Basovizza, Trieste 34149, Italy*

*[6] Department of Physics, Southern University of Science and Technology, Shenzhen 518055, China*

*[7] Quantum Science Center of Guangdong–Hong Kong–Macao Greater Bay Area (Guangdong), Shenzhen, China*

*[8] Anhui Province Key Laboratory of Condensed Matter Physics at Extreme Conditions, High Magnetic Field Laboratory of the Chinese Academy of Science, Hefei 230031, Anhui, China.*

*[9] Hochfeld-Magnetlabor Dresden (HLD-EMFL) and Würzburg-Dresden Cluster of Excellence, Helmholtz-Zentrum Dresden-Rossendorf (HZDR), Dresden 01328, Germany*

*[10] National Research University Higher School of Economics, Moscow 101000, Russia*

*[11] Shanghai Key Laboratory of Material Frontiers Research in Extreme Environments (MFree), Shanghai Advanced Research in Physical Sciences (SHARPS), Pudong, Shanghai 201203, China*

*Corresponding authors: Dmitrii Semenok (dmitrii.semenok@hpstar.ac.cn), Di Zhou (di.zhou@hpstar.ac.cn) and Viktor Struzhkin (viktor.struzhkin@hpstar.ac.cn).

†These authors contributed equally.

## Content

# 1. Methods

## *Experiment*

Crystal structure of La-Sc polyhydrides synthesized in DACs LS-1,3 were studied using the synchrotron X-ray diffraction (XRD) on the ID27 beamline with a beam of wavelength of 0.3738 Å at the European Synchrotron Research Facility (ESRF, proposal MA-5924). Low-temperature XRD in DAC LS-2 was studied on a XPRESS station of Elettra (proposal 20220288) with a beam of wavelength of 0.4957 Å in the interval of 150-300 K.

Mapping of the sample center was carried out on a 4×4 or 5×5 grids with a step of 5 microns, accumulating time was about 10 seconds per image. The experimental XRD images were integrated and analyzed using the Dioptas 0.5 software package [1]. To fit the diffraction patterns and obtain the cell parameter, we analyzed the data using Mercury 2021.2.0 [2] and Jana2006 software [3], employing the Le Bail method [4]. Analysis of the spatial distribution of hydride phases in the sample was performed using the XDI program [5].

The Hall measurements were carried out on a steady water-cooled WM5 magnet (HMFL, Hefei, China) with a maximum magnetic field strength of 33 Tesla. A Lakeshore 370AC resistive bridge was used to measure the resistance of La-Sc superhydride samples in AC mode. The exciting current was 1 mA (RMS, 13.7 Hz), the cooling/heating rate was 2-3 K/min, and the temperature stabilization time was 10 min for each temperature point. When studying the Hall effect (the linear dependence of resistance on the magnetic field), the temperature of the sample was reduced in steps of 5-10 K, at each temperature point the dependence $R(B)$ was measured in the range from –33 T to +33 T, which was achieved by changing the polarity of WM5 electromagnet. To simplify the analysis of the linear dependence $R(B)$ observed in the non-superconducting state after suppression of superconductivity by a magnetic field, we removed the data corresponding to the superconducting transition region and also subtracted the zero-field resistance $R(B = 0)$ from the raw data.

Resistance measurements in pulsed magnetic field were performed in a four-contact van der Pauw scheme using an alternating current of 1 mA with a frequency of 16.66 kHz and 33.33 kHz. To analyze the magnetic phase diagram, we used the real part of the impedance ($Z = U/I$) corrected by the phase factor so that in the superconducting state Re Z = 0. At 33.33 kHz, the residual impedance of the system in the superconducting state reaches 0.5 Ω, and the phase factor was 35 deg. At 16.66 kHz, the small residual impedance (0.06 Ω) was compensated by a phase factor of 26.8 deg.The duration of the magnetic field pulse was 150 ms, the maximum amplitude was 68 Tesla. To control the temperature, a Cernox thermometer was used, glued to the gasket with thermoconductive glue. To stabilize the sample temperature, an external heater, an insulated Ni-Cr wire of 1 m long and 50 microns in diameter, was used.

The radio-frequency (RF) measurements were performed in a DAC LS-6, which had two symmetrical three-stage Lenz lenses consisting of a bimetallic Ta/Au layer deposited on diamond anvils using a JEOL IB-09010CP Cross Section Polisher. The argon ion beam energy was 7.5-8 keV, and the deposition time was 7 minutes for each metal layer. The central part of the culet was then etched using a focused Ga beam (Ga FIB, Thermo Fisher Scientific) until it reached the diamond surface. Then, a 1-2 µm thick LaSc layer was deposited on the anvil using a JEOL IB-09010CP Cross Section Polisher, a LaSc alloy target. The deposition time was 30 minutes at an ion beam energy of 7.5 keV. Finally, re-cutting using Ga FIB allowed the formation of a Lenz lens topology and an annular target made of LaSc.

**Table S1.** Parameters of DACs used in high-pressure experiements.

| DAC number | DAC's diameter, material of the DAC | Culet diameter / Heated at pressure | Initial sample composition, insulating gasket | Experiments |
|---|---|---|---|---|
| LS-1 (electrical cell) | d = 25 mm, BeCu | 50 um / 196 GPa | LaSc/AB, $CaF_2$/epoxy/W | XRD (ESRF, ID27 October 2023) Transport measurements Hall measurements (Hefei high magnetic field center, April 2024) |
| LS-2 (XRD cell) | d = 39 mm, BeCu | 75 um / 150 GPa | LaSc/AB, $CaF_2$/epoxy/W | Low-temperature XRD (Elettra, XPRESS, October 2022) |
| LS-3 (electrical cell) | d = 15.3 mm, NiCrAl | 50 um / 189 GPa | LaSc/AB, $CaF_2$/epoxy/W | XRD (ESRF, ID 27, October 2023) Pulsed magnetic fields (Dresden, October 2023) |
| LS-4 (electrical cell) | d = 25 mm, BeCu | 50 um / 213-220 GPa | LaSc/AB, $CaF_2$/epoxy/W | Only transport measurements (2024) |
| LS-5 (electrical cell) | d = 15.3 mm, NiCrAl | 50 um / 188 GPa | $LaSc_2$/AB, $CaF_2$/epoxy/W | Only transport measurements (HLD HZDR 2024) |
| LS-6 (RF cell) | d = 39 mm, BeCu BX-90 DAC | 50 um / 174 GPa | LaSc/AB, Ga, Ga FIB tretment, $Ta_2O_5$/Ta/W gasket | Radio-frequency measurements (HPSTAR, 2024) |

We performed measurements of the transmission of high-frequency signals through the Lenz lens system in the DAC LS-6 in a circuit with two lock-in amplifiers (Supporting Figure S1). We used a SR844 lock-in amplifier (Stanford Research) as high-frequency signal generator, as well as receiver to measure the transmitted signal at the same frequency. We placed the sample in a low-frequency magnetic field (33 Hz, maximum induction is 40 Gauss), generated by a special solenoid. The current feeding the solenoid was created by an SR830 lock-in amplifier, working in tandem with a Yamaha PX3 power amplifier. The envelope of the high-frequency signal near the superconducting transition temperature contains even harmonics, of which the strongest, the 2$^{nd}$ harmonic, was detected, thus serving as an additional marker of the superconducting transition sensitive to the magnetic field. In high-frequency measurements, the target signal can have two basic forms: (1) A peak, hump, or trough of various shapes; (2) a step. Typical measurement parameters are given in Table S2.

**Table S2.** Parameter settings for the SR844 and SR830 lock-in amplifiers used in our RF measurements.

| | **SR844** | **SR830** |
|---|---|---|
| **Time constant** | 100 µs | 300 ms |
| **Filter** | 18 or 24 dB | 18 or 24 dB |
| **Signal input** | 50 Ω (without capacitors) 30 pF (with capacitors) | AC |
| **Wide Reserve** | Low noise or Normal | Low noise or Normal |
| **Harm #** | 1 or 2 | 2, 4, 6 … |

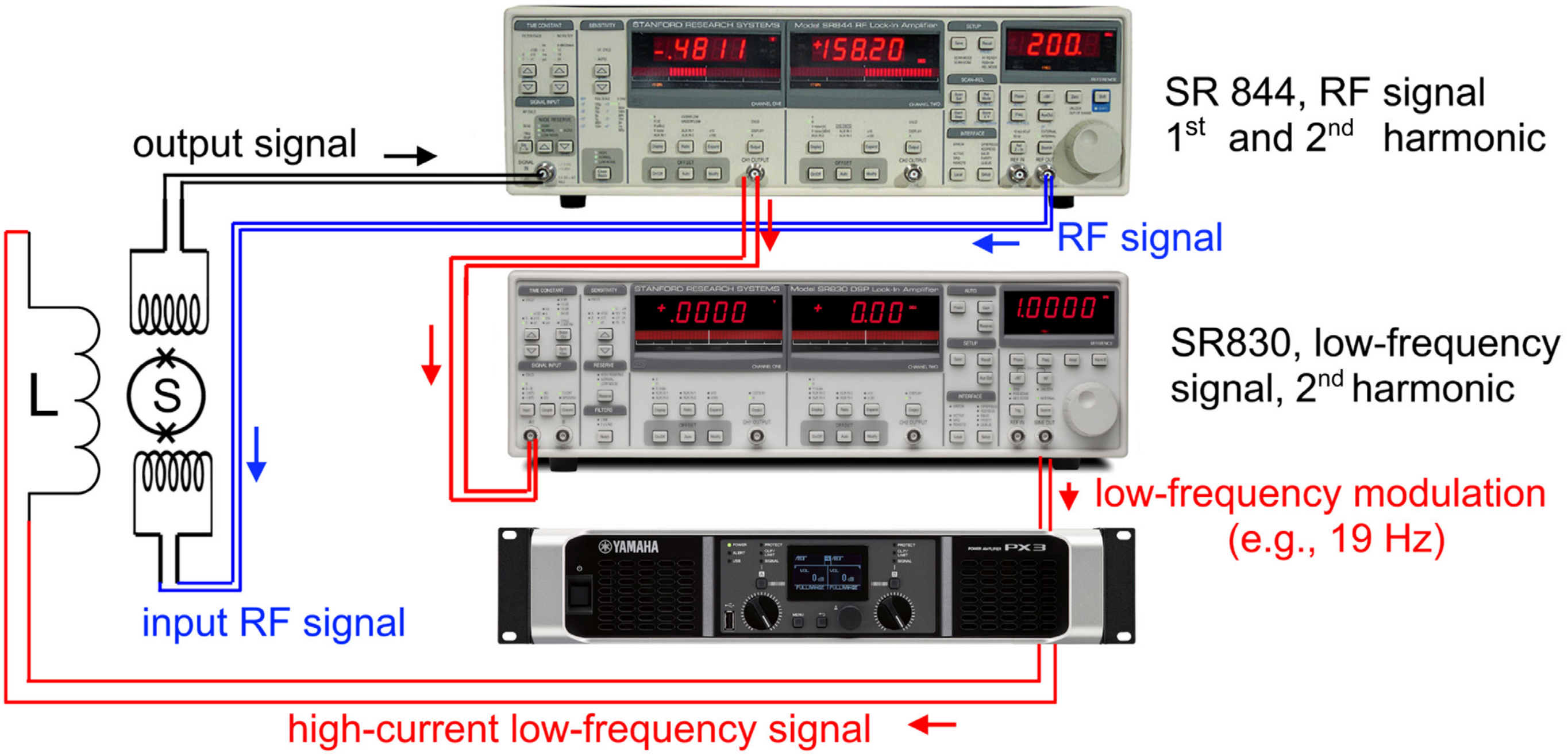

**Figure S1.** Scheme of the setup for radio-frequency transmission studies of the samples in the DACs. "L" represents an external solenoid creating low-frequency modulating magnetic field (usually 19 or 33 Hz). "S" stands for sample.

Lenz lens studies are complementary to those using electrode schemes like the van der Pauw scheme. Mathematically, the electrode system and the Lenz lens system are related to each other as products of a mathematical swap transformation of the polar coordinate system (Supporting Figure S2).Unlike electrodes, where measurements are made at low frequencies and a direct contact with the sample is used, Lenz lens schemes involve fully inductive signal transmission, which requires the operating frequency to be in the megahertz (MHz) range. This makes the high-frequency transmission measurement scheme a method that occupies an intermediate position between optical and electrotransport methods. Unlike optical techniques, the RF method does not impose high requirements on the surface quality and sample thickness.

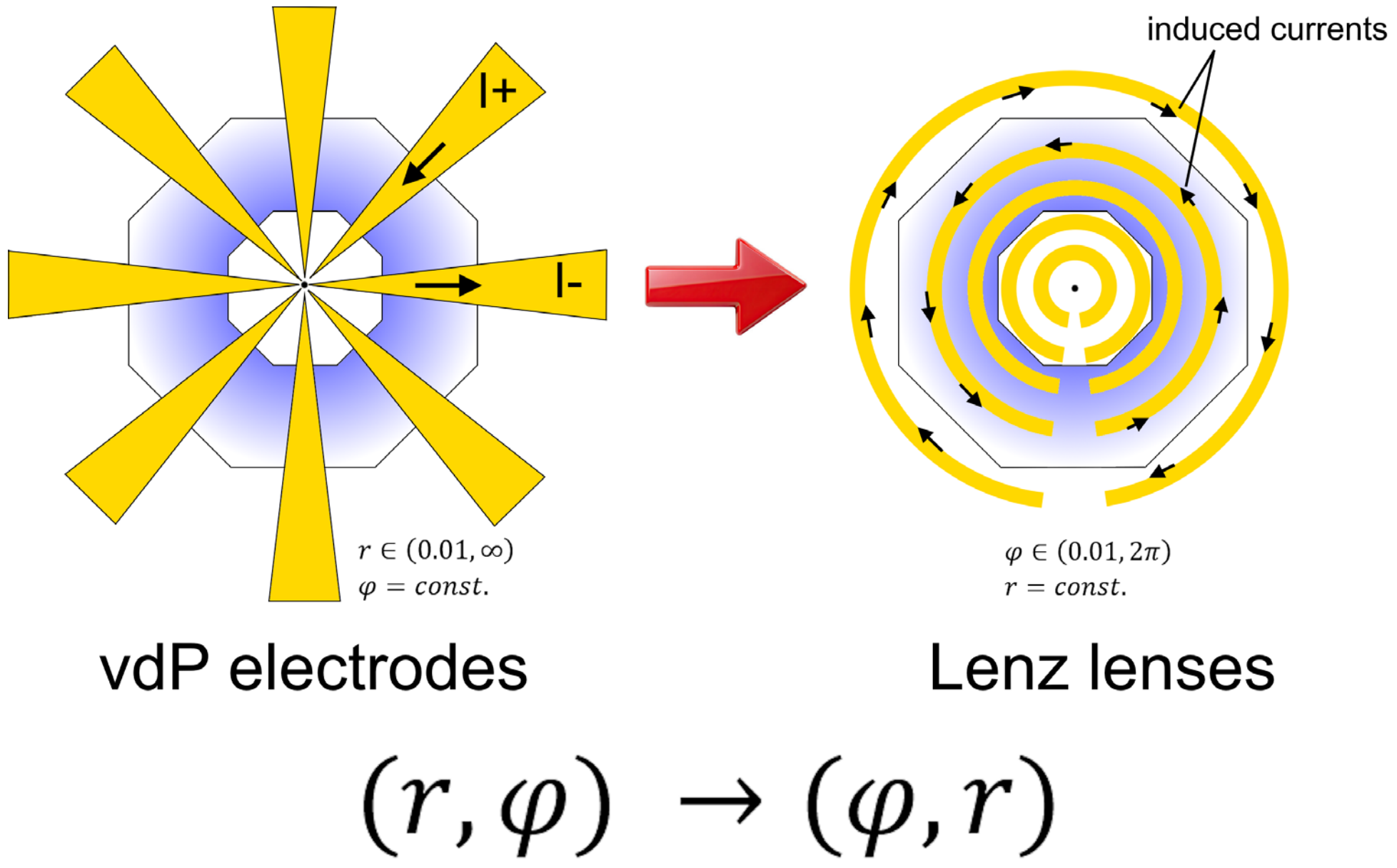

**Figure S2.** Mathematical relationship between the Van der Pauw system and the Lenz lens system as products of swap transformation of polar coordinates.

High-resolution synchrotron X-ray diffraction measurements of initial LaSc alloy were performed at beamline ID31 at the European Synchrotron Radiation Facility (ESRF). The sample powders were loaded into cylindrical slots (approx. 1 mm thickness) held between Kapton windows in a high-throughput sample holder.

Each sample was measured in transmission geometry with an incident X-ray energy of 75.051 keV ($\lambda$ = 0.16520 Å). Measured intensities were collected using a Pilatus CdTe 2M detector (1679×1475 pixels, 172×172 $\mu m^2$ each) positioned with the incident beam in the corner of the detector. The sample-to-detector distance was approximately 1.5 m for the high-resolution measurements. Background measurements for the empty windows were measured and subtracted. NIST SRM 660b (LaB6) was used for geometry calibration performed with the software pyFAI followed by image integration including a flat-field, geometry, solid-angle, and polarization corrections.

## *Theory*

Computational predictions of thermodynamic stability of the (LaSc)–H phases at 150 and 30 GPa were carried out using the variable composition evolutionary algorithm USPEX [6-8]. The maximum number of atoms was 44, range of search was up to $(LaSc)_4H_{40}$. The first generation consisting of 120 structures was produced using the random symmetric [9] and random topological [10] generators, whereas all subsequent generations (80 structures) contained 30% of random structures and 70% of those created using heredity, softmutation, and transmutation operators. The evolutionary searches were combined with structure relaxations using density functional theory (DFT) [11,12] within the Perdew–Burke–Ernzerhof functional (generalized gradient approximation) [13] and the projector augmented wave method [14,15] as implemented in the VASP code [16-18]. The kinetic energy cutoff for plane waves was 600 eV. The Brillouin zone was sampled using Γ-centered *k*-points meshes with a resolution of $2\pi\times0.05$ Å$^{-1}$. This methodology is similar the one used in our previous works [19].

The calculations of the critical temperature of superconductivity $T_c$ and structural relaxation of $LaScH_n$ polyhydrides were carried out using Quantum ESPRESSO (QE) package [20,21]. The phonon frequencies and electron–phonon coupling (EPC) coefficients were computed using density functional perturbation theory [22], employing the plane-wave pseudopotential method and Perdew–Burke–Ernzerhof exchange–correlation functional [14,15]. La.pbe-spfn-kjpaw_psl.1.0.0.UPF, Sc.pbe-spn-kjpaw_psl.1.0.0.UPF and H.pbe-kjpaw_psl.1.0.0.UPF pseudopotentianls were used. In our ab initio calculations of superconductivity within the optimized tetrahedron method [23] in *P*4/*mmm*-$LaScH_{20}$ (200 GPa and 180 GPa) obtained by the replacement La → Sc of 2 La atoms by 2 Sc atoms in $X_4H_{40}$ $Fm\bar{3}m$ structure, the first Brillouin zone was sampled using a 8 × 8 × 8 *k*-points mesh and a denser 16 × 16 × 16 *k*-points mesh was used for electron–phonon coupling calculations. Phonons were computed on a 2× 2 × 2 shifted *q*-points grid. In case of *Pmmm*-$LaScH_{24}$ (193 GPa) obtained from $Pm\bar{3}m$-$Sc_2H_{24}$ [24] via Sc→ La substitution with corresponding relaxation, we applied 12× 12 × 12 *k*-points and 4× 4 × 4 shifted *q*-points grids. $T_c$ was calculated by solving Eliashberg equations [25] using the iterative self-consistent method [26]. More approximate estimates of $T_c$ were made using the Allen–Dynes formula [27].

The most important conclusion from the theoretical analysis is that the $LaScH_{20}$ ($XH_{10}$) structure has lower $T_c$ ~ 200-220 K than structures with a higher hydrogen content like $LaScH_{24}$ ($XH_{12}$) which have a significantly higher $T_c$ of about 260-280 K. Qualitatively this is consistent with experiment even though the theoretical structures have a different symmetry than the experimental phases.

## 2. Elemental analysis of La-Sc alloys

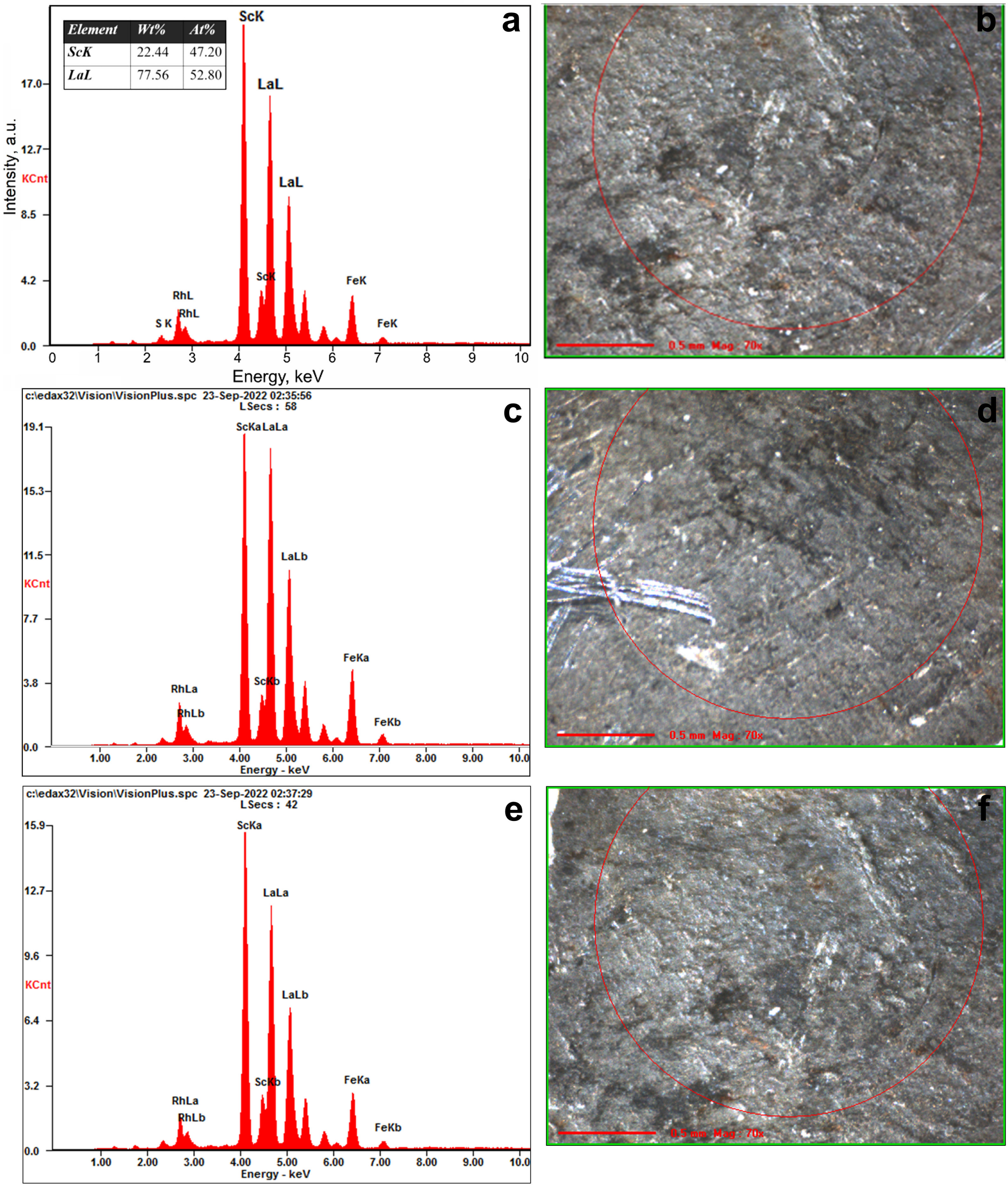

**Figure S3.** XRF analysis and photo of the LaSc 1:1 alloy in the area marked with a red circle (d = 0.5 mm), where the concentration of elements was: (a, b) Sc (K-line) 24.09 (weight, %) 49.50 (atom%), La (L-line) 75.91 (weight, %) 50.50 (atom %); (c, d) Sc (K-line) 19.09 (weight, %) 42.17 (atom %), La (L-line) 80.91 (weight, %) 57.83 (atom %); (e, f) Sc (K-line) 22.76 (weight, %) 47.65 (atom %), La (L-line) 77.24 (weight, %) 52.35 (atom %).

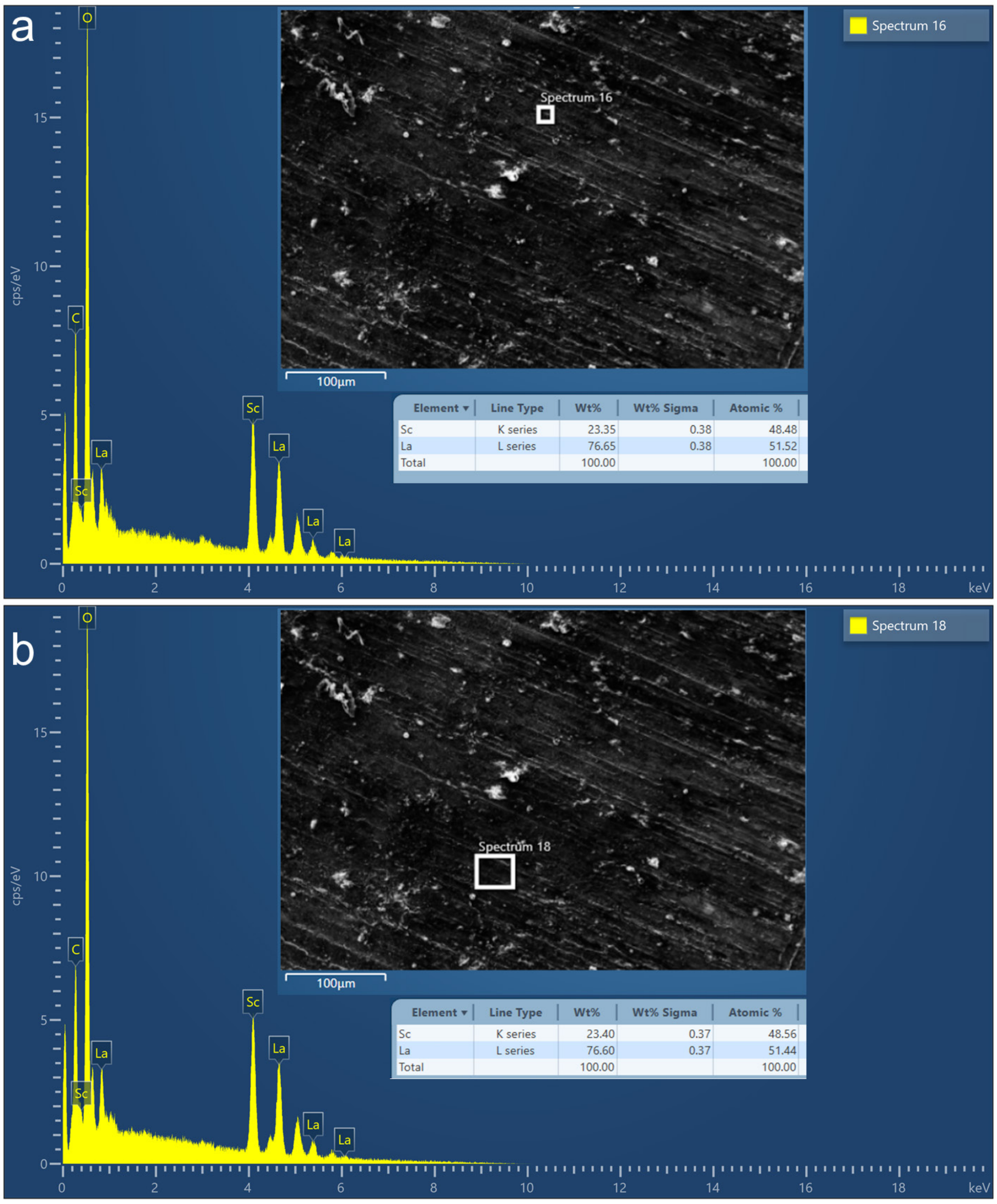

**Figure S4.** Energy dispersive X-ray (EDX) analysis of the La:Sc 1:1 alloy composition performed in two regions (No. 16, 18) the dimensions of which correspond to the dimensions of the loaded samples of DACs LS-1-5. Despite surface oxidation, scandium protects the sample from full degradation. The atomic ratio of La and Sc on the scale of several tens of microns is close to 1:1.

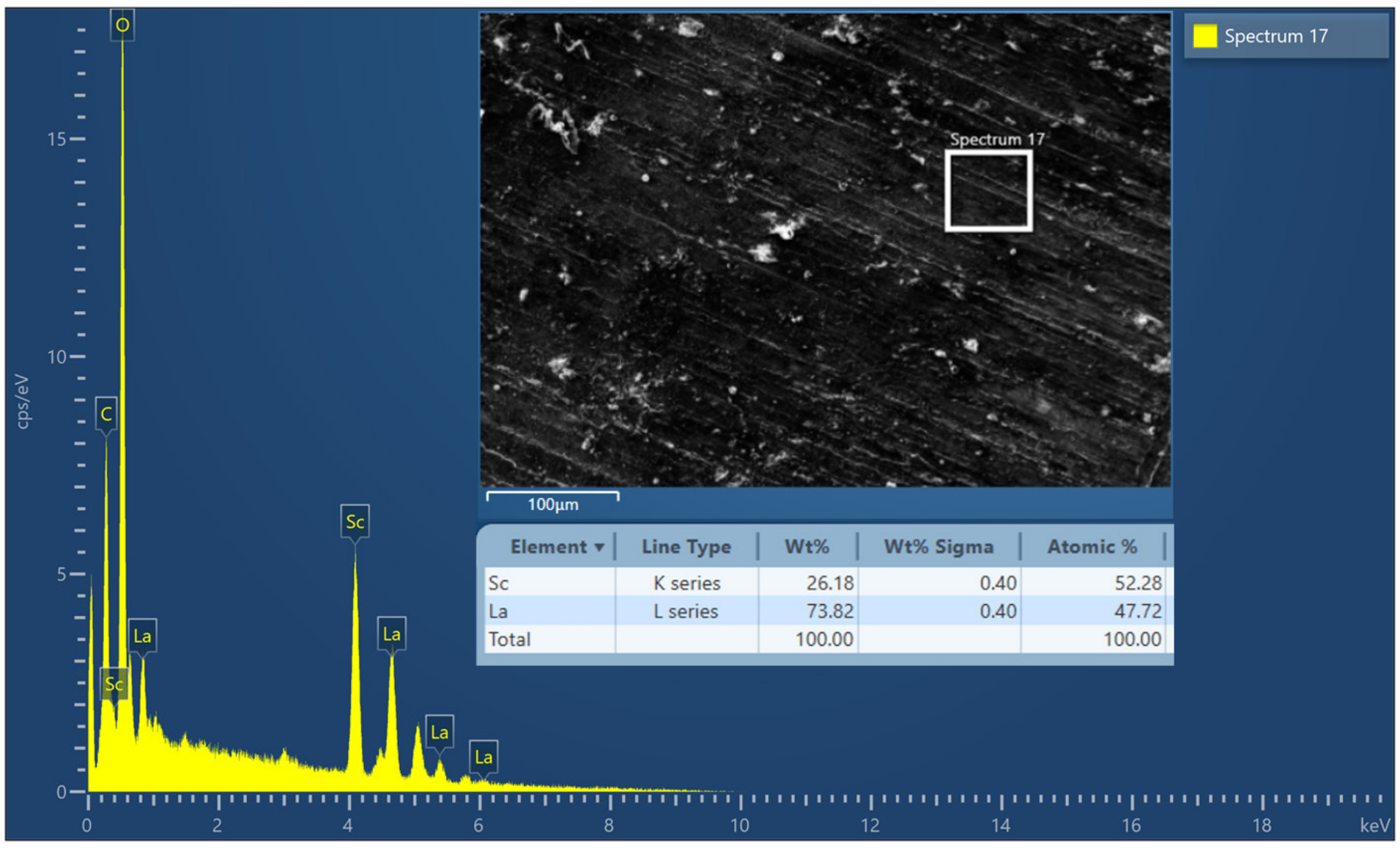

**Figure S5.** Energy dispersive x-ray analysis (EDX) of the composition of the La:Sc 1:1 alloy in the regions (No. 17) of approximately 100 µm size.

## 3. Additional X-ray diffraction data

The initial LaSc alloy (1:1) for the high-pressure synthesis was analyzed using synchrotron radiation of wavelength 0.1652 Å at atmospheric pressure. Scandium and its oxide film effectively protects the LaSc alloy from corrosion in air atmosphere, so that during a year of storage the samples were practically unchanged. We studied it and used without any additional protection. According to XRD analysis (Supporting Figure S6), the LaSc sample consists of two crystalline phases: *dhcp*-(La,Sc) with unit cell parameters somewhat different from *dhcp*-La. For pure lanthanum a = 3.7742 Å, c = 12.171 Å, and V = 37.53 Å$^3$/La, which is about 2 Å$^3$ higher than in our experiment. At the same time, for pure *hcp*-Sc a = 3.3089 Å, c = 5.2680 Å, and V = 24.97 Å$^3$/Sc, which is 0.64 Å$^3$ smaller than in the experimental *hcp*-(La,Sc) structure. Thus, the initial alloy consists of fused crystallites of *dhcp*-(La,Sc) with 15 at.% of Sc, and *hcp*-(La,Sc) with 5 at.% of La. Despite this, EDX and XRF analysis shows that lanthanum and scandium are uniformly distributed throughout the LaSc alloy on scales of 10-50 μm (Supporting Figures S3-S5).

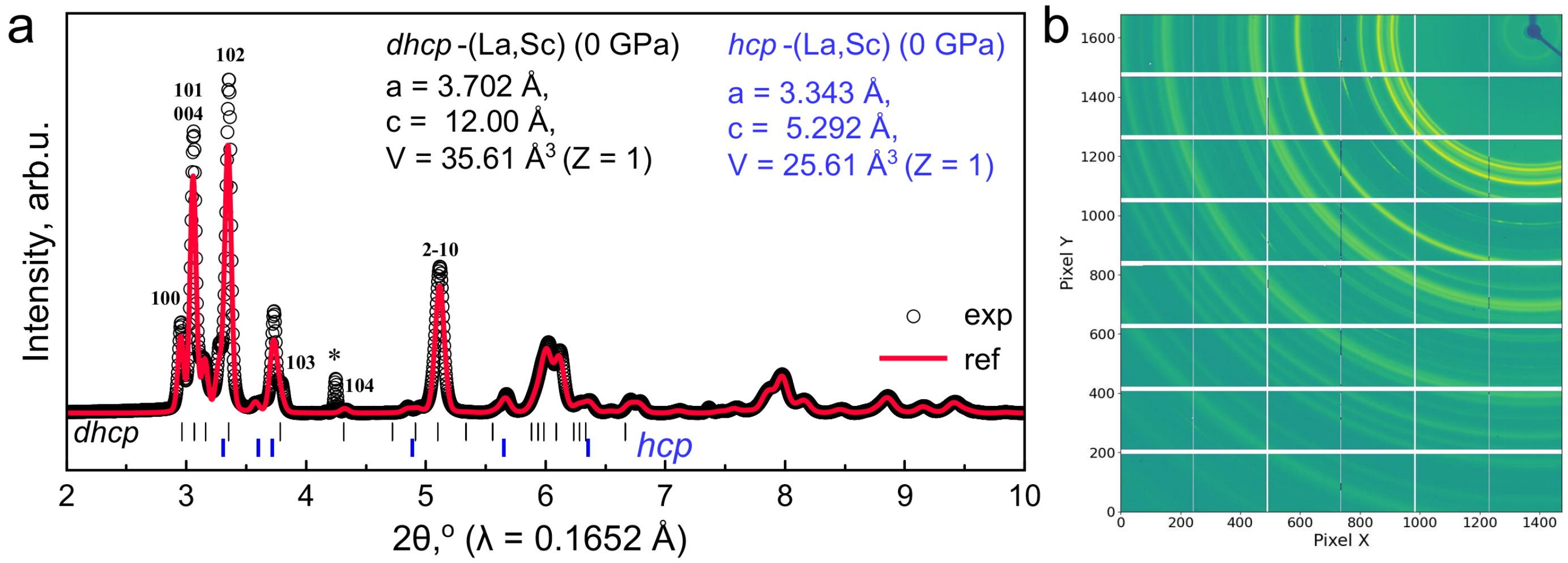

**Figure S6.** Diffraction pattern of the original LaSc alloy (1:1) at 0 GPa. (a) Experimental XRD pattern (black circles), the Le Bail refinement of the *dhcp* and *hcp* structures of (La,Sc) alloy (red curve), and parameters of their crystal lattices. (b) Original XRD image.

The resulting diffraction pattern of DAC LS-1 (Figure 1) qualitatively contains "point" diffraction lines (indicated by "□" in Supporting Figure S7), which correspond to the expected position of cubic reflections of distorted $Fm\bar{3}m$-(La,Sc)$H_{12}$. Along with this ternary hydride phase, there is a redistribution of peak intensity in favor of the 002 reflection, as well as additional diffuse reflections 210 and 211 accompanying the cubic reflections 111 and 002 (Supporting Figure S7a-c, e-h), which may indicate the formation of a distorted *Fmmm*-(La,Sc)$H_{12}$ ternary phase as well. If we include in the analysis a true ternary phase *Pmmm*-$LaScH_{24}$ with ordered La and Sc sublattices, then it shows that for any regular replacement of two La atoms by two Sc atoms in the $La_4H_{40}$ cell, the resulting compound will demonstrate almost the same X-ray diffraction patterns, shown in green in Supporting Figure S7j. As can be seen from this figure, there is a good agreement between the experimental diffraction pattern and the expected diffraction spectrum of the $LaScH_{24}$ with a regular arrangement of La and Sc atoms. However, the absence of a diffraction peak at 6 deg. (Supporting Figure S7j) and the dependence of the relative intensity of individual XRD peaks on the point of study (Supporting Figure S8) make the possibility of $LaScH_{24}$ formation unlikely.

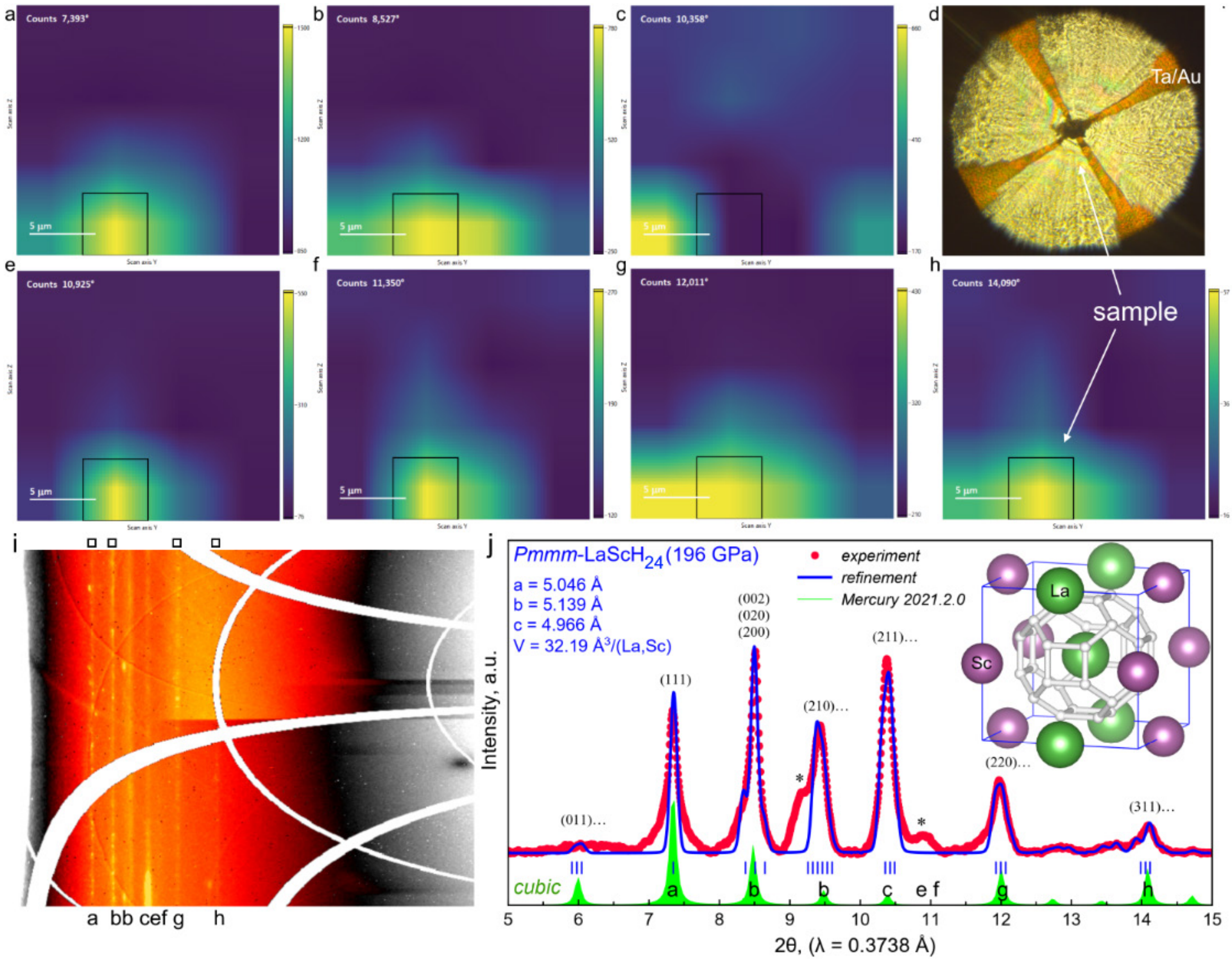

**Figure S7.** Powder X-ray diffraction analysis and the Le Bail refinement of the *Fmmm*-(La,Sc)$H_{12}$ (distorted cubic, pseudo $Fm\bar{3}m$) unit cell in the DAC LS-1 at 196 GPa. XRD map of the sample is shown on pictures (a-c) and (e-h) and indicates that all diffraction peaks belong to the same region of the diamond anvil (left-bottom corner). Panel (d) shows photo of the (La,Sc)$H_{12}$ sample on the culet with four connected Ta/Au electrodes. (i) X-ray diffraction image ("cake") of the sample. (j) Le Bail refinement of the unit cell parameters of *Fmmm*-(La,Sc)$H_{12}$ (JANA2006): red points mark experimental data, the blue line – is the refinement, the green one – is predicted XRD for undistorted cubic (La,Sc)$H_{12}$ obtained by regular substitution of La → Sc in the cubic $Fm\bar{3}m$-$La_2H_{20}$. Asterisks mark unexplained reflections.

The fact that different X-ray reflections belong to two slightly different areas of the sample located 10 µm from each other may indicate that the sample is multiphase and some of the *Fmmm*-(La,Sc)$H_{12}$ reflections actually belong to another phase, for example, to cubic (La,Sc)$H_2$. This is also reflected in Supporting Figure S8. Thus, we come to the conclusion that the most probable composition of the sample includes two $Fm\bar{3}m$ phases:

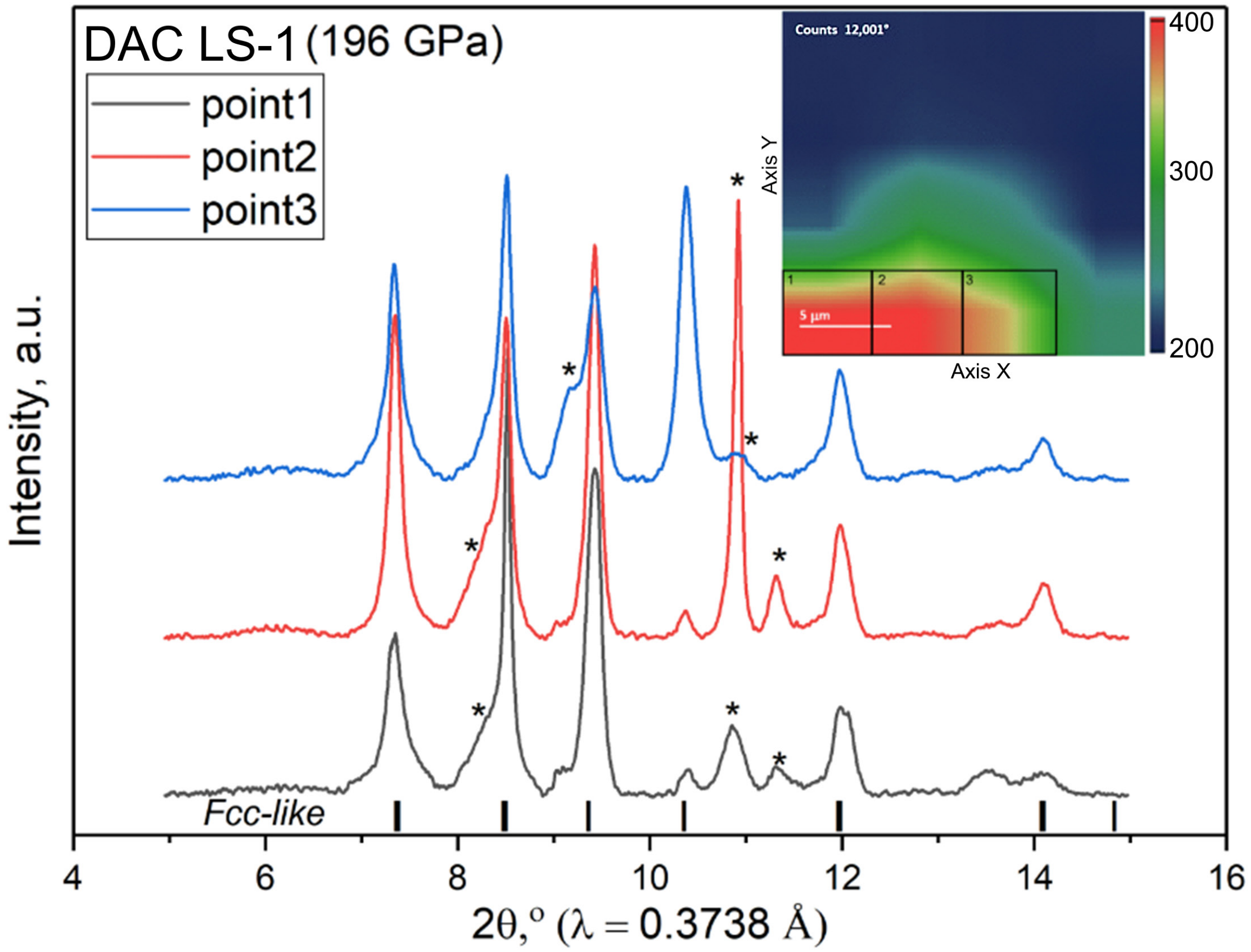

**Figure S8.** Powder X-ray diffraction patterns of the sample in the DAC LS-1 at 196 GPa. A series of integrated diffraction patterns at three points *(#1-3)* shown in the inset. Asterisks mark unexplained reflections. The positions of the expected diffraction peaks from undistorted cubic $LaScH_{20}$ (marked "*Fcc-like*") obtained by regular substitution of La → Sc in the $Fm\bar{3}m$-$La_4H_{40}$ are shown by black dashes. Inset: intensity distribution map of the peak at 12 deg., related to the main $Fm\bar{3}m$ phase.

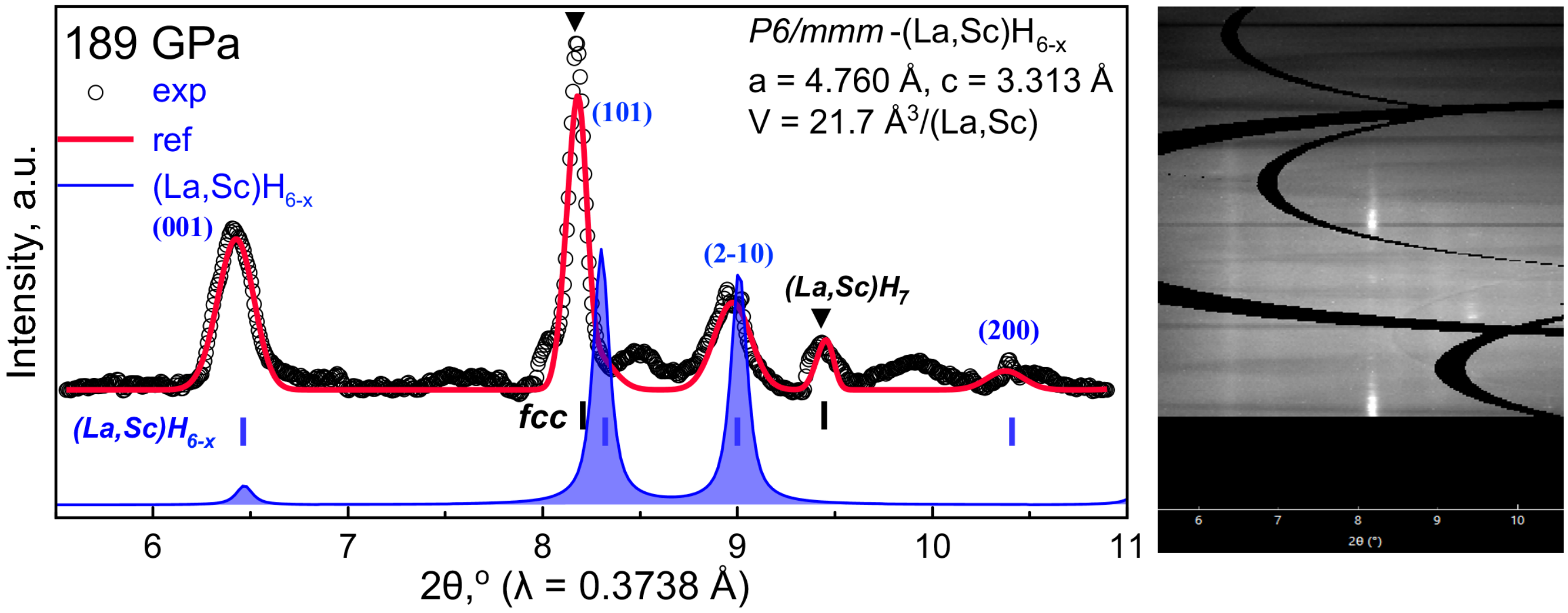

**Figure S9.** Powder X-ray diffraction analysis and Le Bail refinement of $P6/mmm$-$(La,Sc)H_{6-x}$ and $Fm\bar{3}m$-$(La,Sc)H_{\sim 7}$ in the DAC LS-3 at 189 GPa. Black points mark experimental data, the red line – is the refinement, the blue one – is predicted XRD (Mercury 2021 [2]). As can be seen, the assumption of the presence of a cubic phase $(La,Sc)H_{\sim 7}$ with a = 4.525 Å, V = 23.16 Å$^3$/(La,Sc) at 189 GPa makes it possible to explain some additional diffraction peaks, but significantly reduces the volume of the $P6/mmm$-$(La,Sc)H_{6-x}$ phase.

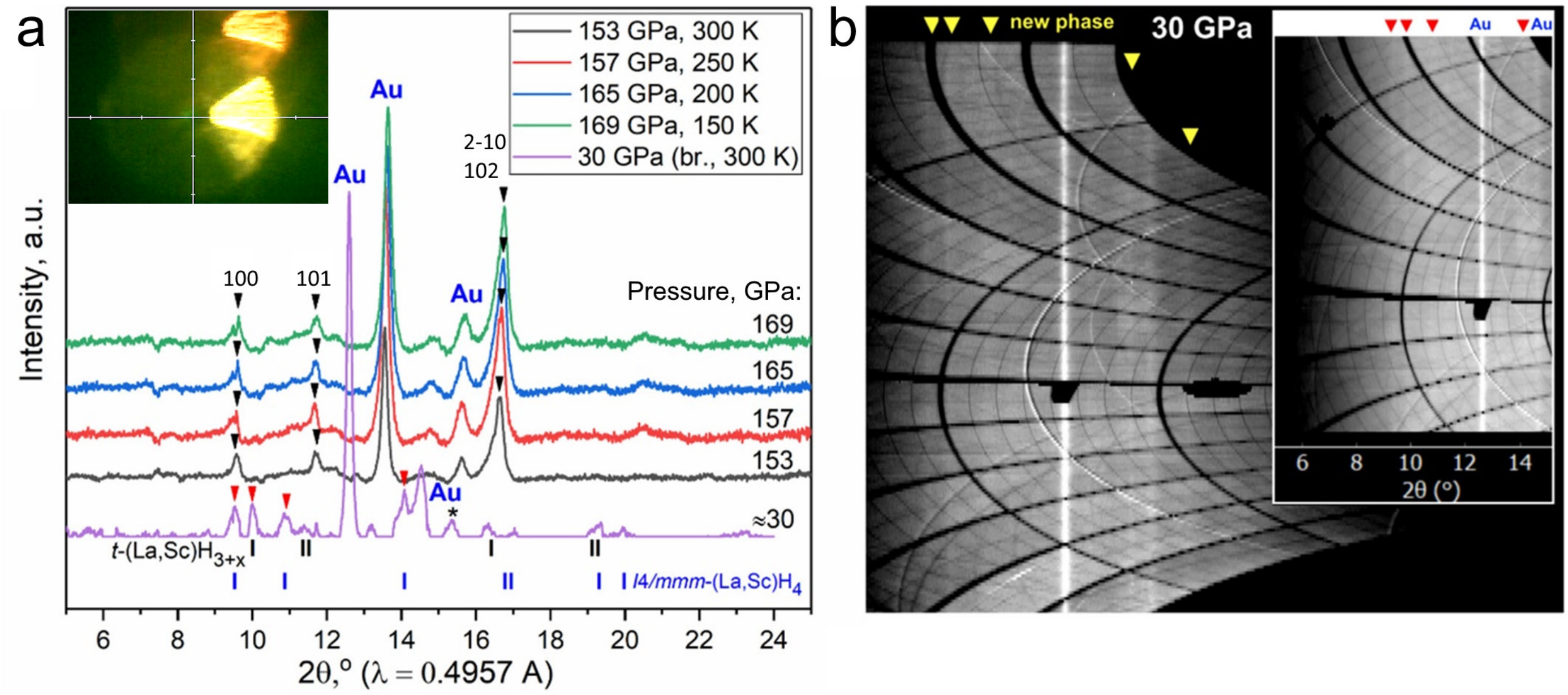

**Figure S10.** (a) A series of X-ray diffraction patterns obtained during the sample cooling from 300 to 150 K, as well as one XRD pattern measured after DACs failure (30 GPa). Inset: optical image of the sample in the cryostat. (b) Diffraction image obtained after the DAC collapse. The yellow triangles show reflections from the new (La,Sc)$H_x$ phases formed when the pressure was dropped to 30 GPa.

**Table S3.** Unit cell parameters of *I*4/*mmm*-(La, Sc)$H_{4-x}$ (Z = 4) and (La, Sc)$H_3$ (Z = 8) refined from the low-temperature XRD experiment. Due to the low intensity of the diffraction pattern, we do not expect to detect superstructural peaks corresponding to the scandium sublattice. Calculated unit cell volumes relate to 1:1 La:Sc ratio.

| Pressure, GPa | Possible structural solution | a, Å | c, Å | $V_{exp}$, Å$^3$/metal atom | $V_{DFT}$, Å$^3$/metal atom |
|---|---|---|---|---|---|
| 30 | *I*4/*mmm*-(La, Sc)$H_{4-x}$ | 2.890 | 6.010 | 25.08 | 28.88<br>29.01* |
| 30 | *P*4$_1$-(La, Sc)$H_{3-x}$<br>*Fm*$\bar{3}$*m*-(La, Sc)$H_3$ | 4.903<br>4.930 | 9.950<br>- | 29.89<br>29.94 | 33.09<br>28.24** |
| 153 (300 K) | *P*6$_3$/*mmc*-(La,Sc)$H_{6-x}$ | 3.424 | 4.306 | 21.85 | 21.0 ($ScH_6$) |
| 157 (250 K) | *P*6$_3$/*mmc*-(La,Sc)$H_{6-x}$ | 3.489 | 4.157 | 21.92 | 20.86 ($ScH_6$) |
| 165 (200 K) | *P*6$_3$/*mmc*-(La,Sc)$H_{6-x}$ | 3.450 | 4.171 | 21.50 | 20.6 ($ScH_6$) |
| 169 (150 K) | *P*6$_3$/*mmc*-(La,Sc)$H_{6-x}$ | 3.446 | 4.161 | 21.39 | 20.43 ($ScH_6$) |

*Averaged value: 0.5×V($ScH_4$) + 0.5×V($LaH_4$)

**Averaged value: 0.5×V($ScH_3$) + 0.5×V($LaH_3$)

Note that in the low-temperature diffraction patterns (Figure 2e), the 100 reflection looks split below 250 K, which could indicate some low-symmetry distortion. Considering the changes in lattice parameters in Table S3 for *P6₃/mmc*-(La,Sc)$H_{6-x}$, first the unit cell volume increases noticeably even though the temperature is reduced from 300 K to 250 K and the pressure is increased. When the temperature is further reduced to 150 K, the parameter remains almost unchanged, but there is a visible reduction in volume due to the compression along the c-axis. This behavior may indirectly indicate a phase transition accompanied by some structural distortion.

# 4. Transport measurements

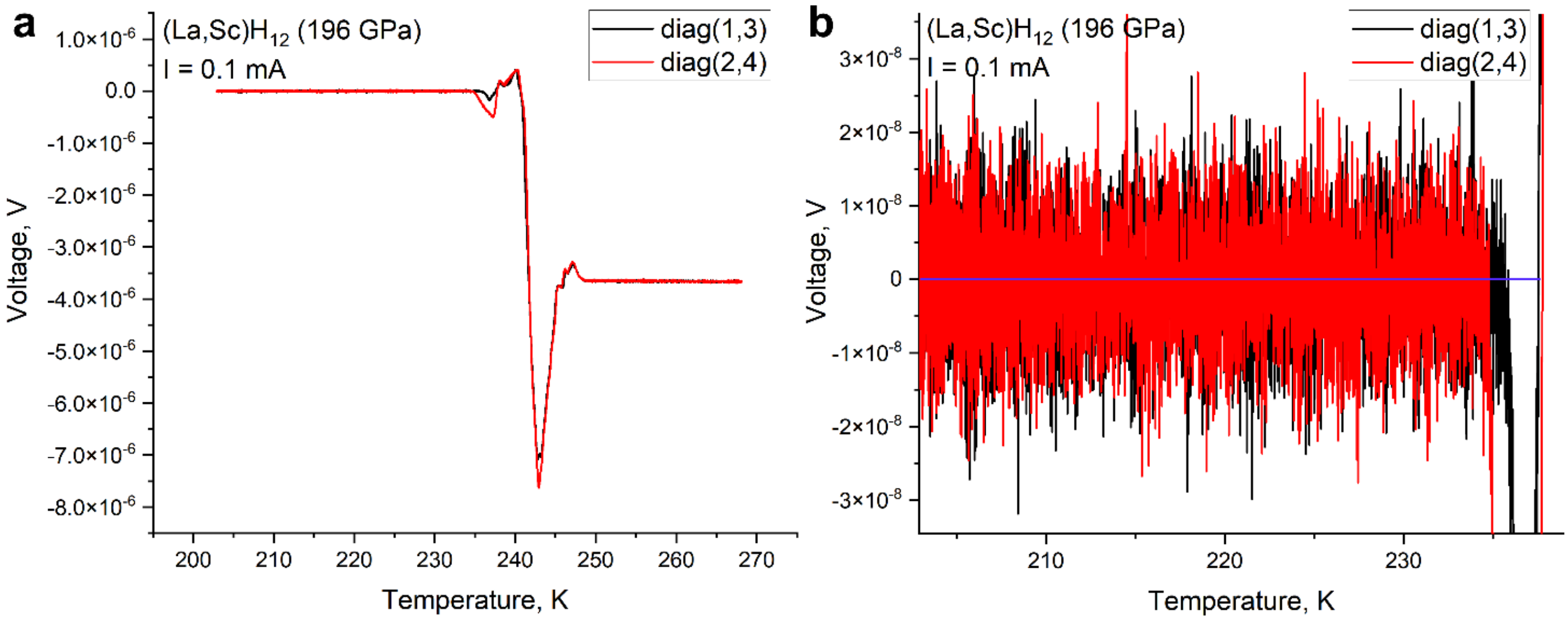

**Figure S11.** (a) The voltage drop across the (La,Sc)$H_{12}$ sample at 196 GPa (DAC LS-1) measured in the diagonal connection of the van der Pauw circuit *diag(1,3)* and *diag(2,4)*. (b) Residual resistance of the sample after transition to the superconducting state. The current was 0.1 mA.

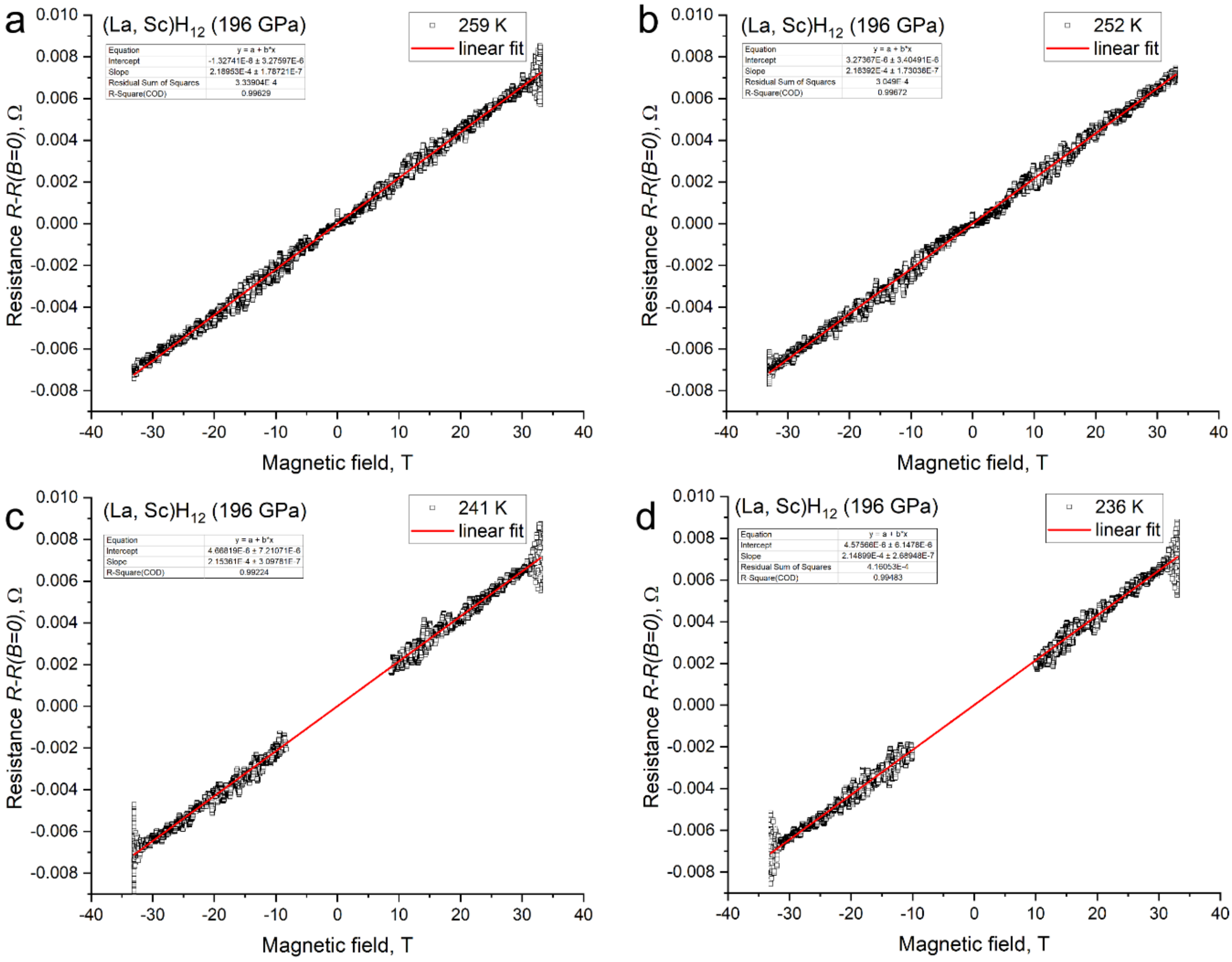

**Figure S12.** Hall effect in (La, Sc)$H_{12}$ at 196 GPa (DAC LS-1) at temperatures: (a) 259 K, (b) 252 K, (c) 241 K, (d) 236 K in steady magnetic fields. The measured Hall coefficient has a negative absolute sign. Red line – is a linear fit. The central part of the *R(B)* curve corresponds to the region of the superconducting transition. It has been removed for clarity.

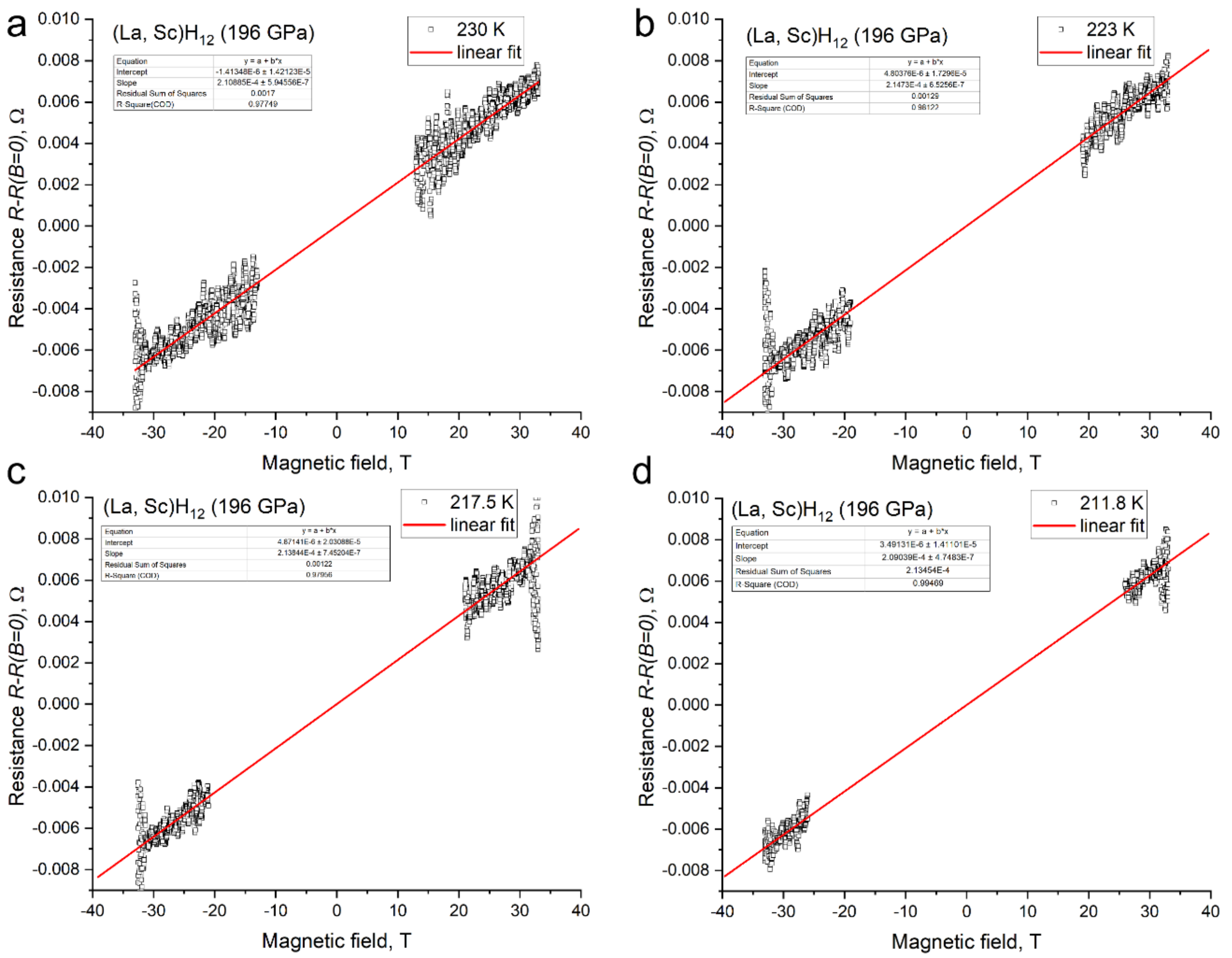

**Figure S13.** Hall effect in (La, Sc)$H_{12}$ at 196 GPa (DAC LS-1) at temperatures: (a) 230 K, (b) 223 K, (c) 217.5 K and (d) 211.8 K in steady magnetic fields. The measured Hall coefficient has a negative absolute sign. Red line – is a linear fit. The central part of the *R(B)* curve corresponds to the region of the superconducting transition. It has been removed for clarity.

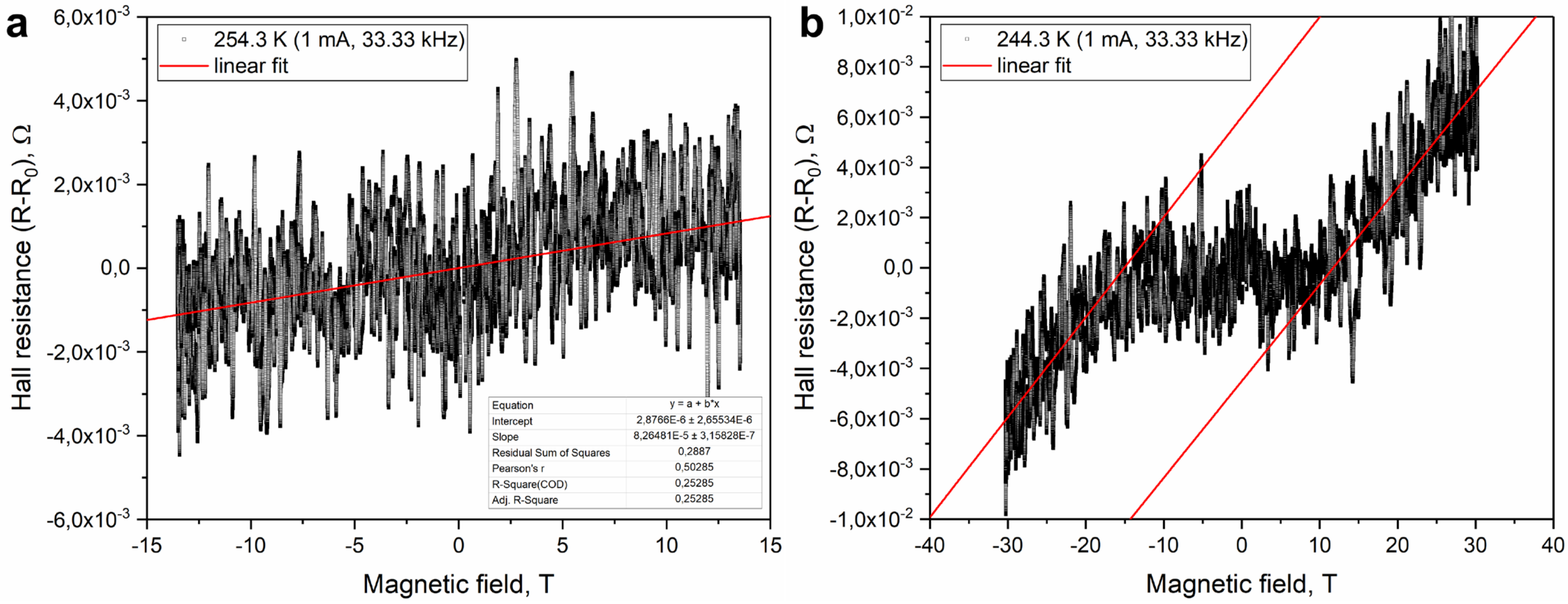

**Figure S14.** Hall effect measurements in (La,Sc)$H_{12}$ at 189 GPa (DAC LS-3) in AC mode in pulsed magnetic field at two temperature points: (a) 254.3 K and (b) 244.3 K. In general, pulsed measurements of the Hall coefficient are much noisier and less reliable than measurements in a steady magnetic field. The nonlinear behavior of the Hall coefficient at 244 K may be due to the appearance of an anomaly in the electrical resistance of the LS-3 sample (Supporting Figure S21).

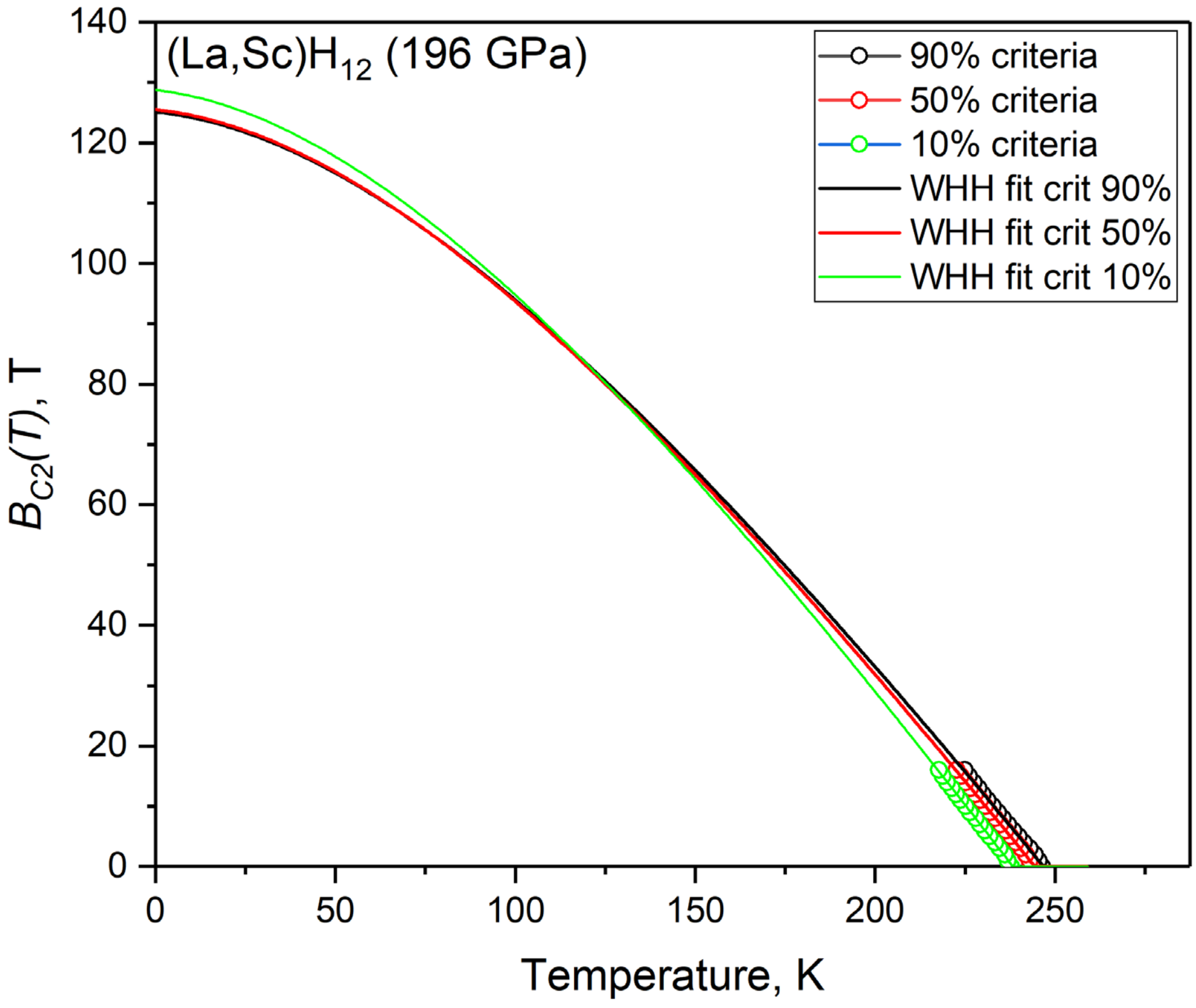

**Figure S15.** Extrapolation of experimental $B_{c2}(T)$ dependences to the region of strong magnetic fields for the (La,Sc)$H_{12}$ sample in DAC LS-1 using the simplified WHH model [28]. All used criteria of 90, 50, 10% of residual resistance give approximately the same $B_{c2}(0)$ = 125-130 T. The study was performed in fields up to 16 T immediately after sample synthesis (DAC LS-1, 196 GPa). This shows that the quality of the LS-1 sample is higher than that of LS-3, where the extrapolated upper magnetic field can reach 200 T and higher.

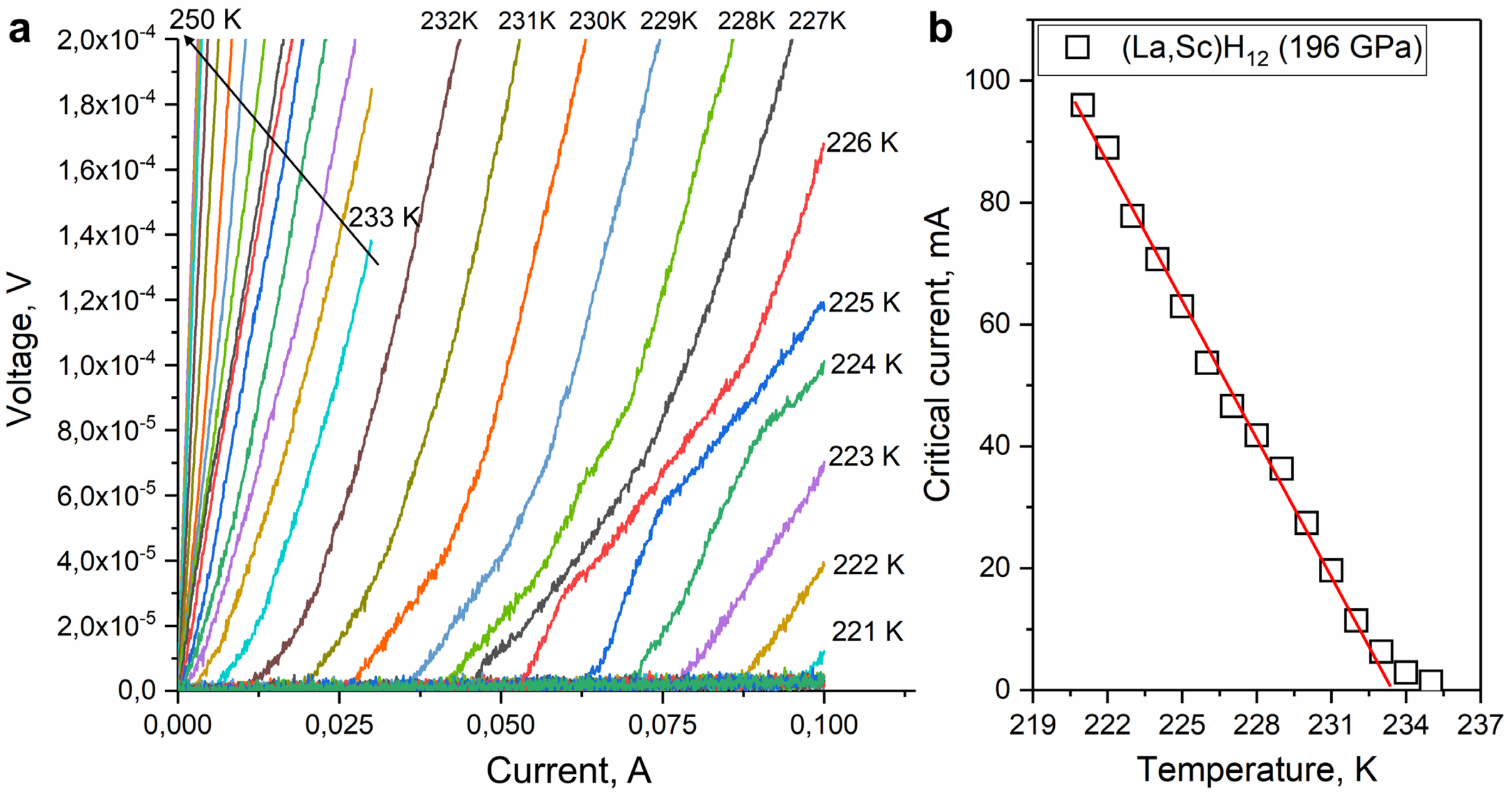

**Figure S16.** Current-voltage (V-I) characteristics and critical current of the $Fm\bar{3}m$-(La,Sc)$H_{12}$ sample in DAC LS-1 at 196 GPa. (a) Current-voltage characteristics in the temperature range from 221 K to 250 K. V-I were taken before the degradation of the sample, which led to a broadening of the superconducting transition and a decrease in the critical current. (b) Dependence of the critical current in the (La,Sc)$H_{12}$ on temperature. The dependence is quasi-linear.

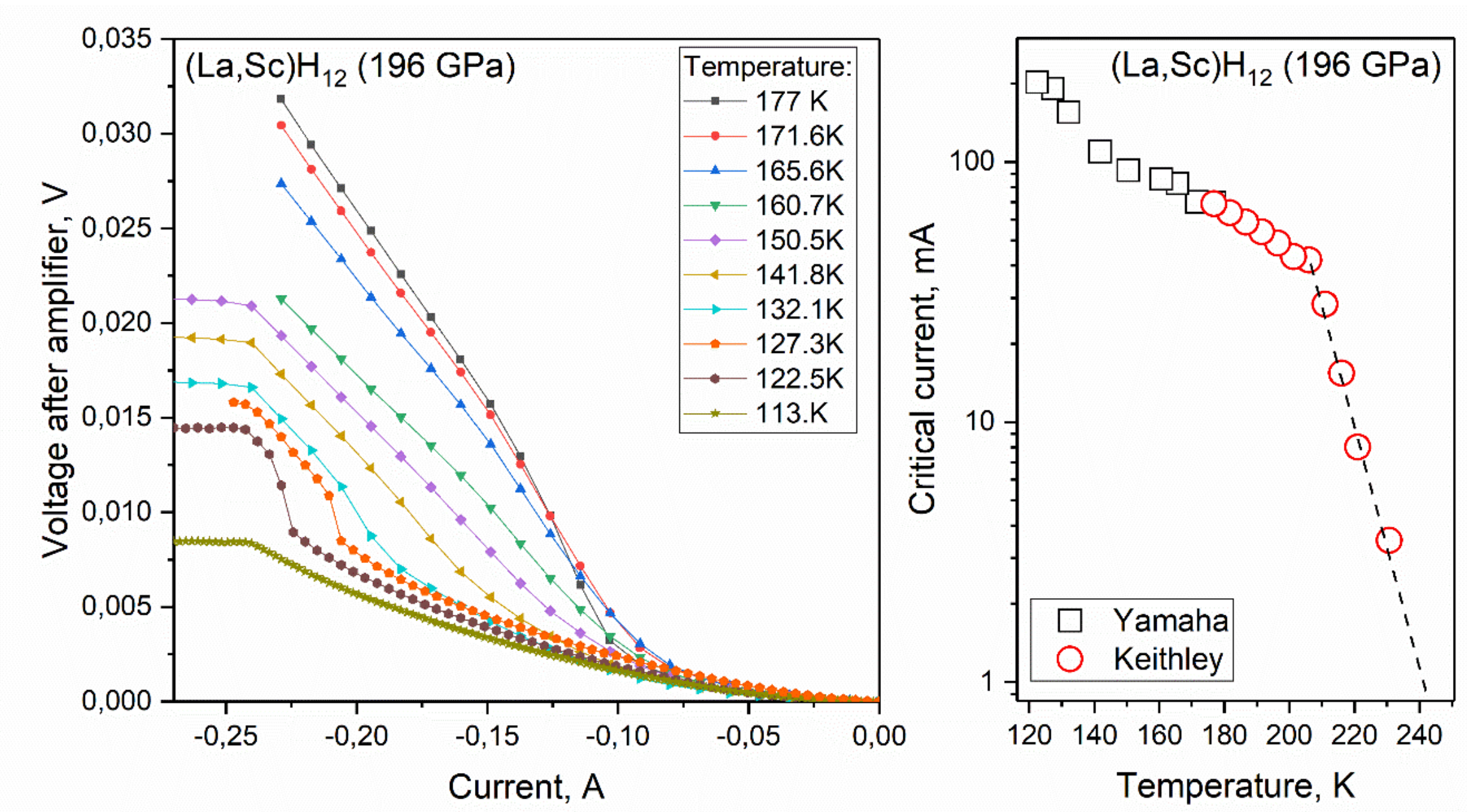

**Figure S17.** Current-voltage characteristics of the sample $Fm\bar{3}m$-(La, Sc)$H_{12}$ (DAC LS-1) studied using a Yamaha PX-3 amplifier of pulsed П-shaped signals generated by a Keithley 6221 source at a temperature of 113 – 177 K. The experiment was carried out after X-ray diffraction studies (degraded sample). The duration of the current pulse was 2 ms, the interval between pulses was 4-5 seconds, the result was averaged over 6 voltage values for each pulse current value. As the current increases, the sample degrades, which leads to the disappearance of the region of sharp linear increase in resistance below 122 K.

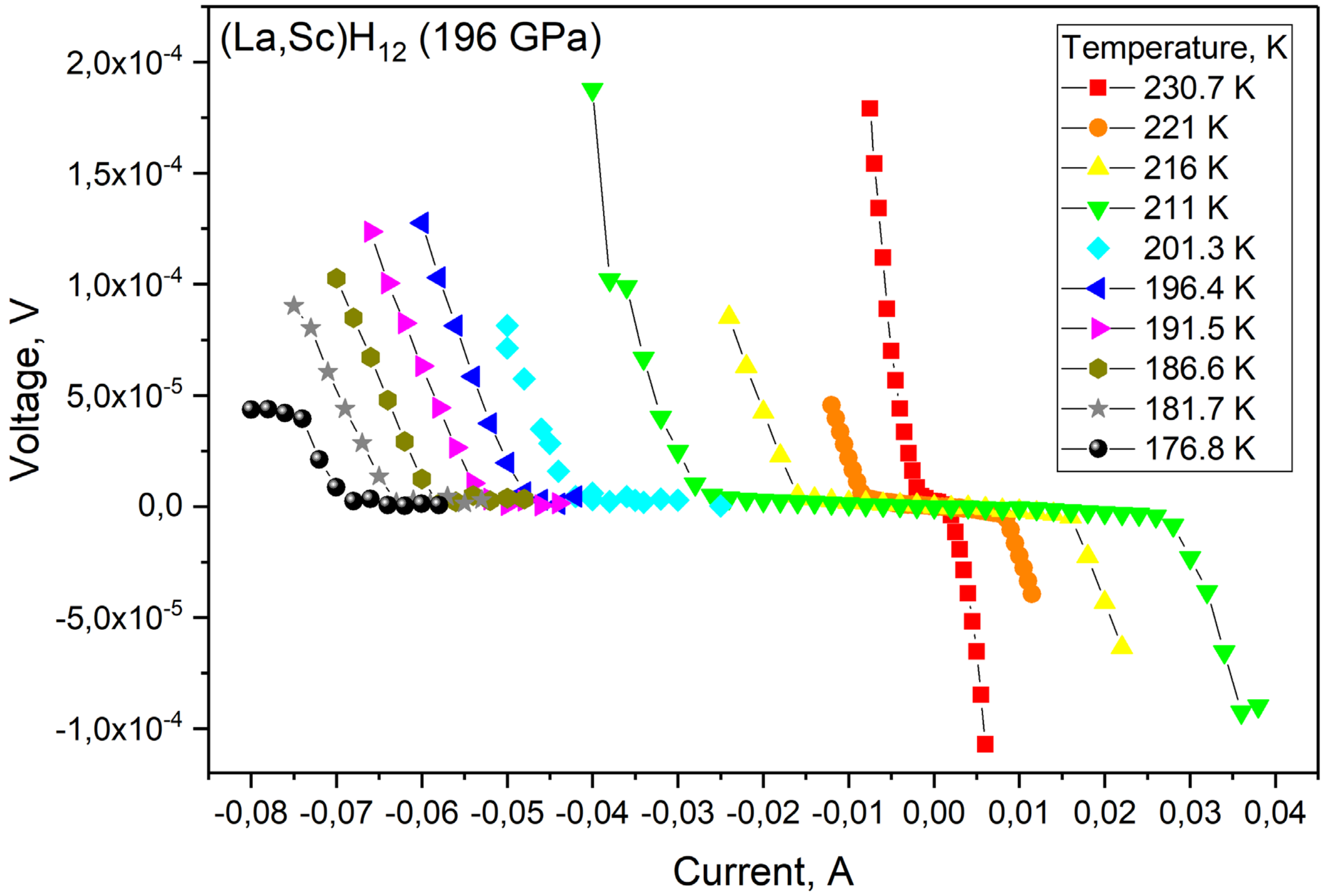

**Figure S18.** Current-voltage characteristics of $Fm\bar{3}m$-(La, Sc)$H_{12}$ (DAC LS-1) taken in a pulsed mode using a Keithley 6221 current source. The pulse length was chosen 5 ms, averaging was carried out over 10 points, the interval between which was 1 second, and the interval between changes in the maximum pulse current value was 3 seconds. The sign of the voltage or current depends on the connection of the contacts to the sample and has no physical meaning.

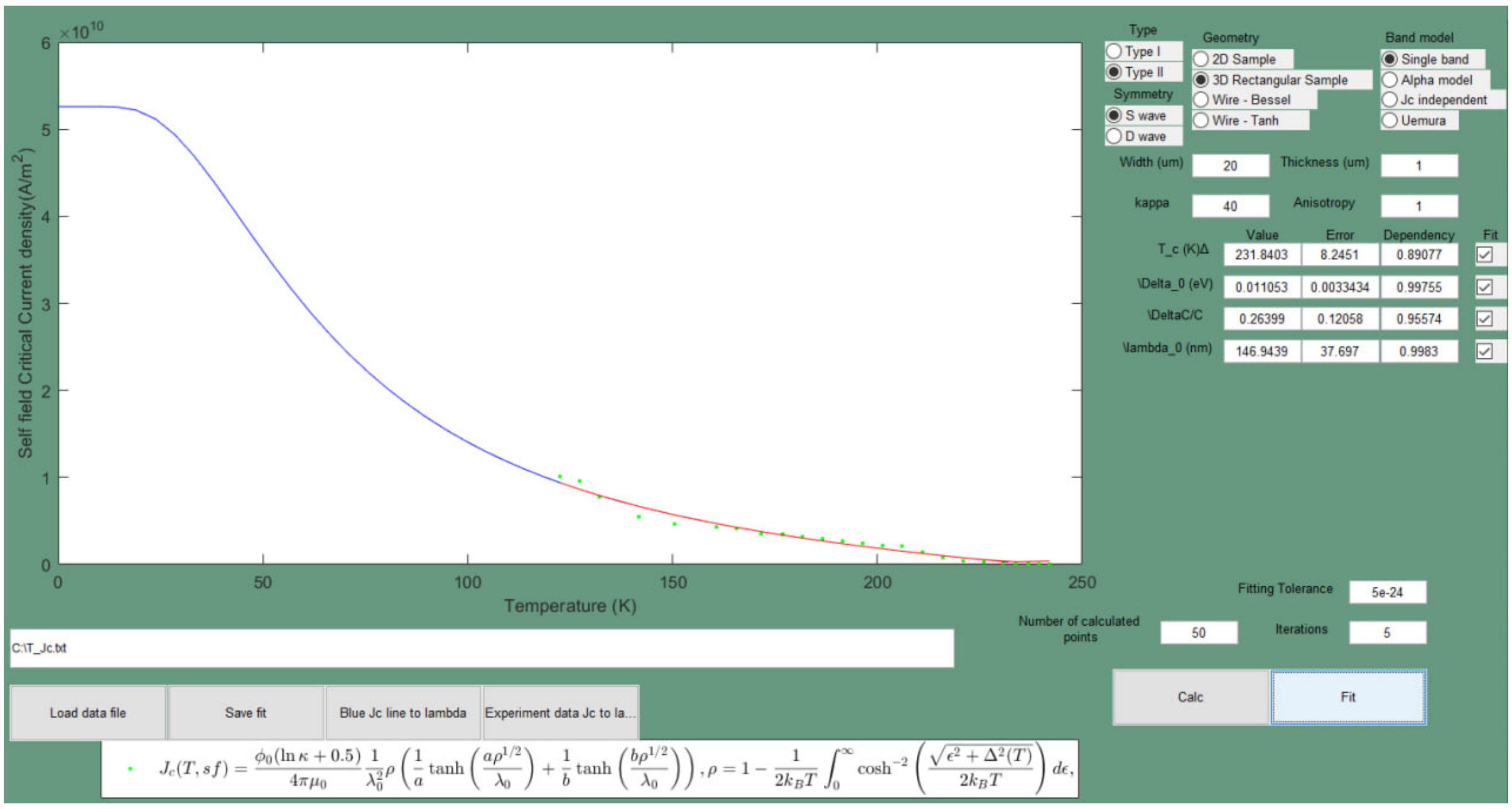

**Figure S19.** An attempt to extract the value of the superconducting gap from the critical current measurements using the Talantsev-Tallon method [29,30]. The obtained value of the superconducting gap is significantly lower than expected: *Δ(0)* = 11 ± 3 meV.

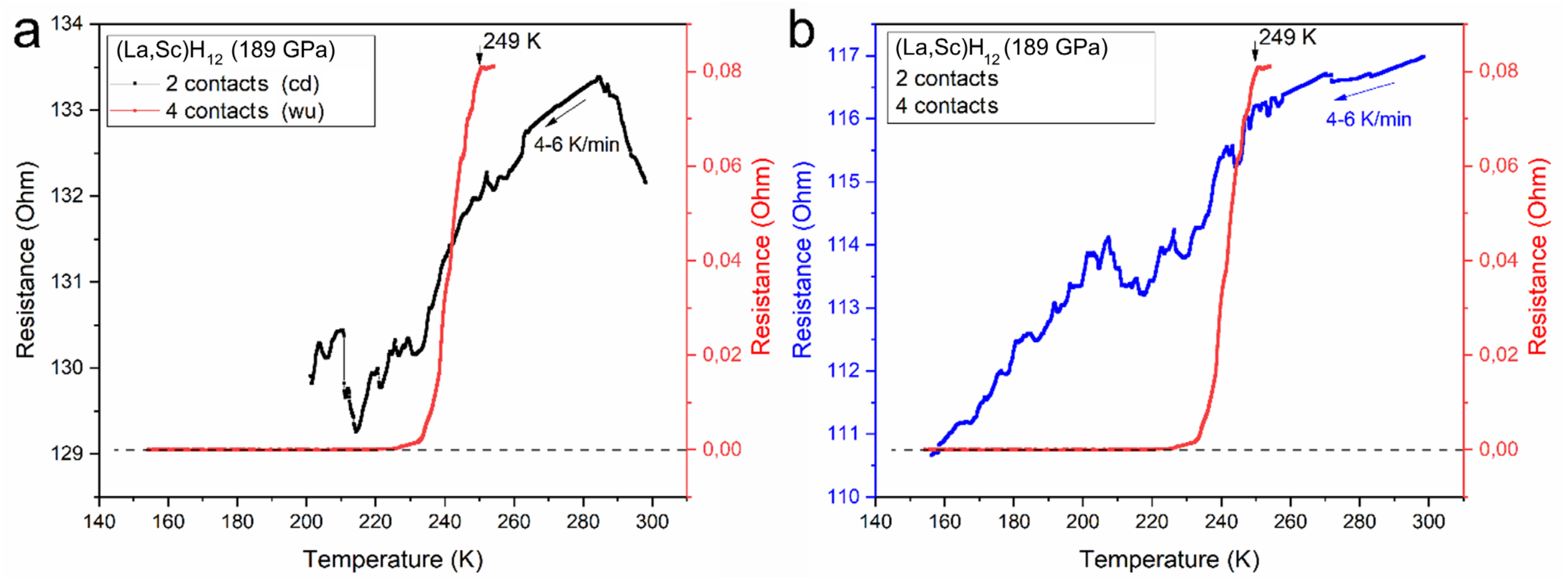

**Figure S20.** Comparison of the dependence of electrical resistance in two-contact ("12" and "34") and four-contact modes (van der Pauw scheme, vdP). The delta mode measurements were used, the excitation current was $I$ = 1 mA. "cd" means cooling down, "wu" means warming up. The cooling rate was chosen to be 4-6 K/min. Due to different heating and cooling rates, the superconducting transitions may be offset by ± 5 K. Measurements in "a" and "b" panels are for different runs. It is easy to see that in two-contact scheme the superconducting transition is almost impossible to detect: *R(T)* behaves unpredictably.

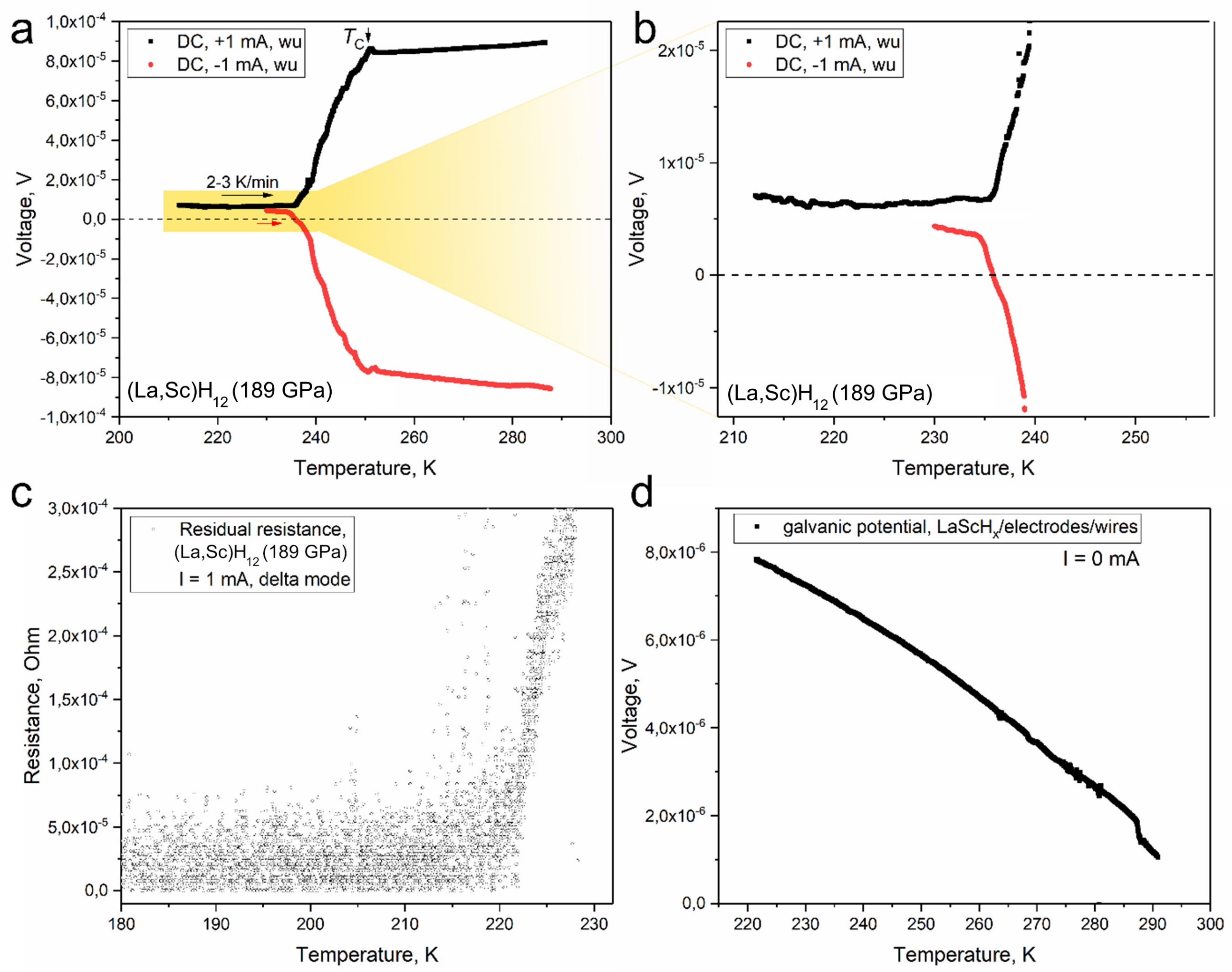

**Figure S21.** Comparison of the dependence of electrical resistance of (La,Sc)$H_{12}$ in DAC LS-3 in a four-probe scheme (van der Pauw, vdP) in DC mode (a, b) without changing the current direction, and in the delta (c) mode. (a) Dependence of the voltage drop across the sample when applying a constant current of different directions (+1 mA and -1 mA). "wu" corresponds to warming up cycle at a rate of 2-3 K/min. (c) Residual resistance of the sample measured in delta mode ($I$ = 1 mA). At 230K the resistance is almost the same as 0.5×[V(+I)-V(-I)]/|I| from graphs (a, b). Figure (d) shows the voltage drop at zero external current, caused by the difference in contact potentials.

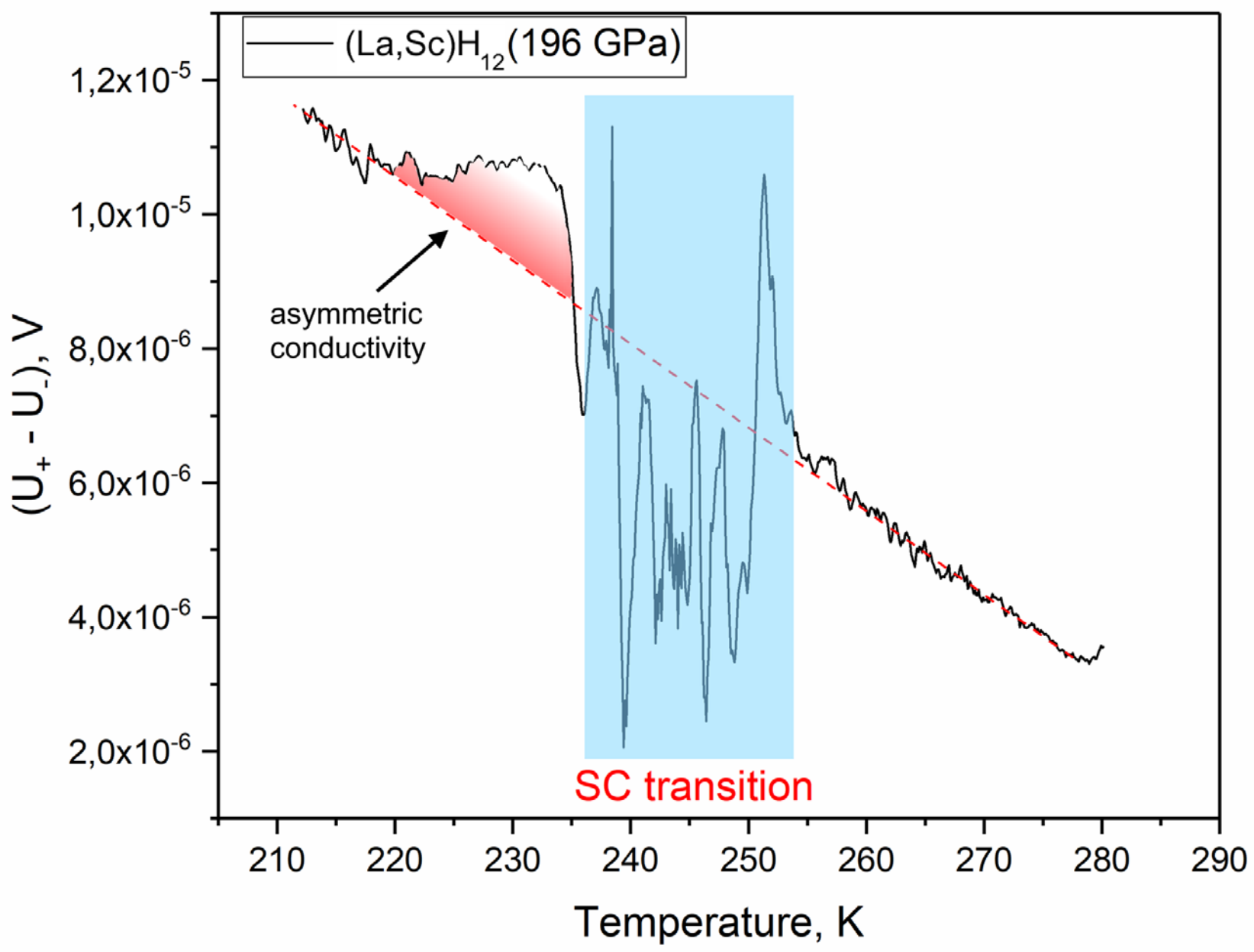

**Figure S22.** This figure shows that subtracting voltage measurements made using DC current ($I = \pm$ 1 mA, $U_+$ means voltage which is measured when +1 mA current passes through) indicates the presence of deviations (conductivity asymmetry) from the linear dependence in the vicinity of the superconducting transition associated with the diode properties of the sample.

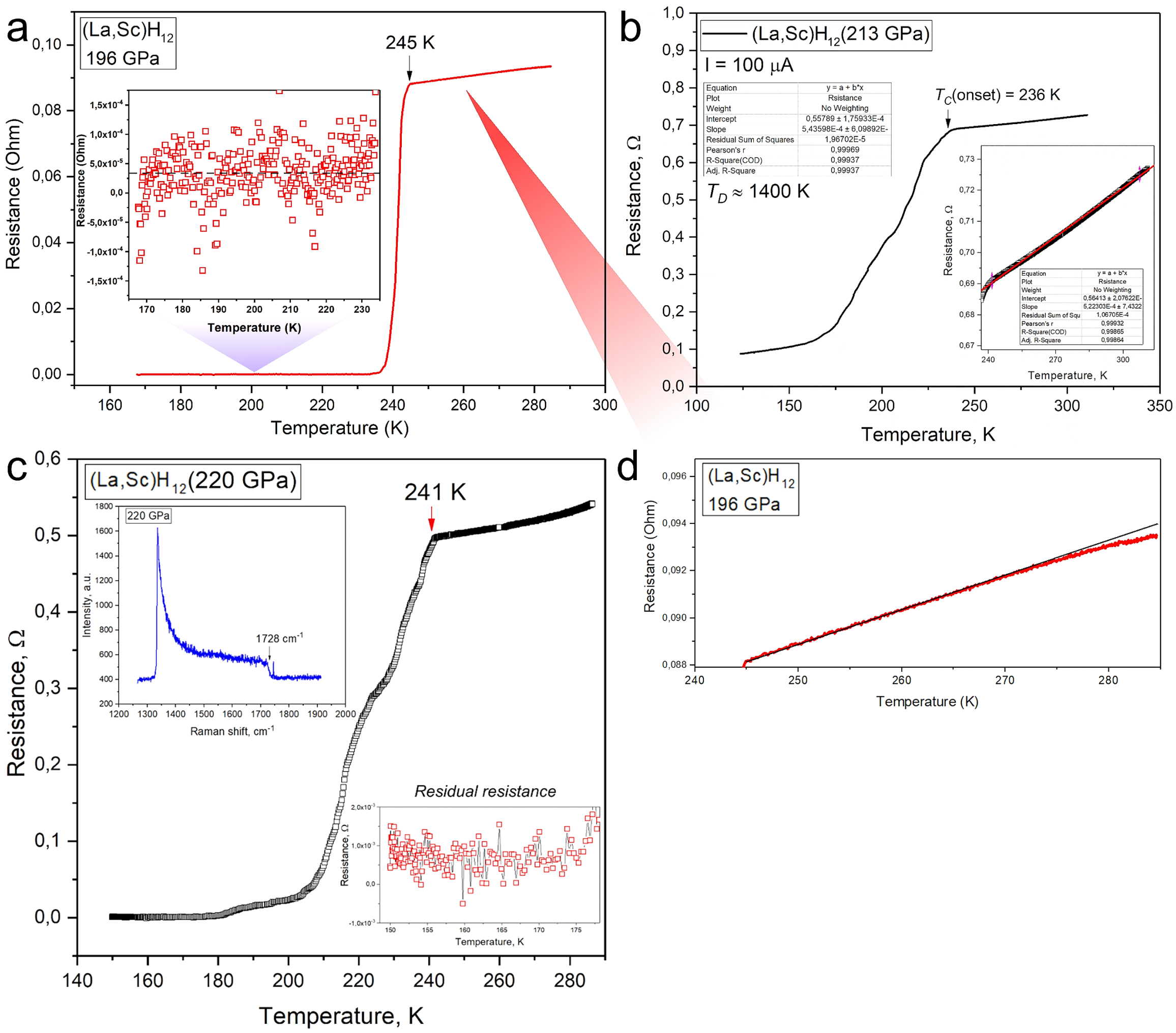

**Figure S23.** Superconducting transitions and temperature dependence of the electrical resistance of (La,Sc)$H_{12}$ at different pressures: (a) 196 GPa (DAC LS-1), (b) 213 GPa, (c) 220 GPa. In panel (b) we can see a superconducting transition in a (La,Sc)$H_x$ sample synthesized at 213 GPa (DAC LS-4). The sample has not completely reacted with hydrogen. As a result, $T_c$ is lower than previously found value (245 K) and there is a residual resistance below 150 K. Also, a large transition width indicates the heterogeneity of the sample. In panel (c) there is a dependence of electrical resistance on temperature for the (La,Sc)$H_x$ sample after laser heating at 220 GPa (DAC LS-4). The Debye temperature cannot be determined reliably from this dependence since Bloch–Grüneisen (BG) fit gives $T_D$ > 10000 K. Finally, in panel (d) a quasi-linear dependence of *R(T)* and its deviation from the BG behavior in the normal state is shown.

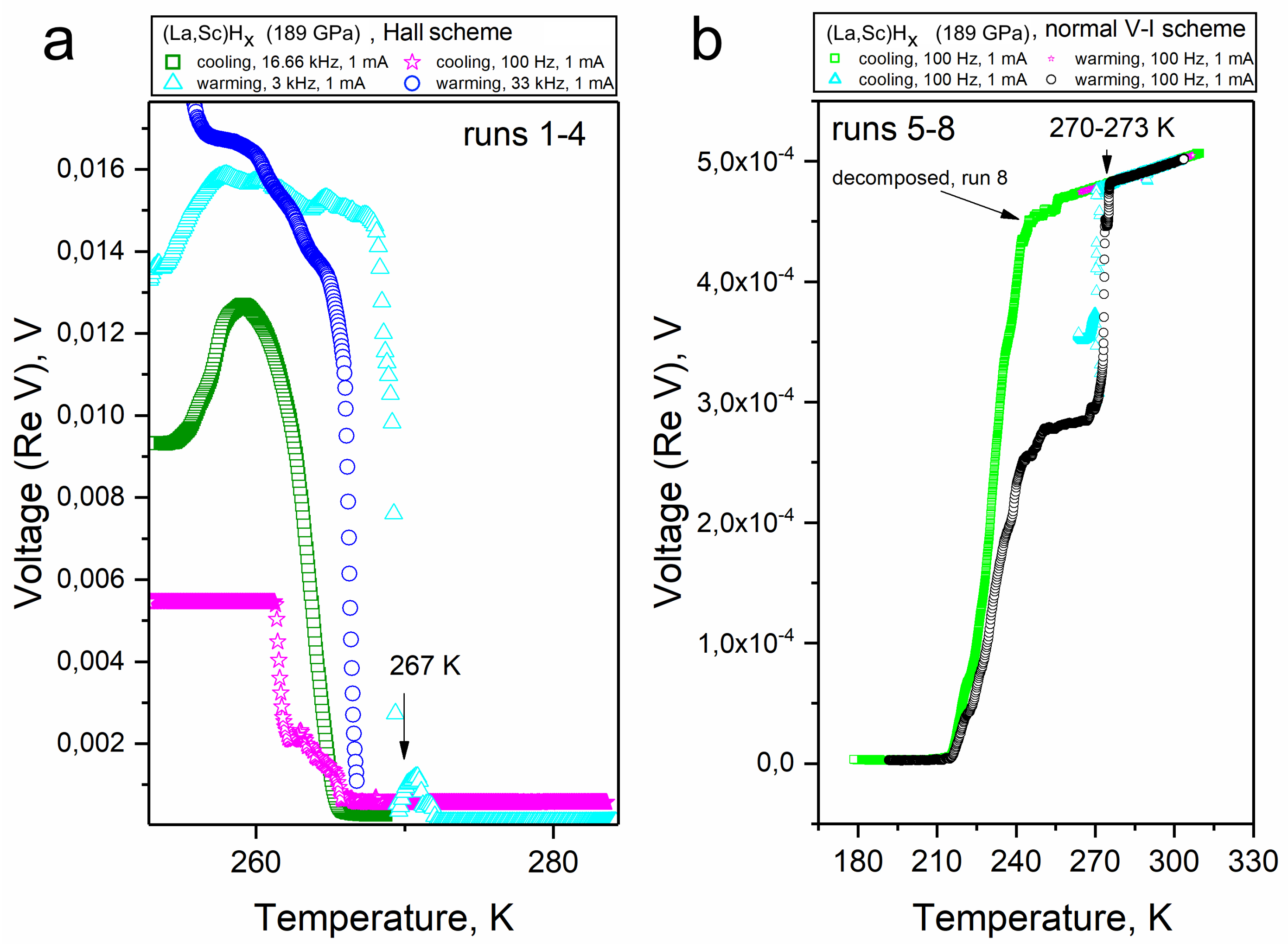

**Figure S24.** High temperature anomaly in the electrical resistivity of the sample at 189 GPa (DAC LS-3). (a) It was originally recorded during the study of the Hall effect, so the first 4 heating and cooling cycles (runs 1-4) were carried out in this configuration at different current frequencies (33 kHz, 16.66 kHz, 3 kHz and 100 Hz). The variation in the starting point of the anomaly may be due to the high rate of cooling and heating and the accompanying thermal hysteresis. (b) Warming up the sample in the usual V-I circuit. Current frequency 100 Hz. After a series of cooling and warming cycles, the additional drop in *R(T)* disappeared with the restoration of the previous resistance.

There are several possible explanations for the observed resistance anomalies:

(1) Breaks or short circuit of electrodes. In this case, we would observe a significant change in resistance, its drop to zero or large increase. Such false transitions have a limited point density.

In our case, we are dealing with only a partial change ~ 40 % in the resistance of the sample, as is the case with two-phase samples (e.g., $CeH_{9-10}$ case [31]). Moreover, a high density of *R(T)* points can be achieved; the beginning and end of the anomalous resistance jump are gentle (Figure S21), which corresponds to current views on the dynamics of the vortex lattice in superhydrides [32,33].

(2) A chemical, electrochemical [34] reaction or phase transition in the vicinity of one of the electrodes: $(La,Sc)H_{12} \pm nH_2 \rightarrow$ phase X, with its subsequent decomposition to the initial $(La,Sc)H_{12}$.

One of the arguments in favor of this version is the detection of traces of other hexagonal phases in DAC samples LS-2, LS-3 (Figures 1e, 2c). Some of such phases were predicted to be room-temperature superconductors [35].

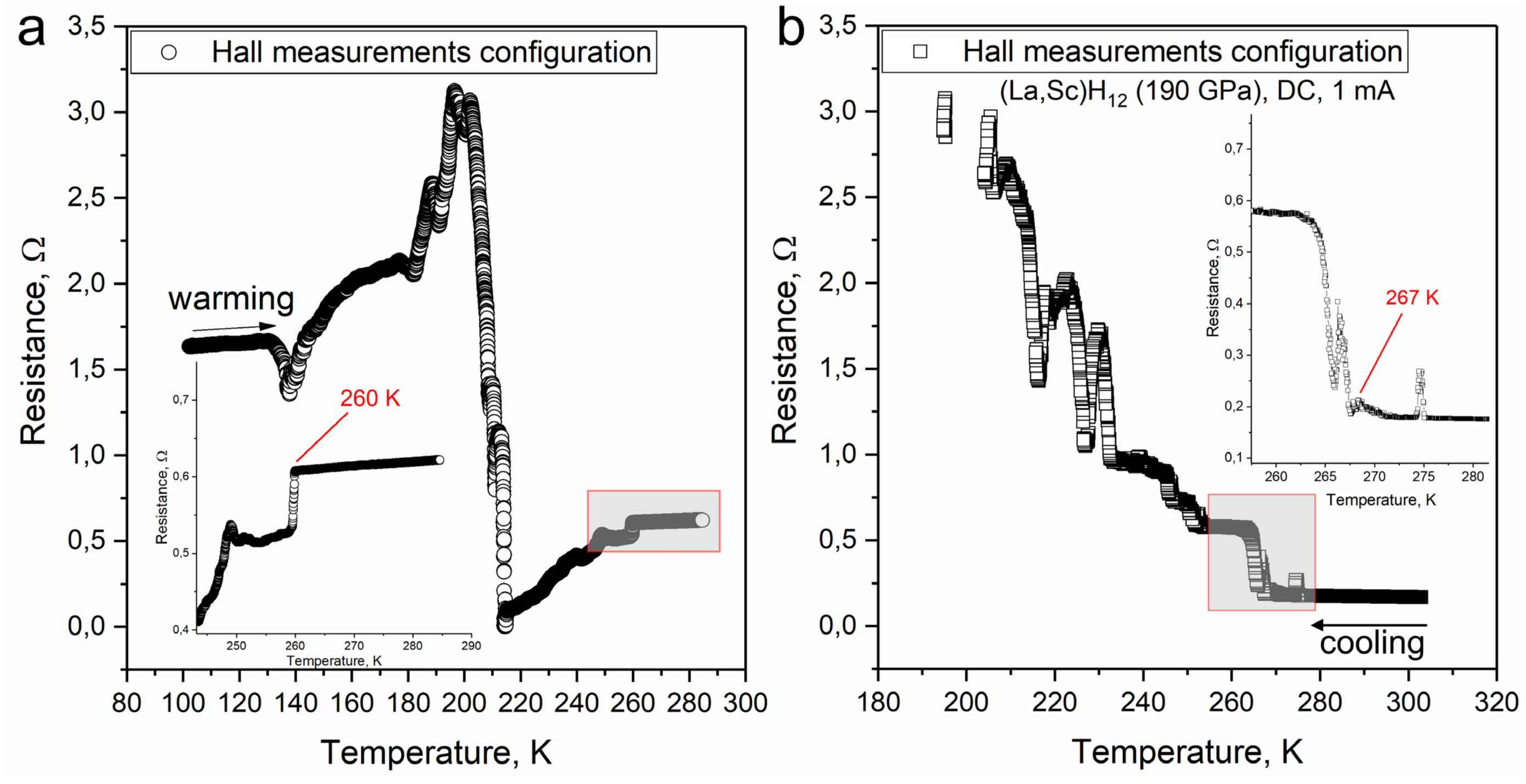

**Figure S25.** Hall resistance measurements in two (a, b) different diagonal connections approximately a month after the anomaly was first detected (Figure S21). As before, the beginning of the anomaly can be traced in the region of 260-267 K. After these measurements, the chamber collapsed. Because of this, it was not possible to study the behavior of this resistance anomaly in magnetic fields.

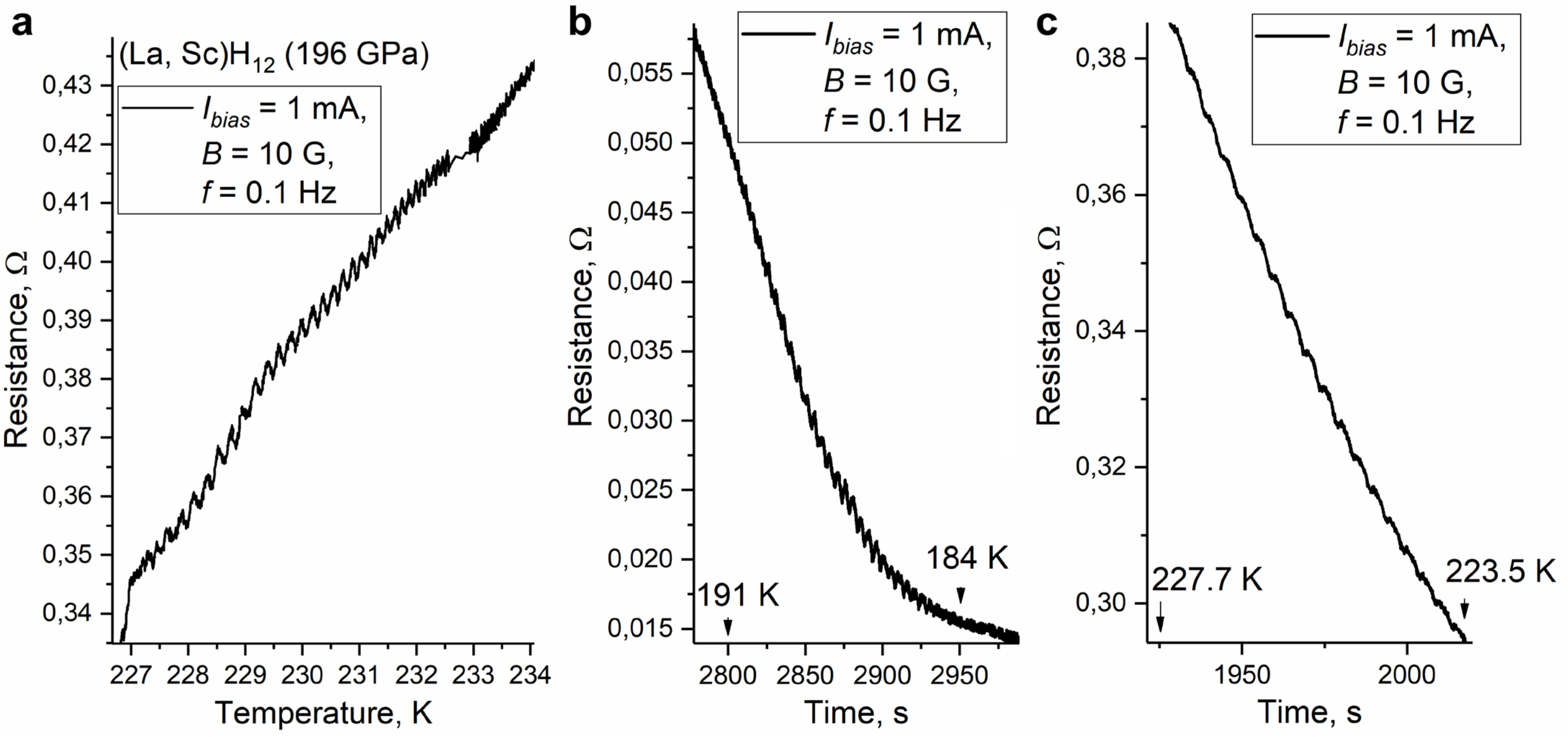

**Figure S26.** A SQUID-like effect in the DAC LS-3 sample upon cooling in a periodic magnetic field ~10 G with a frequency of 0.1 Hz. The bias current is 1 mA. The delta mode of resistance measurements was used. (a) The appearance of oscillations of electrical resistance on the *R(T)* dependence in the range of 227 – 232 K. (b) Similar oscillations on the dependence of resistance on time between 191 K and 184 K. (c) Weak oscillations of electrical resistance between 227.7 K and 223.5 K.

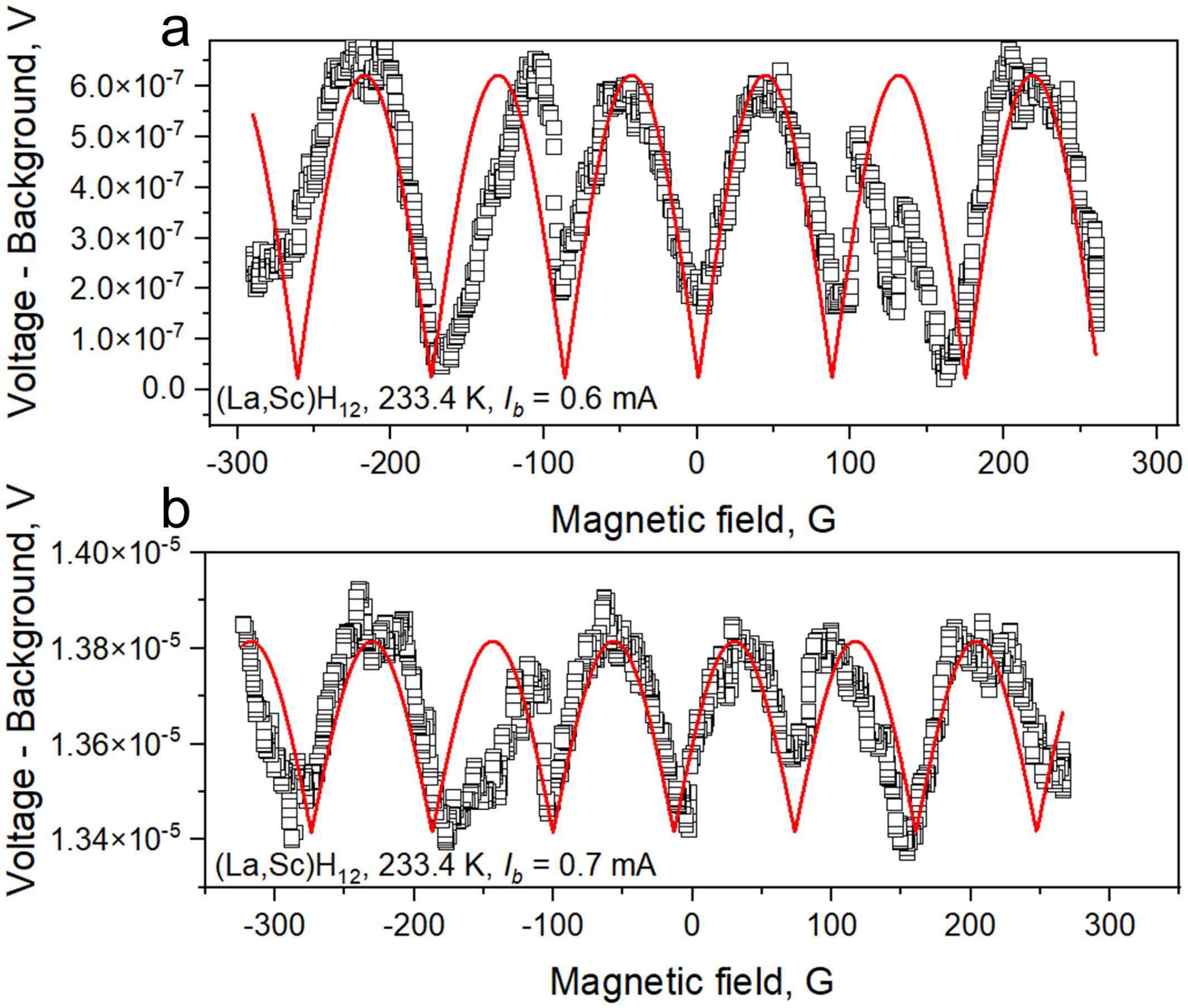

**Figure S27.** Voltage drop across the (La,Sc)$H_{12}$ sample (DAC LS-1) as a function of the applied external magnetic field at 233.4 K. Detection was performed in AC mode, at a frequency of 77 Hz with a bias current ($I_b$) amplitude of (a) 0.6 mA, (b) 0.7 mA. For ease of fitting using the RSJ-model, we subtracted the parabolic background from the signal ($V \propto H^2$). Red line corresponds to the fitting by function $R = \frac{U}{I_{bias}} = \frac{R_J I_c}{I_{bias}}\sqrt{\left(\frac{I_{bias}}{2I_c}\right)^2 - \cos^2\left(\pi\frac{\Phi}{\Phi_0}\right)}$, where $\Phi$ - is a magnetic flux passing through a SQUID circuit calculated for a ring radius of 300 nm, $R_J$ - is a resistance of the Josephson junction and $I_C$ – its critical current, while $I_{bias}$ – is a current used for measurements. Deviations from this model are due to the presence of other SQUID circuits in the sample.

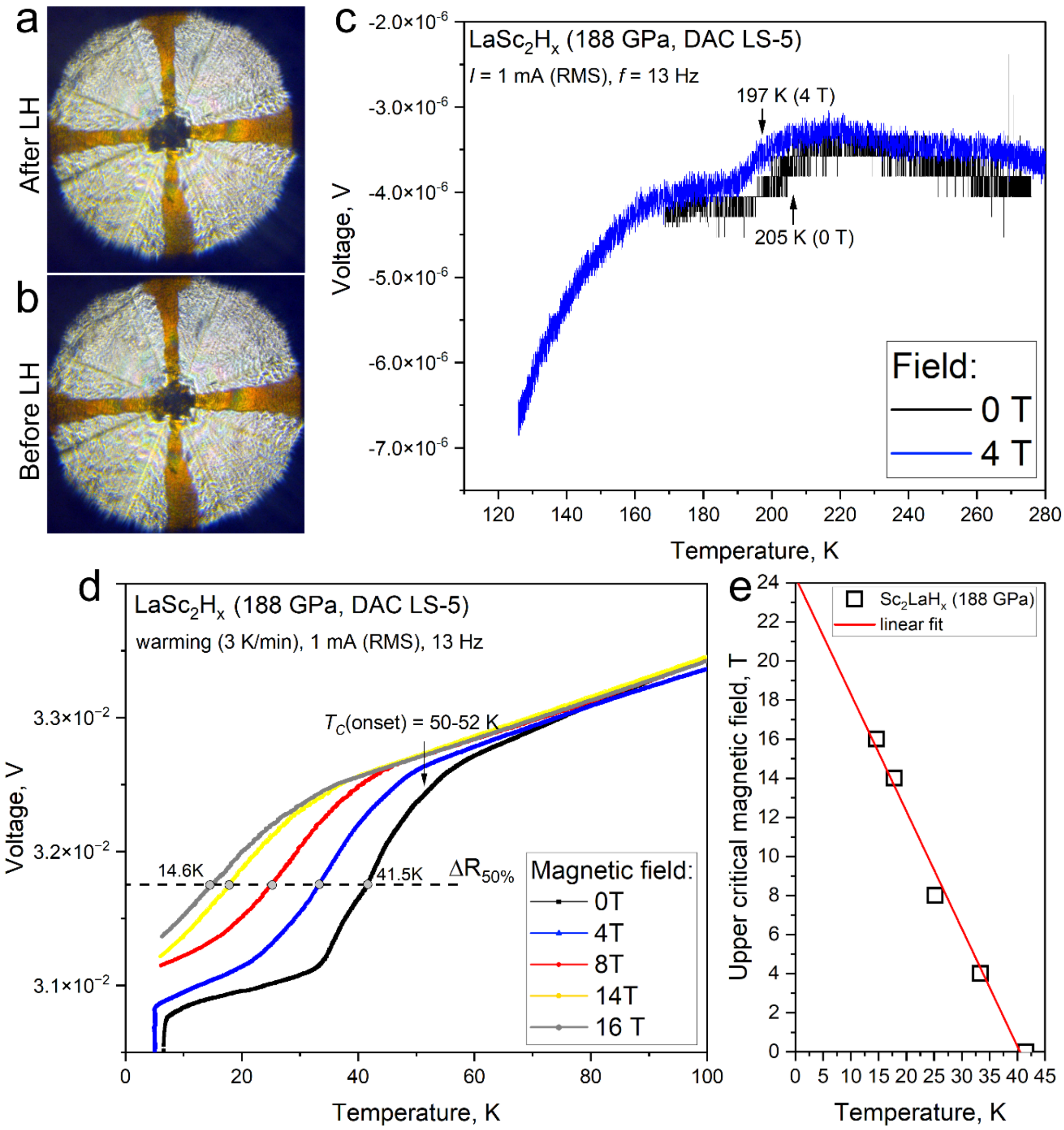

**Figure S28.** Study of transport properties of a sample in DAC SL-5 synthesized from the $LaSc_2$ alloy and ammonium borane ($NH_3BH_3$) at 188 GPa. (a, b) Photographs of the sample and electrical contacts before (before LH, 186 GPa) and after (after LH, 188 GPa) laser heating. (c) Temperature dependence of the voltage drop across the sample measured in AC mode at 13 Hz (lock-in SR830). A current of 1 mA (RMS) and an external magnetic field of 0 and 4 T were used. (d) Temperature dependence of the voltage drop across the sample measured on another pair of contacts at low temperature in magnetic fields from 0 to 16 T. (e) Dependence of $T_c^{50\%}$ of the low-temperature transition on the applied magnetic field.

## 5. Radio-frequency measurements

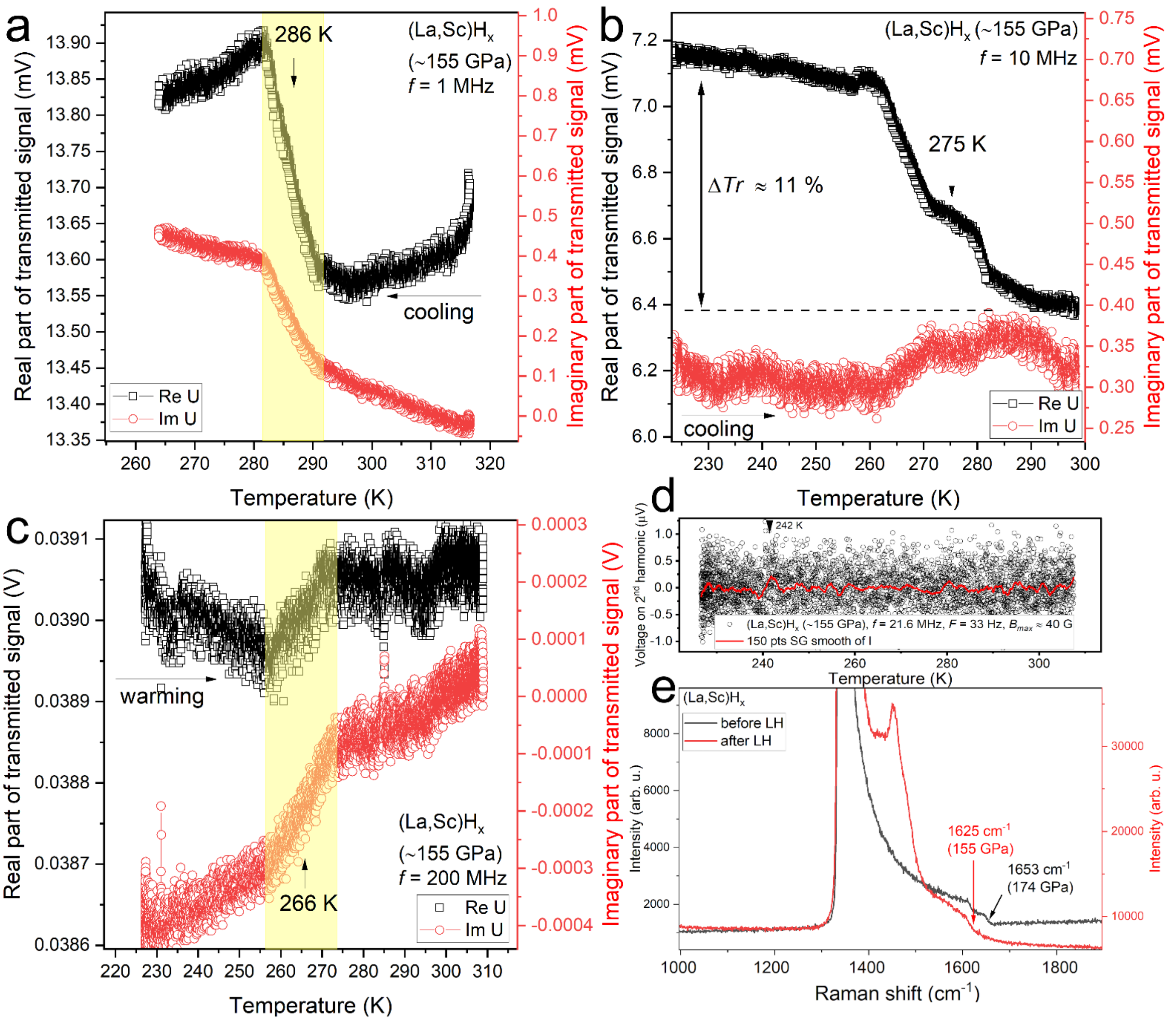

**Figure S29.** Radio-frequency transmission through the sample in DAC LS-6 at 155 GPa at different carrier frequencies $f$ in cooling and warming cycles. Real and imaginary components of the transmitted signal at: (a) $f = 1$ MHz, (b) $f = 10$ MHz, (c) $f = 200$ MHz. In all cases, a pronounced feature is observed in the transmission of the sample between 270 K and 290 K. The change in transmission at some frequencies can reach 11% (see panel "b"). (d) A weak signal at the 2nd harmonic, visible after smoothing the data. This signal at 242 K corresponds to the modulation of the high-frequency transmission when the sample is placed in a sinusoidal magnetic field of $B_{max} \approx 40$ G with a frequency of 33 Hz. (e) Raman spectra of the culet of DAC LS-6 before (black line) and after (red line) laser heating (LH) showing a significant drop in pressure due to the formation of a crack.

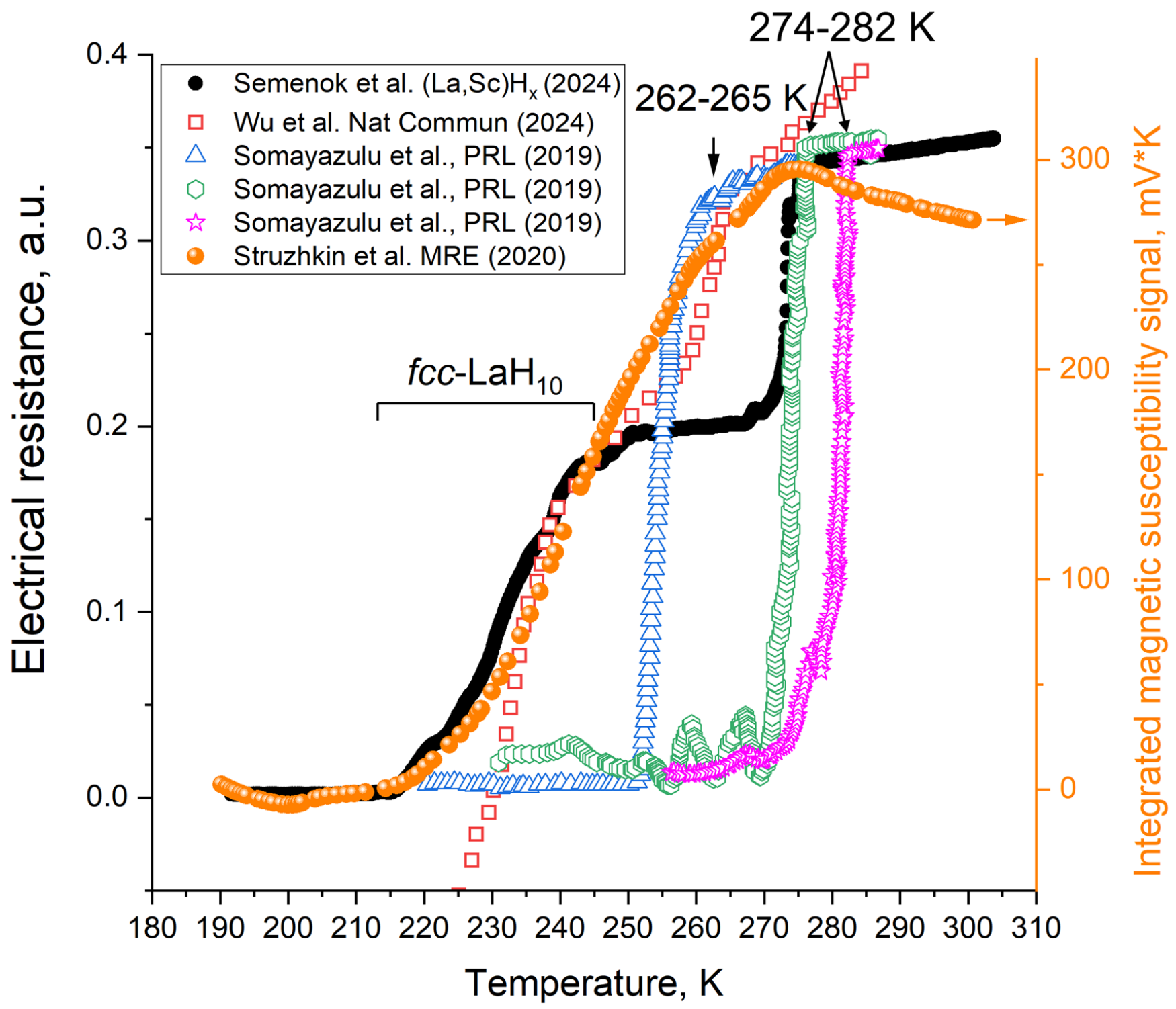

**Figure S30**. Review data on high-temperature (> 250 K) transitions in La-H and La-Sc-H systems at pressures of about 150-200 GPa, obtained by different groups for the period from 2019 to 2024. As can be seen, there are at least two groups of such high-$T_c$ transitions: (1) at 262-265 K [36] and (2) at 274-282 K [37,38], which may be related to the manifestation of superconductivity in the previously unknown lanthanum polyhydride $LaH_x$. Similar transitions are observed in the radio-frequency transmission measurements of the $(Ls,Sc)H_x$ sample in DAC LS-6 (Supporting Figure S29).

## 6. Theoretical calculations

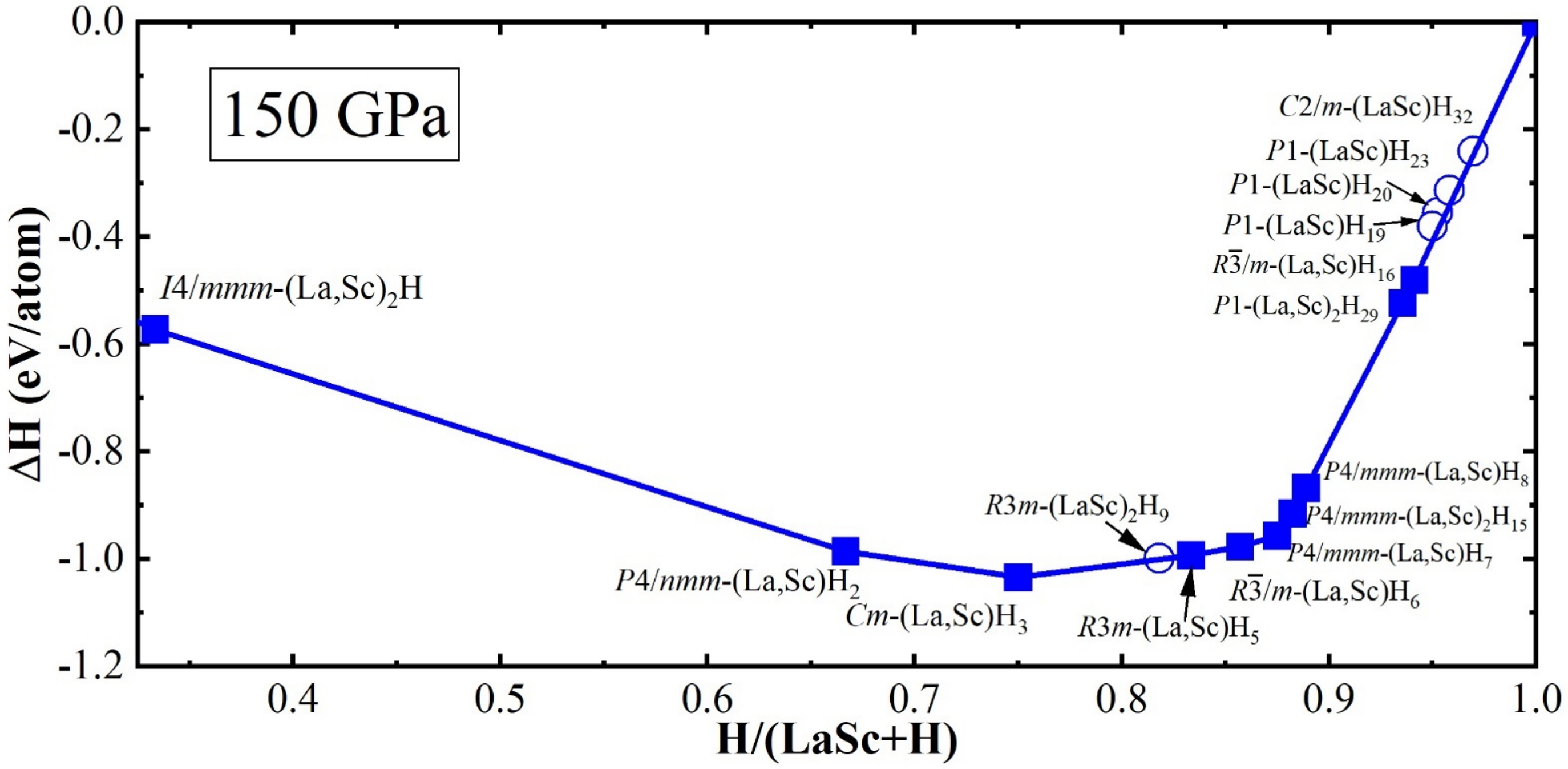

**Figure S31.** Convex hull of (LaSc)-H system at 150 GPa and 0 K calculated without accounting of zero-point energy effects in harmonic approximation.

**Table S4.** Results of an evolutionary search by USPEX 2023.0 code (python version) for stable structures in the (LaSc)-H system at 150 GPa and 0 K after 63 generations. All structures given in the Table lie on the convex hull.

| ID | Origin | Formula | Enthalpy, eV | Volume, Å$^3$ |
|---|---|---|---|---|
| 597 | Heredity | $LaScH_5$ | 11.303 | 34.427 |
| 986 | Permutation | $LaScH_{16}$ | 3.342 | 54.236 |
| 1097 | Heredity | $La_2Sc_2H_{15}$ | 16.348 | 78.703 |
| 1768 | Permutation | $La_2Sc_2H_{29}$ | 8.371 | 101.814 |
| 1912 | Heredity | $LaScH_7$ | 8.562 | 38.935 |
| 2100 | Permutation | LaSc | 0.0 | 24.42 |
| 2504 | Transmutation | $La_2Sc_2H_6$ | 28.34 | 62.246 |
| 3191 | Transmutation | $LaScH_8$ | 7.882 | 39.984 |
| 4203 | Permutation | $La_2Sc_2H_1$ | 37.514 | 50.986 |
| 4997 | Permutation | $La_2Sc_2H_4$ | 31.749 | 57.239 |
| 4998 | Permutation | $LaScH_6$ | 9.898 | 37.143 |
| 5075 | RandSymPyXtal | $H_{24}$ | -12.559 | 34.427 |

**Table S5.** Structures of the most thermodynamically stable (LaSc)-H phases found by USPEX at 150 GPa in the POSCARS format.

| | |
|---|---|
| EA597 | EA986 |
| 1.0000000000000000 | 1.0000000000000000 |
| 2.668835 -0.668172 0.097448 | 6.406570 -0.269267 -0.217726 |
| -0.024904 5.475358 -0.032237 | 2.039868 2.870803 -0.448799 |
| -1.337060 0.694375 2.307773 | 0.032796 -1.753643 3.087161 |
| H La Sc | H La Sc |
| 5 1 1 | 16 1 1 |
| Direct | Direct |
| 0.990937 0.659182 0.782584 | 0.612789 0.490501 0.015401 |
| 0.672738 0.731842 0.067006 | 0.112675 0.685290 0.867703 |
| 0.272191 0.580923 0.455271 | 0.416844 0.762801 0.656432 |
| 0.783584 0.083400 0.957256 | 0.938054 0.882364 0.957541 |
| 0.115508 0.075431 0.626459 | 0.242689 0.841667 0.929501 |
| 0.548628 0.363730 0.212518<br>0.387589 0.888737 0.345047 | 0.794125 0.416004 0.223776<br>0.419896 0.768684 0.144970<br>0.236922 0.326543 0.396982<br>0.796163 0.952893 0.770890<br>0.247032 0.313620 0.944880<br>0.791696 0.936069 0.228844<br>0.620017 0.479216 0.500999<br>0.104169 0.348493 0.187743<br>0.616608 0.992743 0.003717<br>0.915610 0.590387 0.304904<br>0.422345 0.254610 0.155043<br>0.027939 0.109925 0.573076<br>0.517152 0.127020 0.580492 |
| EA1097<br>1.0000000000000000<br>5.171060 0.318179 0.233481<br>-0.251795 3.862650 0.164362<br>-0.166062 -0.178357 3.909404<br>H La Sc<br>15 2 2<br>Direct<br>0.671455 0.274826 0.511113<br>0.232132 0.032151 0.270222<br>0.929148 0.766685 0.521268<br>0.024183 0.289942 0.028582<br>0.113852 0.782533 0.517027<br>0.223255 0.533944 0.776385<br>0.234524 0.523447 0.268910<br>0.221042 0.023628 0.768052<br>0.364815 0.280541 0.531332<br>0.810726 0.528163 0.780820<br>0.815171 0.518652 0.261781<br>0.810063 0.030807 0.265478<br>0.817913 0.036549 0.783093<br>0.671058 0.791367 0.029314<br>0.367565 0.787910 0.015040<br>0.528139 0.785038 0.522824<br>0.530794 0.292157 0.018947<br>0.023808 0.772180 0.022473<br>0.023163 0.276832 0.520719 | EA1768<br>1.0000000000000000<br>5.339820 0.343504 -0.002025<br>-0.241826 3.697004 0.009448<br>0.036292 -0.037618 5.135654<br>H La Sc<br>29 2 2<br>Direct<br>0.894097 0.916884 0.273566<br>0.306533 0.931676 0.687301<br>0.629878 0.648803 0.634121<br>0.719024 0.400374 0.111986<br>0.689515 0.384030 0.766685<br>0.630672 0.654462 0.254839<br>0.443717 0.656674 0.320755<br>0.286807 0.378016 0.181500<br>0.886480 0.435410 0.259531<br>0.955495 0.059363 0.546299<br>0.970397 0.227924 0.330171<br>0.447802 0.661873 0.569694<br>0.182201 0.170516 0.557617<br>0.292554 0.941535 0.203397<br>0.954616 0.642036 0.761682<br>0.602287 0.110880 0.023207<br>0.768342 0.670118 0.446561<br>0.974381 0.674893 0.125648<br>0.830806 0.660590 0.940513<br>0.156095 0.655371 0.061872<br>0.887290 0.406893 0.622895<br>0.396173 0.181033 0.790449<br>0.674066 0.967878 0.774710<br>0.433135 0.156824 0.038347<br>0.317968 0.393120 0.687400<br>0.216989 0.130386 0.297290<br>0.728988 0.906350 0.107085<br>0.158243 0.657832 0.815689<br>0.877262 0.891329 0.626250<br>0.575652 0.152496 0.442838<br>0.020766 0.166374 0.930135<br>0.486348 0.654871 0.942409<br>0.110349 0.668510 0.444720 |
| EA1912<br>1.0000000000000000<br>5.590672 -1.319490 -0.510131<br>0.645850 2.688712 0.139706<br>-1.096384 0.131270 2.542350<br>H La Sc<br>7 1 1<br>Direct<br>0.217875 0.540717 0.557568<br>0.797993 0.536291 0.139413<br>0.661208 0.537833 0.483743<br>0.356208 0.509129 0.195337 | EA2100<br>1.0000000000000000<br>2.555205 0.603244 0.004092<br>-0.813713 3.446689 0.002238<br>-0.003625 -0.002612 2.626398<br>La Sc<br>1 1<br>Direct<br>0.282126 0.708674 0.069292<br>0.783583 0.207064 0.558263 |
| 0.215553 0.011337 0.036547<br>0.000500 0.031014 0.337335<br>0.801129 0.035765 0.618422<br>0.502434 0.027681 0.820827<br>0.000837 0.535091 0.834229 | |
| EA2504<br>1.0000000000000000<br>9.210557 -0.939627 1.352709<br>-0.182925 1.564494 2.258593<br>-0.240218 -2.766571 0.210050<br>H La Sc<br>6 2 2<br>Direct<br>0.890846 0.264615 0.102631<br>0.291086 0.261590 0.100364<br>0.165998 0.569621 0.767439<br>0.043099 0.898563 0.435377<br>0.352968 0.567262 0.731169<br>0.097453 0.250710 0.114841<br>0.502249 0.224749 0.043495<br>0.739220 0.916087 0.390588<br>0.984083 0.585298 0.786155<br>0.233485 0.905222 0.429921 | EA3191<br>1.0000000000000000<br>5.224338 0.100430 0.030453<br>-0.053569 2.761278 -0.133541<br>-0.022066 0.133248 2.764108<br>H La Sc<br>8 1 1<br>Direct<br>0.710620 0.577797 0.009334<br>0.575794 0.082457 0.023771<br>0.128153 0.073201 0.522578<br>0.003103 0.605764 0.541516<br>0.823821 0.589129 0.526580<br>0.137460 0.595814 0.005531<br>0.704791 0.094468 0.526948<br>0.267766 0.077620 0.034066<br>0.419898 0.583175 0.529256<br>0.923803 0.095907 0.031336 |
| EA4203<br>1.0000000000000000<br>3.607190 0.407292 0.091988<br>-2.220233 2.965481 2.805752<br>0.031734 -1.043071 3.464114<br>H La Sc<br>1 2 2<br>Direct<br>0.139060 0.968623 0.936552<br>0.872182 0.418519 0.208082<br>0.430806 0.531492 0.646574<br>0.651309 0.976957 0.924069<br>0.137305 0.964030 0.434434 | EA4997<br>1.0000000000000000<br>2.639027 -0.022500 0.022870<br>0.022260 2.638321 0.014768<br>0.012840 -0.077000 8.220089<br>H La Sc<br>4 2 2<br>Direct<br>0.085919 0.625339 0.228064<br>0.596937 0.122542 0.978242<br>0.618507 0.613524 0.094750<br>0.109137 0.116899 0.093650<br>0.578777 0.098376 0.719039<br>0.100260 0.615196 0.474928<br>0.598488 0.103831 0.234713<br>0.107623 0.596112 0.958754 |
| EA4998<br>1.0000000000000000<br>5.700874 0.218452 0.389112<br>-0.037105 2.735489 -0.565684<br>-0.813919 -0.911795 2.506160<br>H La Sc<br>6 1 1<br>Direct<br>0.023251 0.468701 0.613412<br>0.459208 0.955897 0.605884<br>0.194548 0.203850 0.080806<br>0.546320 0.319912 0.300200<br>0.624441 0.691859 0.057385<br>0.106994 0.851443 0.379367<br>0.828866 0.088113 0.829725<br>0.328683 0.569186 0.825364 | EA5075<br>1.0000000000000000<br>4.511691 -0.172302 0.009738<br>-0.132127 3.291453 0.002085<br>-0.009422 -0.002850 3.066034<br>H<br>24<br>Direct<br>0.045833 0.788389 0.992944<br>0.535071 0.290058 0.490189<br>0.547251 0.301617 0.738417<br>0.547805 0.781566 0.683748<br>0.292671 0.008582 0.235425<br>0.665927 0.944687 0.028332<br>0.911600 0.639541 0.644892<br>0.045776 0.278226 0.940151<br>0.304839 0.536549 0.410459<br>0.795308 0.548264 0.266860<br>0.542723 0.775479 0.449776<br>0.404089 0.132809 0.126182<br>0.798749 0.046225 0.902294<br>0.297091 0.043318 0.757675<br>0.046324 0.291132 0.166917<br>0.033820 0.769753 0.228049<br>0.783606 0.534531 0.760339<br>0.172486 0.442442 0.528858<br>0.278923 0.511175 0.925941 |
| | 0.664539 0.454020 0.124699<br>0.174059 0.940793 0.626759<br>0.923266 0.128210 0.539486<br>0.777507 0.035126 0.417195<br>0.407389 0.618651 0.019577 |
| EA2135<br>1.0000000000000000<br>2.783468 -0.000573 -0.054394<br>0.023217 5.740146 -0.012544<br>-1.351080 -0.680404 2.351399<br>H La Sc<br>6 1 1<br>Direct<br>0.216604 0.356096 0.438530<br>0.071646 0.015603 0.184147<br>0.725218 0.924837 0.472314<br>0.345952 0.835837 0.737945<br>0.833598 0.276968 0.708296<br>0.590004 0.435886 0.170958<br>0.963561 0.640425 0.933328<br>0.461377 0.136933 0.940386 | EA5053<br>1.0000000000000000<br>5.206659 -0.031199 0.462914<br>0.001504 2.760409 0.166160<br>-0.244101 -0.165397 2.750400<br>H La Sc<br>8 1 1<br>Direct<br>0.763406 0.965988 0.352996<br>0.615048 0.460961 0.358658<br>0.056580 0.974519 0.365588<br>0.489759 0.476275 0.855125<br>0.200763 0.968671 0.853785<br>0.616971 0.977261 0.846453<br>0.313279 0.471303 0.874043<br>0.193330 0.461231 0.371683<br>0.903681 0.464904 0.850085<br>0.413935 0.960799 0.351722 |
| EA3221<br>1.0000000000000000<br>5.228619 0.047873 0.024311<br>-0.020819 2.762623 -0.138640<br>-0.017667 0.138277 2.761665<br>H La Sc<br>8 1 1<br>Direct<br>0.745456 0.808925 0.925810<br>0.166650 0.294026 0.420374<br>0.032665 0.812911 0.418660<br>0.732483 0.294872 0.432284<br>0.158802 0.805552 0.936892<br>0.294050 0.322654 0.931693<br>0.859604 0.823242 0.434429<br>0.593881 0.310132 0.930742<br>0.446166 0.807302 0.424149<br>0.949546 0.307751 0.911089 | EA821<br>1.0000000000000000<br>2.669865 -0.671925 0.099322<br>-0.036864 5.480006 -0.035464<br>-1.335973 0.690189 2.305548<br>H La Sc<br>5 1 1<br>Direct<br>0.972642 0.597579 0.809494<br>0.411076 0.918860 0.359562<br>0.257883 0.516197 0.507026<br>0.552502 0.427091 0.178344<br>0.080291 0.931279 0.691238<br>0.807816 0.215664 0.913633<br>0.693312 0.740640 0.105075 |

**Table S6.** Calculated unit cell volume data used in this work to estimate the hydrogen content of synthesized La-Sc polyhydrides. Most of the calculations were done in the Quantum ESPRESSO, with the exception of one calculation done using the VASP code. The protypes used were $I4/mmm$-$LaH_4$ for (La, Sc)$H_4$, $Fm\bar{3}m$-$LaH_3$ for (La, Sc)$H_3$, and $P6_3/mmc$-$ScH_6$ for (La, Sc)$H_6$.

| **Compound** | **Pressure, GPa** | **Unit cell volume, Å$^3$/(La,Sc)** | **Compound** | **Pressure, GPa** | **Unit cell volume, Å$^3$/(La,Sc)** |
|---|---|---|---|---|---|
| Tetragonal (La,Sc)$H_4$ | 30 | 29.17 | sc-$ScH_{12}$ | 200 | 28.7 |
| | 150 | 20.09 | $P6/mmm$-(La,Sc)$H_8$ | 190 | 25.10 |
| Cubic (La,Sc)$H_3$ | 30 | 27.77 | $P6_3/mmc$-(La,Sc)$H_6$ | 190 | 22.11 (VASP) |
| | 150 | 19.32 | Tetragonal (La,Sc)$H_4$ | 190 | 18.72 |
| $P6_3/mmc$-(La,Sc)$H_6$ | 150 | 23.64 | Cubic (La,Sc)$H_3$ | 200 | 17.69 |
| | | | | 190 | 17.98 |

**Table S7.** CIF files of the most thermodynamically stable structures in the (LaSc)-H system found by USPEX code at 30 GPa.

| ***P*4₁-LaScH₆** | ***P*4₁-LaScH₆** |
|---|---|
| 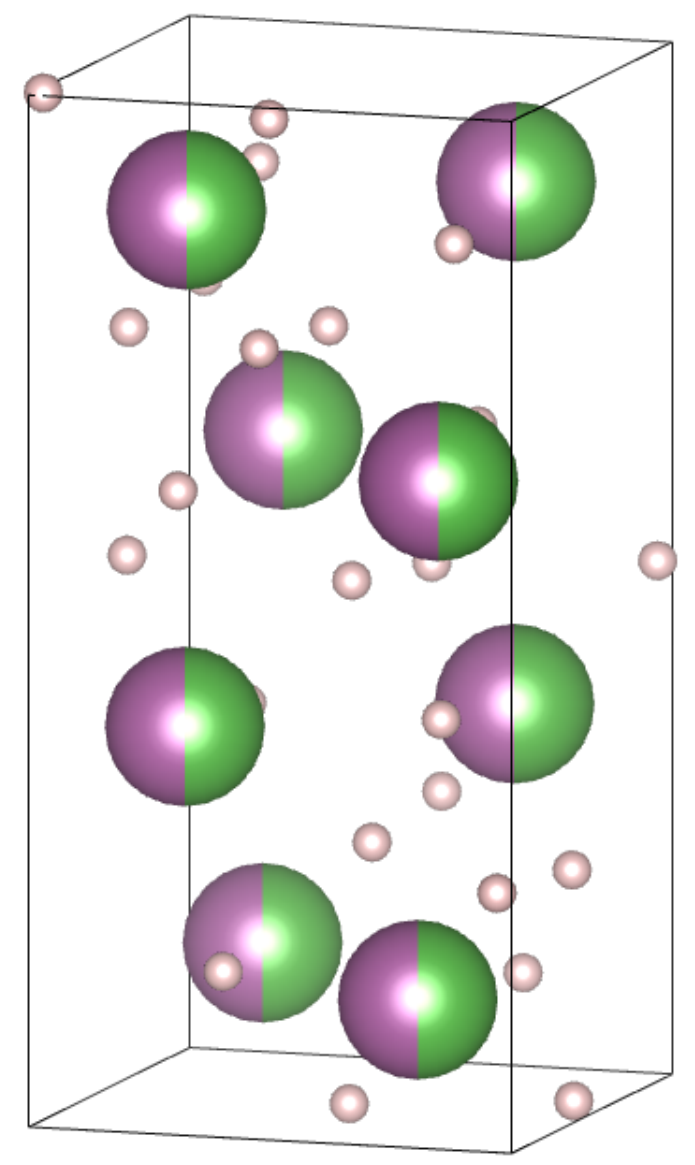<br>$La_4Sc_4H_{24}$<br>Space Group: P 41 (#76-1)<br>a = 4.78257 Å α = 90.0000°<br>b = 4.78257 Å β = 90.0000°<br>c = 9.83966 Å γ = 90.0000°<br>V = 225.0621 Å$^3$ | # generated using pymatgen<br>data_LaScH6<br>_symmetry_space_group_name_H-M P4_1<br>_cell_length_a 4.78256806<br>_cell_length_b 4.78256806<br>_cell_length_c 9.83965941<br>_cell_angle_alpha 90.00000000<br>_cell_angle_beta 90.00000000<br>_cell_angle_gamma 90.00000000<br>_symmetry_Int_Tables_number 76<br>_chemical_formula_structural LaScH6<br>_chemical_formula_sum 'La4 Sc4 H24'<br>_cell_volume 225.06210918<br>_cell_formula_units_Z 4<br>loop_<br>_symmetry_equiv_pos_site_id<br>_symmetry_equiv_pos_as_xyz<br>1 'x, y, z'<br>2 '-y, x, z+1/4'<br>3 '-x, -y, z+1/2'<br>4 'y, -x, z+3/4'<br>loop_<br>_atom_site_type_symbol<br>_atom_site_label<br>_atom_site_symmetry_multiplicity<br>_atom_site_fract_x<br>_atom_site_fract_y<br>_atom_site_fract_z<br>_atom_site_occupancy<br>La La0 4 0.23330400 0.75419200 0.87643200 0.5<br>Sc Sc0 4 0.23330400 0.75419200 0.87643200 0.5<br>La La1 4 0.23606500 0.27271500 0.62377100 0.5<br>Sc Sc1 4 0.23606500 0.27271500 0.62377100 0.5<br>H H2 4 0.01839000 0.03586500 0.74979600 1<br>H H3 4 0.02676200 0.53132800 0.26402200 1<br>H H4 4 0.03606600 0.51411700 0.48635500 1<br>H H5 4 0.19942600 0.75731300 0.10734400 1<br>H H6 4 0.25168400 0.68070100 0.63941600 1<br>H H7 4 0.45890000 0.48883200 0.50039600 1 |

| ***Pm*-La₃Sc₃H₁₆** | ***Pm*-La₃Sc₃H₁₆** |
|---|---|

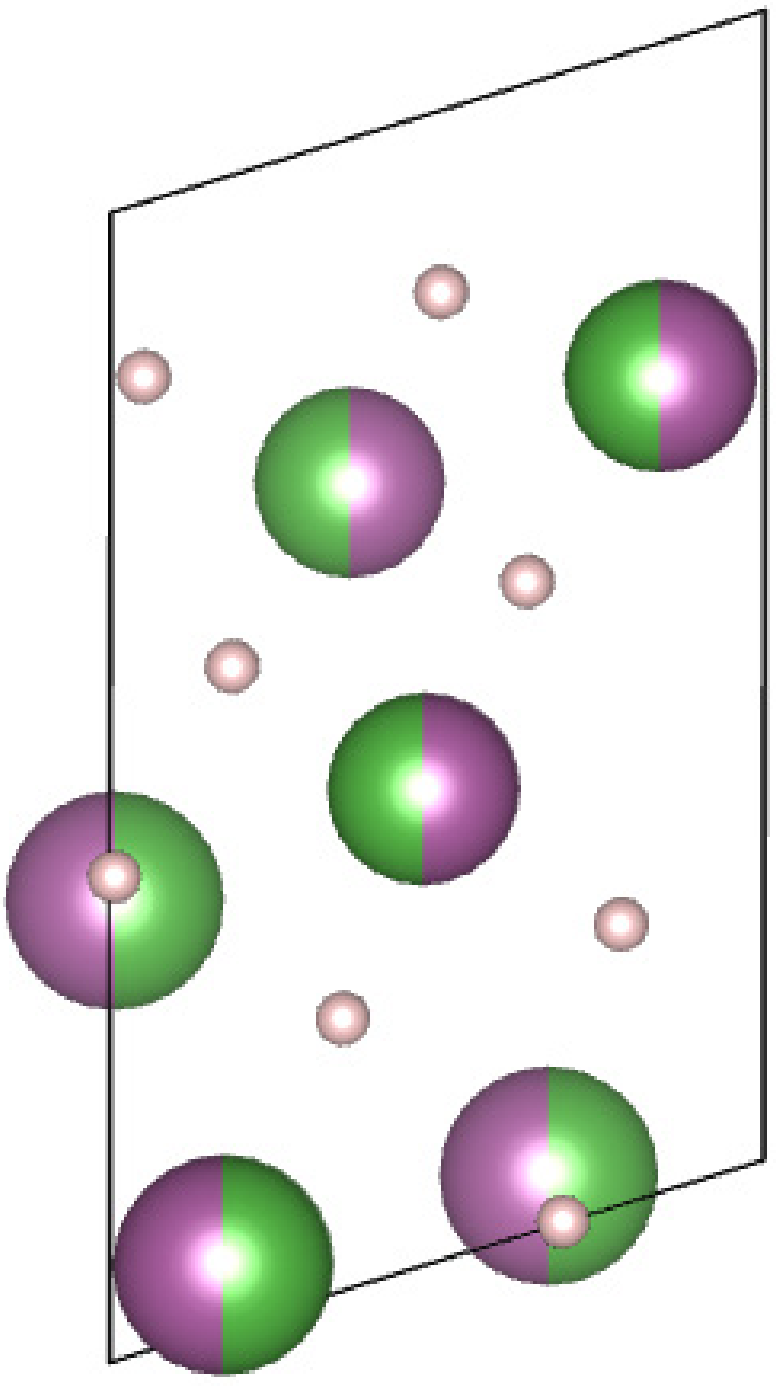

$La_3Sc_3H_{16}$
Space Group: P m (#6-1)
a = 4.69234 Å α = 90.0000°
b = 4.69614 Å β =107.2486°
c = 7.88047 Å γ = 90.0000°
V = 165.8433 $Å^3$

```
# generated using pymatgen
data_La3Sc3H16
_symmetry_space_group_name_H-M   Pm
_cell_length_a   4.69234051
_cell_length_b   4.69613509
_cell_length_c   7.88046544
_cell_angle_alpha   90.00000000
_cell_angle_beta   107.24855117
_cell_angle_gamma   90.00000000
_symmetry_Int_Tables_number   6
_chemical_formula_structural   La3Sc3H16
_chemical_formula_sum   'La3 Sc3 H16'
_cell_volume   165.84325821
_cell_formula_units_Z   1
loop_
 _symmetry_equiv_pos_site_id
 _symmetry_equiv_pos_as_xyz
  1  'x, y, z'
  2  'x, -y, z'
loop_
 _atom_site_type_symbol
 _atom_site_label
 _atom_site_symmetry_multiplicity
 _atom_site_fract_x
 _atom_site_fract_y
 _atom_site_fract_z
 _atom_site_occupancy
  La  La0  1  0.00646800  0.50000000  0.59943400  0.5
  Sc  Sc0  1  0.00646800  0.50000000  0.59943400  0.5
  La  La1  1  0.17363800  0.00000000  0.94586600  0.5
  Sc  Sc1  1  0.17363800  0.00000000  0.94586600  0.5
  La  La2  1  0.67075700  0.50000000  0.95585600  0.5
  Sc  Sc2  1  0.67075700  0.50000000  0.95585600  0.5
  Sc  Sc3  1  0.36523500  0.50000000  0.29901000  0.5
  La  La3  1  0.36523500  0.50000000  0.29901000  0.5
  Sc  Sc4  1  0.47918400  0.00000000  0.58538400  0.5
  La  La4  1  0.47918400  0.00000000  0.58538400  0.5
  Sc  Sc5  1  0.84141600  0.00000000  0.29007800  0.5
  La  La5  1  0.84141600  0.00000000  0.29007800  0.5
  H  H6  2  0.05047700  0.23141100  0.15154700  1
  H  H7  2  0.18608200  0.23378700  0.42756300  1
  H  H8  2  0.35597600  0.23723200  0.76350600  1
  H  H9  2  0.50531600  0.23190100  0.15811500  1
  H  H10  2  0.63692000  0.22710000  0.43301000  1
  H  H11  2  0.78160100  0.22297000  0.75653600  1
  H  H12  1  0.00785400  0.00000000  0.57813000  1
  H  H13  1  0.15429400  0.50000000  0.93470300  1
  H  H14  1  0.69459200  0.00000000  0.99877400  1
  H  H15  1  0.82041600  0.50000000  0.29216100  1
```

| ***Amm*2-$La_6Sc_6H_{32}$** | ***Amm*2-$La_6Sc_6H_{32}$** |
|---|---|

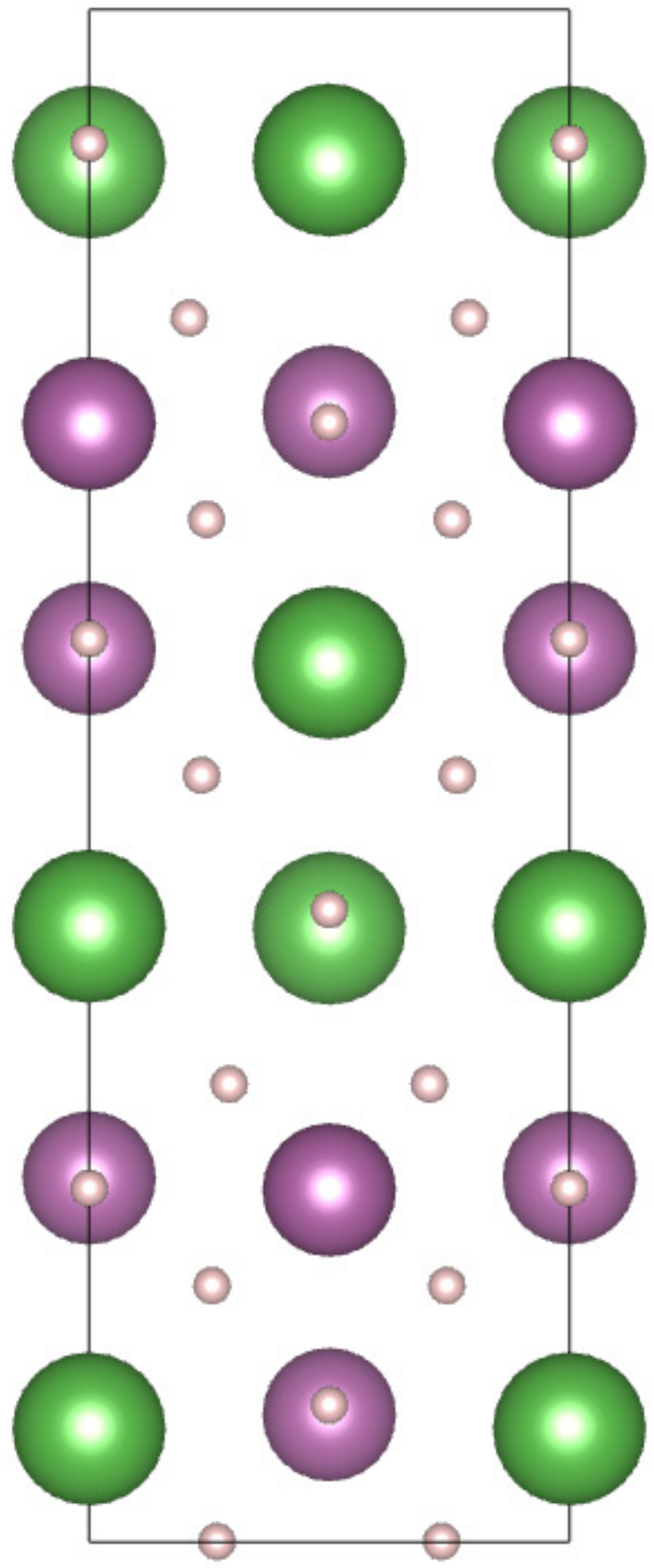

$La_6Sc_6H_{32}$
Space Group: *Amm*2 (#38-1)
a = 4.68095 Å  α = 90.0000°
b = 4.71263 Å  β = 90.0000°
c = 15.04397 Å  γ = 90.0000°
V = 331.8634 Å$^3$

```
# generated using pymatgen
data_La3Sc3H16
_symmetry_space_group_name_H-M   Amm2
_cell_length_a   4.68095040
_cell_length_b   4.71262662
_cell_length_c   15.04396747
_cell_angle_alpha   90.00000000
_cell_angle_beta   90.00000000
_cell_angle_gamma   90.00000000
_symmetry_Int_Tables_number   38
_chemical_formula_structural   La3Sc3H16
_chemical_formula_sum   'La6 Sc6 H32'
_cell_volume   331.86347479
_cell_formula_units_Z   2
loop_
 _symmetry_equiv_pos_site_id
 _symmetry_equiv_pos_as_xyz
  1  'x, y, z'
  2  '-x, -y, z'
  3  '-x, y, z'
  4  'x, -y, z'
  5  'x, y+1/2, z+1/2'
  6  '-x, -y+1/2, z+1/2'
  7  '-x, y+1/2, z+1/2'
  8  'x, -y+1/2, z+1/2'
loop_
 _atom_site_type_symbol
 _atom_site_label
 _atom_site_symmetry_multiplicity
 _atom_site_fract_x
 _atom_site_fract_y
 _atom_site_fract_z
 _atom_site_occupancy
  La  La0  2  0.00000000  0.00000000  0.07383700  1
  La  La1  2  0.00000000  0.00000000  0.40186000  1
  La  La2  2  0.50000000  0.00000000  0.90057050  1
  Sc  Sc3  2  0.00000000  0.00000000  0.72976650  1
  Sc  Sc4  2  0.50000000  0.00000000  0.23767650  1
  Sc  Sc5  2  0.50000000  0.00000000  0.58316000  1
  H  H6  8  0.24773000  0.20837700  0.79882200  1
  H  H7  8  0.27887200  0.24438650  0.66727150  1
  H  H8  8  0.28194600  0.23386850  0.50037150  1
  H  H9  2  0.00000000  0.00000000  0.23086950  1
  H  H10  2  0.00000000  0.00000000  0.58936500  1
  H  H11  2  0.00000000  0.00000000  0.91269000  1
  H  H12  2  0.50000000  0.00000000  0.37699600  1
```

| **$P\bar{4}m2$-$La_4Sc_4H_9$** | **$P\bar{4}m2$-$La_4Sc_4H_9$** |
|---|---|

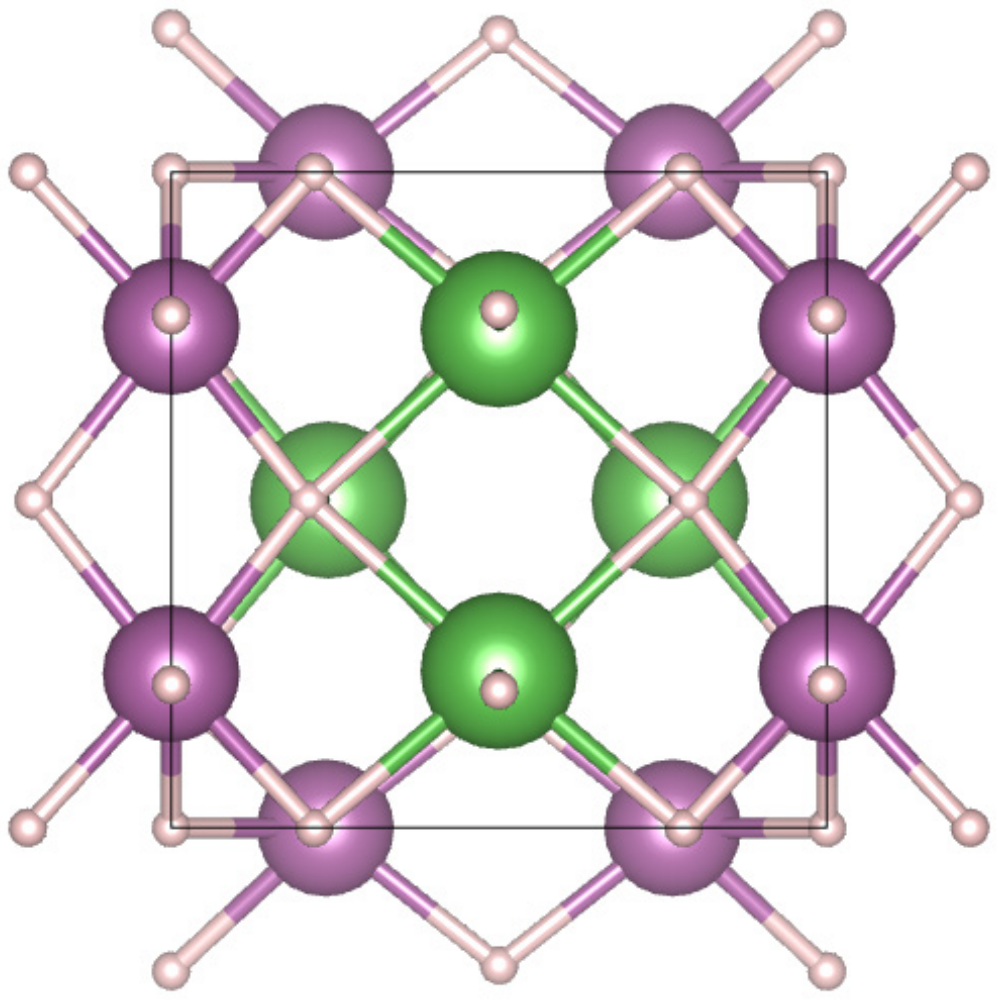

$La_4Sc_4H_9$

```
# generated using pymatgen
data_La4Sc4H9
_symmetry_space_group_name_H-M   P-4m2
_cell_length_a   6.27001833
_cell_length_b   6.27001833
_cell_length_c   4.57199096
_cell_angle_alpha   90.00000000
_cell_angle_beta   90.00000000
_cell_angle_gamma   90.00000000
_symmetry_Int_Tables_number   115
_chemical_formula_structural   La4Sc4H9
_chemical_formula_sum   'La4 Sc4 H9'
_cell_volume   179.73927433
_cell_formula_units_Z   1
loop_
 _symmetry_equiv_pos_site_id
 _symmetry_equiv_pos_as_xyz
  1  'x, y, z'
```

| | |
|---|---|
| Space Group: *P*-4*m*2 (#115-1)<br>a = 6.27002 Å α = 90.0000°<br>b = 6.27002 Å β = 90.0000°<br>c = 4.57199 Å γ = 90.0000°<br>V = 179.7393 Å³ | 2 'y, -x, -z'<br>3 '-x, -y, z'<br>4 '-y, x, -z'<br>5 '-x, y, z'<br>6 '-y, -x, -z'<br>7 'x, -y, z'<br>8 'y, x, -z'<br>loop_<br>_atom_site_type_symbol<br>_atom_site_label<br>_atom_site_symmetry_multiplicity<br>_atom_site_fract_x<br>_atom_site_fract_y<br>_atom_site_fract_z<br>_atom_site_occupancy<br>La La0 4 0.23929750 0.50000000 0.25605450 1<br>Sc Sc1 4 0.00000000 0.23522250 0.75785650 1<br>H H2 4 0.00000000 0.21842150 0.18872750 1<br>H H3 4 0.20951850 0.50000000 0.74912250 1<br>H H4 1 0.00000000 0.00000000 0.50000000 1 |

**Table S8.** CIF files of structures of La-Sc polyhydrides used in this work for superconductivity calculations. $LaScH_{24}$ is derivative of $ScH_{12}$ [24], while $LaScH_{20}$ is derivative of canonical cubic $LaH_{10}$.

| ***Pmmm*-LaScH₂₄ (193 GPa)** | ***Pmmm*-LaScH₂₄ (193 GPa)** |
|---|---|
| 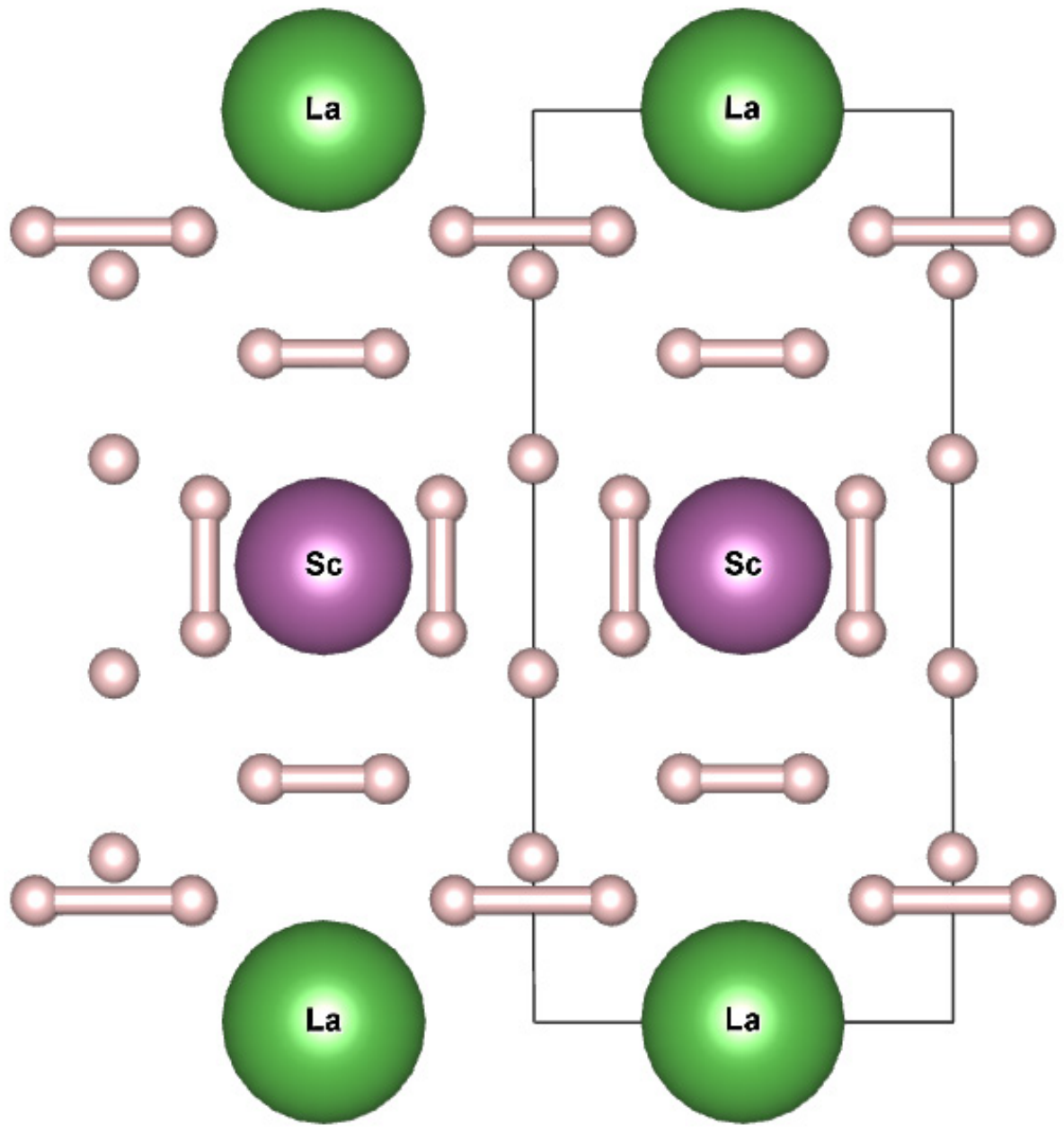<br><br>Space Group: *Pmmm* (#47-1)<br>a = 3.039 Å α = 90.0°<br>b = 6.747 Å β = 90.0°<br>c = 3.102 Å γ = 90.0°<br>V = 63.60 Å³ | data_findsym-output<br><br>_symmetry_space_group_name_H-M 'P 2/m 2/m 2/m'<br>_symmetry_Int_Tables_number 47<br><br>_cell_length_a 3.03900<br>_cell_length_b 6.74700<br>_cell_length_c 3.10200<br>_cell_angle_alpha 90.00000<br>_cell_angle_beta 90.00000<br>_cell_angle_gamma 90.00000<br><br>loop_<br>_space_group_symop_operation_xyz<br>x,y,z<br>x,-y,-z<br>-x,y,-z<br>-x,-y,z<br>-x,-y,-z<br>-x,y,z<br>x,-y,z<br>x,y,-z<br><br>loop_<br>_atom_site_label<br>_atom_site_type_symbol<br>_atom_site_fract_x<br>_atom_site_fract_y<br>_atom_site_fract_z<br>_atom_site_occupancy<br>La1 La 0.50000 0.00000 0.50000 1.00000<br>Sc1 Sc 0.50000 0.50000 0.50000 1.00000<br>H1 H 0.35778 -0.18042 0.00000 1.00000<br>H2 H 0.33876 0.38256 0.00000 1.00000<br>H3 H 0.00000 -0.13321 0.18612 1.00000<br>H4 H 0.00000 0.42783 0.21838 1.00000<br>H5 H 0.22739 -0.26665 0.35533 1.00000 |

| *P4/mmm*-$LaScH_{20}$ (200 GPa) | *P4/mmm*-$LaScH_{20}$ (200 GPa) |
|---|---|
| 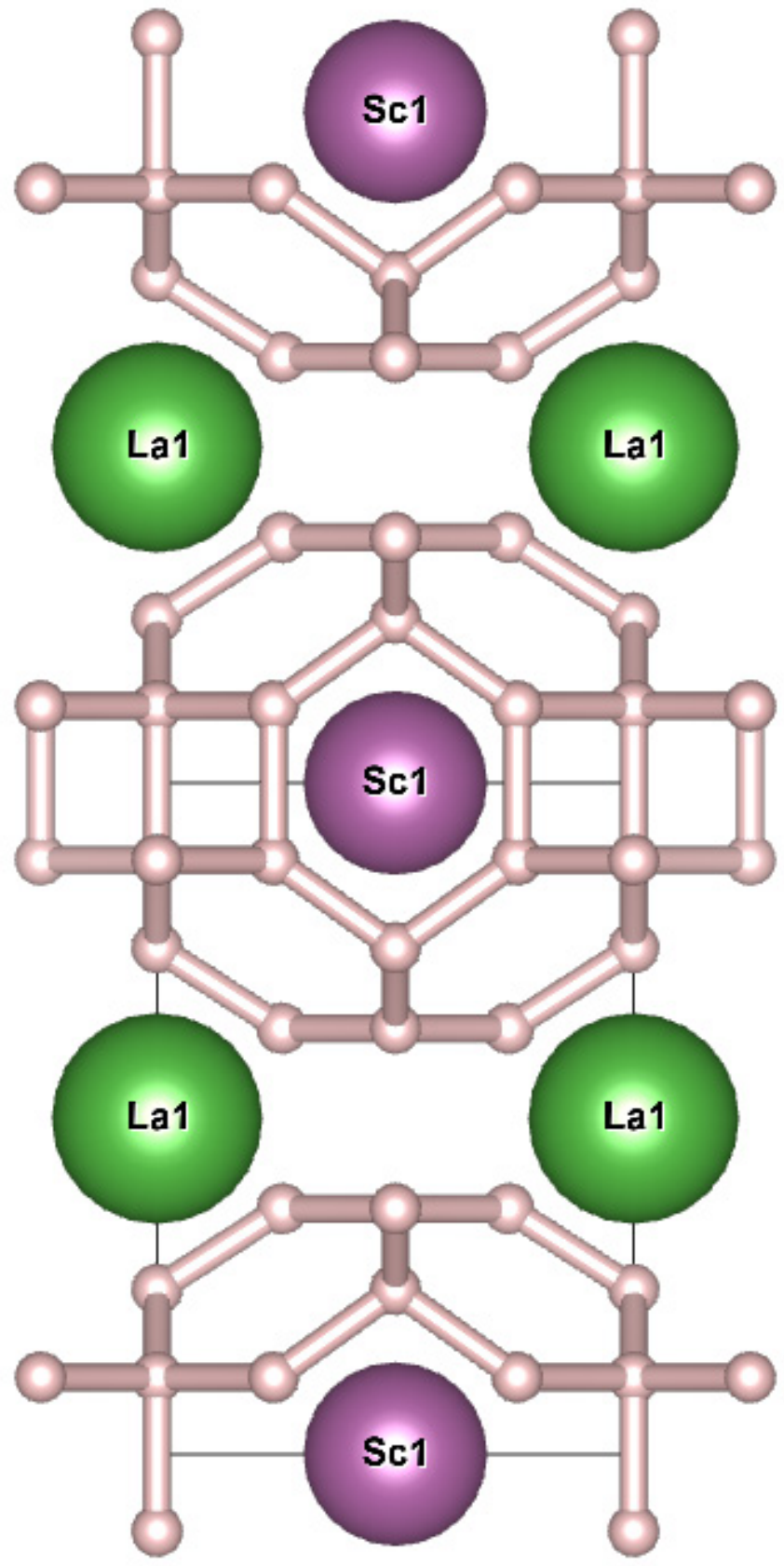<br><br>Space Group: P 4/m m m (#123-1)<br>a = 3.401 Å α = 90.0000°<br>b = 3.401 Å β = 90.0000°<br>c = 4.788 Å γ = 90.0000°<br>V = 55.41 Å$^3$ | data_findsym-output<br><br>_symmetry_space_group_name_H-M 'P 4/m 2/m 2/m'<br>_symmetry_Int_Tables_number 123<br><br>_cell_length_a 3.40189<br>_cell_length_b 3.40189<br>_cell_length_c 4.78800<br>_cell_angle_alpha 90.00000<br>_cell_angle_beta 90.00000<br>_cell_angle_gamma 90.00000<br><br>loop_<br>_space_group_symop_operation_xyz<br>x,y,z<br>x,-y,-z<br>-x,y,-z<br>-x,-y,z<br>-y,-x,-z<br>-y,x,z<br>y,-x,z<br>y,x,-z<br>-x,-y,-z<br>-x,y,z<br>x,-y,z<br>x,y,-z<br>y,x,z<br>y,-x,-z<br>-y,x,-z<br>-y,-x,z<br><br>loop_<br>_atom_site_label<br>_atom_site_type_symbol<br>_atom_site_fract_x<br>_atom_site_fract_y<br>_atom_site_fract_z<br>_atom_site_occupancy<br>La1 La 0.00000 0.00000 0.50000 1.00000<br>Sc1 Sc 0.50000 0.50000 0.00000 1.00000<br>H1 H -0.24448 0.00000 0.11461 1.00000<br>H2 H 0.26219 0.50000 -0.36558 1.00000<br>H3 H 0.00000 0.50000 0.24546 1.00000 |

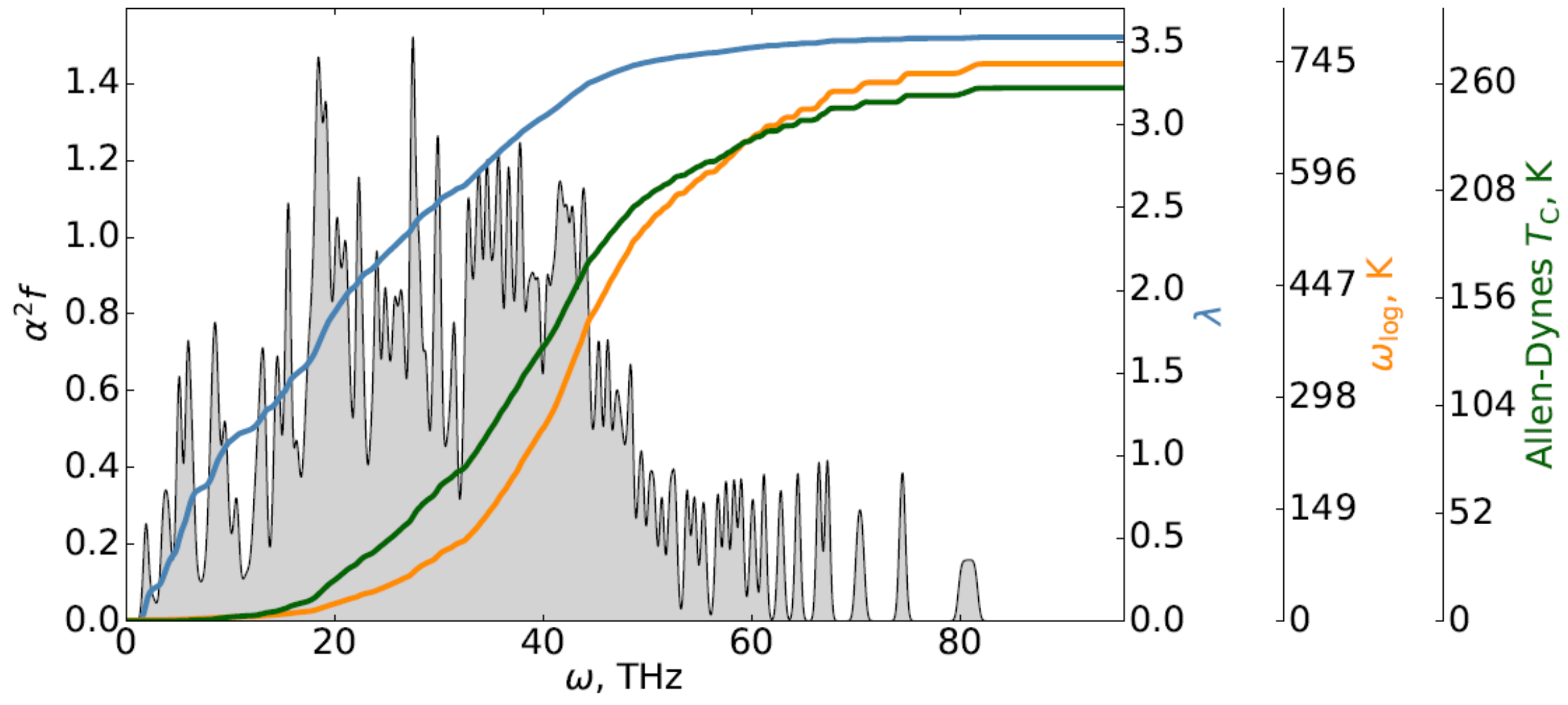

**Figure S32.** Eliashberg function, electron-phonon interaction parameter, logarithmically averaged frequency and critical temperature calculated using the Allen-Dean formula [27] with μ* = 0.1 for the *Pmmm*-$LaScH_{24}$ structure at 193 GPa. The

*q*-mesh density was 4 × 4 × 4, and the *k*-mesh density was chosen to be 12 × 12 × 12. To construct the graphs, we used the script proposed by G. Shutov [39].

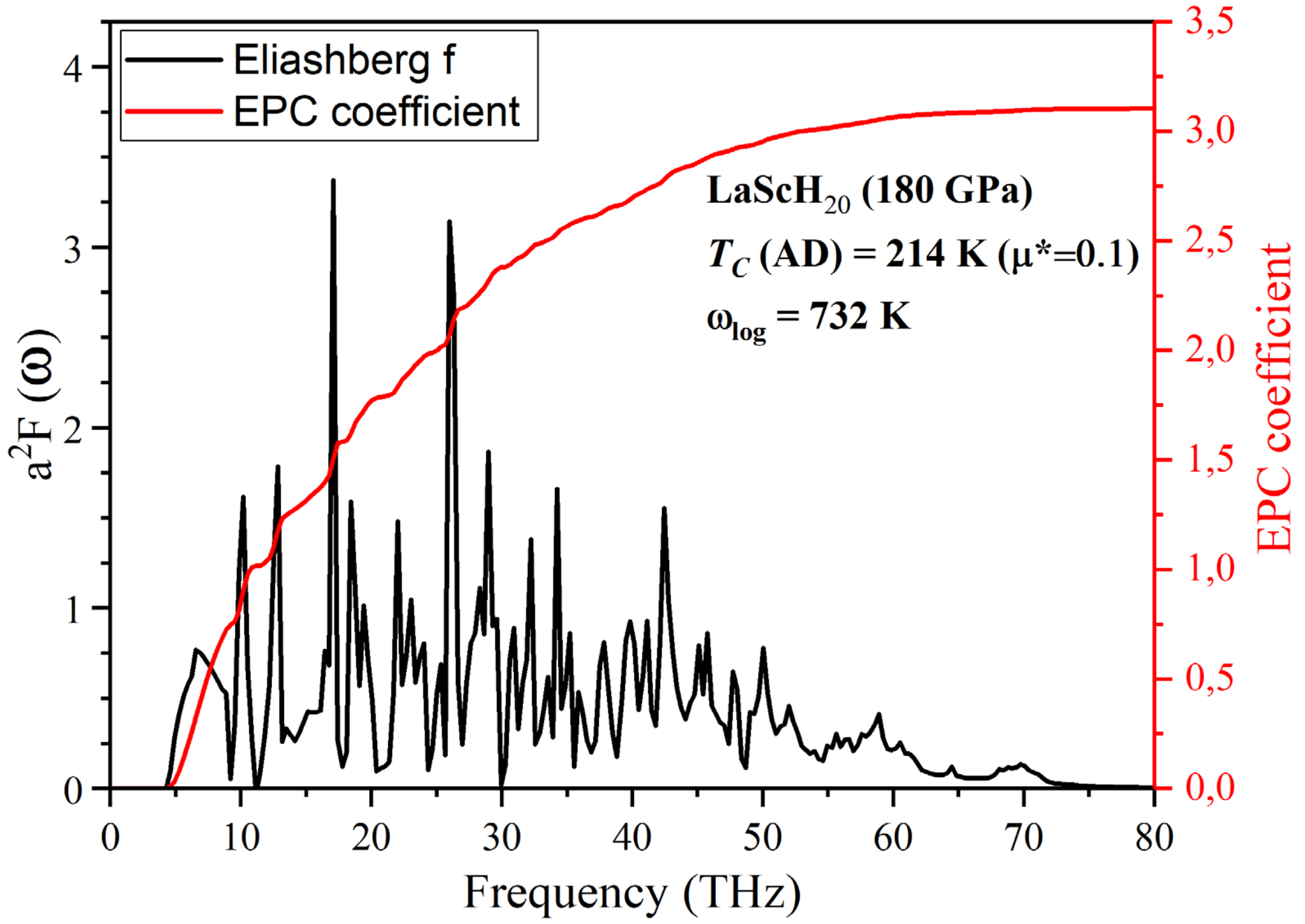

**Figure S33.** Eliashberg function, electron-phonon interaction parameter, logarithmically averaged frequency and critical temperature calculated using the Allen-Dean formula [27] with μ* = 0.1 for the *P*4/*mmm*-$LaScH_{20}$ structure at 180 GPa. The *q*-mesh density was 3 ×3 × 2, and the *k*-mesh density was chosen to be 12 × 12 × 12 for electronic wavefunctions and 18 × 18 × 18 for electron-phonon calculations.

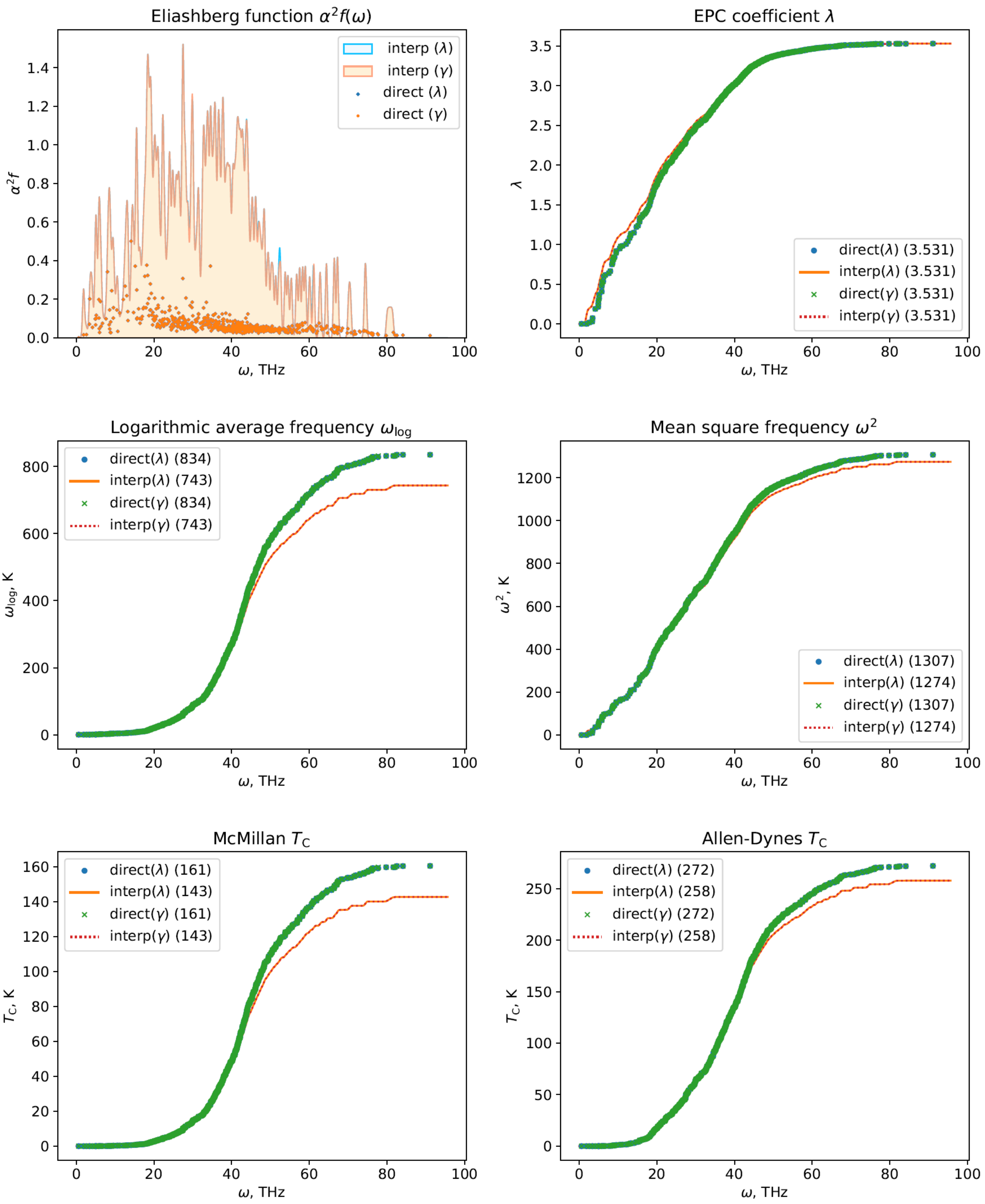

**Figure S34.** Convergence, Eliashberg and integral functions of frequency of various parameters of electron-phonon interaction in $LaScH_{24}$ at 193 GPa. Smoothing was used in the region ω = 0…3 THz. Coulomb pseudopotential is μ* = 0.1. To construct the graphs, a script written by G. Shutov [39] was used. Results are as follows: λ= 3.53, $\omega_{log}$ = 743-834 K, $\omega_2$ = 1274-1306 K, calculated Allen-Dynes $T_c$ = 272 K and the Eliashberg $T_c$ = 282 K.

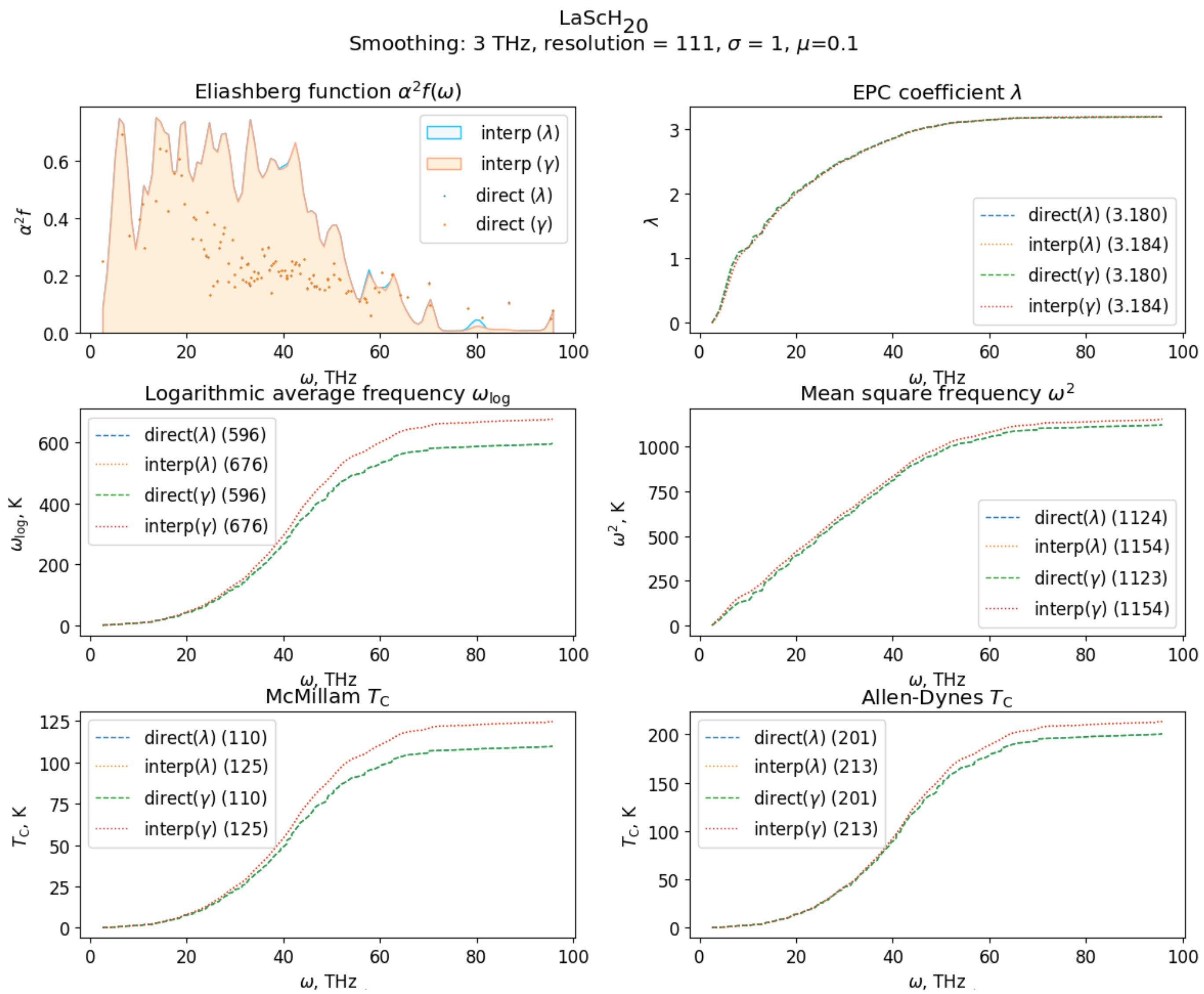

**Figure S35.** Convergence, Eliashberg and integral functions of frequency of various parameters of electron-phonon interaction in *P*4/*mmm*-$La_2Sc_2H_{40}$ at 200 GPa. Smoothing was used in the region $\omega < 3$ THz, Coulomb pseudopotential is $\mu^* = 0.1$. To construct the graphs, a script written by G. Shutov [39] was used. Results are as follows: $\lambda = 3.18$, $\omega_{\log} = 596\text{-}676$ K, $\omega_2 = 1123\text{-}1154$ K, calculated Allen-Dynes $T_c = 201$ K and the Eliashberg $T_c = 221$ K.